\documentclass[12pt, fullpage]{amsart}
\usepackage{amssymb}
\usepackage{amsmath}
\usepackage{amsfonts}
\usepackage{mathtools}
\usepackage{mathrsfs}
\usepackage{graphicx}
\usepackage{placeins}
\usepackage{color}
\usepackage[onehalfspacing]{setspace}
\usepackage{caption}
\AtBeginDocument{\captionsetup{width=\textwidth}}
\usepackage{subcaption}
\usepackage{natbib}
\usepackage{enumerate}
\usepackage[utf8]{inputenc}
\usepackage{tikz-cd}
\usepackage[charter,cal=cmcal]{mathdesign}
\usepackage{epigraph}
\usepackage[page,header]{appendix}
\usepackage{titletoc}
\usepackage{lipsum}
\usepackage{bibunits}
\usepackage{comment}
\usepackage[colorlinks=true,citecolor=blue,urlcolor=blue,pdfpagemode=UseNone,pdfstartview=FitH]{hyperref}

\newcommand{\eqdef}{\coloneqq}

\usepackage[
	textcolor=red,
	textsize=scriptsize,
	linecolor=red,
	bordercolor=red,
	backgroundcolor=white,
]{todonotes}

\makeatletter
\renewcommand{\@todonotes@useSizeCommand}{%
	\ifcsname \expandafter\string\@todonotes@sizecommand\endcsname
		\csname \expandafter\string\@todonotes@sizecommand\endcsname%
	\else
		\@todonotes@sizecommand
	\fi
	\baselineskip=\f@baselineskip\relax
}
\makeatother

\makeatletter
\renewenvironment{proof}[1][\proofname]{\par
	\normalfont \topsep6\p@\@plus6\p@\relax
	\trivlist
	\item[\hskip\labelsep
		\textbf{#1\@addpunct{:}}]\ignorespaces
}{%
	$\blacksquare$\endtrivlist\@endpefalse
}
\makeatother

\DeclareMathOperator\supp{supp}

\newcommand{\CI}{\mathrel{\perp\mspace{-10mu}\perp}}

\DeclareTextFontCommand{\bi}{%
	\fontseries\bfdefault 
	\itshape
}
\newcommand{\csdot}{\,\cdot\,}

\makeatletter
\def\section{\@startsection{section}{1}
	\z@{0.6\linespacing\@plus\linespacing}{.6\linespacing}{\Large}}

\def\subsection{\@startsection{subsection}{2}
	\z@{.5\linespacing\@plus.7\linespacing}{.5\linespacing}{\large}}

\def\subsubsection{\@startsection{subsubsection}{3}
	\z@{.5\linespacing\@plus.7\linespacing}{-.5em}{\normalfont\bfseries}}
\makeatother

\newtheorem{theorem}{Theorem}[section]
\newtheorem{proposition}{Proposition}[section]
\newtheorem{lemma}{Lemma}[section]
\newtheorem{corollary}{Corollary}[section]

\theoremstyle{definition}

\theoremstyle{definition}
\newtheorem{assumption}{Assumption}[section]

\theoremstyle{definition}

\theoremstyle{definition}
\newtheorem{examplex}{Example}[section]
\newenvironment{example}
  {\begin{examplex}}
  {$\square$\end{examplex}}

\theoremstyle{remark}

\calclayout

	\title{}
	
\begin{document}
			\vspace*{5ex minus 1ex}
		\begin{center}
			\Large \textsc{A Decomposition Approach to Counterfactual Analysis in Game-Theoretic Models}
			\bigskip
		\end{center}
		
		\date{%
			\today%
		}
		
		\vspace*{3ex minus 1ex}
		\begin{center}
			Nathan Canen and Kyungchul Song\\
			\textit{University of Warwick and CEPR, and University of British Columbia}\\
			\bigskip
		\end{center}
		
		\fontsize{12}{14} \selectfont

\begin{bibunit}[econometrica]		
\begin{abstract}
	Decomposition methods extrapolate the observed relationship between an outcome and an explanatory variable and are often used for counterfactual predictions in non-strategic settings. When the outcome arises from a game where agents gain by deviating from their strategies after a new policy, such predictions, though practically simple, are hard to justify. Our main result provides formal conditions in Bayesian games under which decomposition-based predictions coincide with equilibrium-based ones. This follows from an invariance condition on equilibrium selection rules, automatically satisfied in games with a unique equilibrium. We illustrate by revisiting \cite{Ciliberto/Tamer:09:Eca}'s empirical analysis of airline entry decisions.
\medskip

\medskip

\end{abstract}
\thanks{We thank Aimee Chin, Sukjin Han, Chinhui Juhn, Arvind Magesan, \'Aureo de Paula, Daniela Scur, Eduardo Souza-Rodrigues, Ko Sugiura, Andrea Szabo, Xun Tang, and participants at many seminars and conferences for their helpful comments. We are grateful to Elaheh Fatemipour for excellent research assistance. All errors are ours. Song acknowledges that this research was supported by the Social Sciences and Humanities Research Council of Canada.\\ Corresponding address: Kyungchul Song, Vancouver School of Economics, University of British Columbia, Vancouver, BC, Canada. Email address: kysong@mail.ubc.ca.}
\maketitle

\newpage

\section{Introduction}

One of the key aims of empirical research in economics is to predict the impact of new policies. Examples include studying how an increase in minimum wage affects labor market outcomes, how new healthcare legislation influences health outcomes, and how a change in market characteristics impacts firm entry. To provide counterfactual predictions in strategic environments, researchers typically specify a game-theoretic model and estimate structural parameters of the model, the results of which are used for generating post-policy predictions. This approach ensures that its predictions are based on agents' strategies that are incentive compatible, that is, agents have no incentive to deviate from the strategies yielding the post-policy predictions. However, it requires a detailed specification of the game (e.g., parametrization of payoffs or unobserved heterogeneity), often made for computational feasibility. Computational challenges frequently stem from multiple equilibria, which may result in identified sets too large for meaningful policy analysis.\footnote{See \cite{Eizenberg:14:ReStud}, Section 5 for a discussion on demand estimation, and \cite{Giacomini/Kitagawa:21:ECMA}, pp.1543-1547, on policy analysis in macroeconomic models. In some contexts, such large sets produce conflicting conclusions (see, e.g., \cite{Aguirregabiria/Nevo:12:WP}, p.111).} 

Decomposition methods offer an alternative approach for counterfactual analysis (\cite{Kitagawa:55:JASA}, \cite{Oaxaca:73:IER} and \cite{Blinder:73:JHR}). These methods extrapolate observed outcome-policy relationships to counterfactual scenarios and are common in labor economics. For example, when studying minimum wage effects on wage inequality, researchers estimate the relationship from current data, then apply these estimates to counterfactual minimum wages (e.g., \cite{DiNardo/Fortin/Lemieux:96:Eca}). This approach is computationally simple, provides point-identified predictions, requires less detailed specifications, and allows straightforward statistical inference through bootstrap. However, decomposition methods have primarily been used in non-strategic settings.

What prevents us from using decomposition methods for counterfactual analysis in a strategic setting? These methods assume that the causal relationship between the outcome variable and the policy variable remains the same in the counterfactual environment. This assumption fails when agents have incentives to deviate from the original equilibrium strategies under new policies.

This paper's main contribution is to provide conditions where an equilibrium-based prediction coincides with its decomposition-based counterpart in strategic settings. For a formal analysis, we consider a generic game, which includes games with various information structures, with the solution concept of Bayes Correlated Equilibria (BCE, hereafter) of \cite{Bergemann/Morris:16:TE} (which includes Nash equilibria and Bayes Nash equilibria as special cases). Throughout the paper, we examine policies that alter a publicly observed payoff component, taking as the prediction target the agents' responses or a known function of them. While this setting accommodates many policy analyses, it rules out interventions that change the set of players or their available strategies. We clarify the scope of our results later.

When the game admits a unique equilibrium, common in settings spanning auctions, social interactions, and networks,\footnote{See \cite{Guerre/Perrigne/Vuong:09:ECMA}, \cite{Calvo/Patacchini/Zenou:ReStud:09}, \cite{Blume/Brock/Durlauf/Jayaraman:15:JPE}, \cite{Xu:18:IER} and \cite{Eraslan/Tang:25:WP}.} sufficient conditions for the validity of decomposition-based predictions are: (a) the policy alters only a publicly observed payoff component and (b) the policy keeps the payoff state within its support in the pre-policy game. Condition (a) is satisfied by many policies, including those that affect taxes, tariffs or laws, which are often observed by all agents. Most of all, complete information games satisfy this condition immediately. Condition (b) is also satisfied in many empirical applications. Even when the condition is not met, we show below that we can obtain decomposition-based bounds for the equilibrium-based prediction. These bounds are generally informative and easy to estimate. 

The validity of decomposition-based predictions extends to games with multiple equilibria under one additional condition: the equilibrium selection rule at each payoff state must remain unchanged \textit{when the set of equilibrium actions at the same payoff state is unaffected by the policy.} This condition embodies the principle that the predictive content of a game lies entirely in the equilibrium action sets at each payoff state, leaving no separate role for the equilibrium selection rule. If the prediction at a payoff state were to change while its equilibrium action set at the payoff state stayed the same after the policy, the change must be attributed to the equilibrium selection rule, that is, to factors outside the game.

This invariance condition on the equilibrium selection rule is not new. It is implicitly invoked whenever one selects a specific equilibrium, such as a Pareto superior equilibrium or the most profitable one for a given firm \citep{Jia:08:Eca}, and assumes that the same selection persists after the policy \citep{Aguirregabiria/Mira:10:JOE}. It is likewise invoked whenever one adopts a parametrized equilibrium selection rule for counterfactual analysis \citep{Bajari/Hong/Ryan:10:Eca}. The condition is not innocuous, but it is reasonable for small policy variations.

The basic intuition behind our result may appear straightforward. Nevertheless, our contribution is novel, as no formal analysis of decomposition-based predictions in a general game-theoretic setting has previously appeared in the literature. The generality of our setting shows that the coincidence between equilibrium-based and decomposition-based predictions applies to both complete and incomplete information games, regardless of whether the solution concept is Bayes Nash equilibrium or correlated equilibrium.

Our results apply to many empirical settings, including counterfactual exercises in entry games where policies change market characteristics (e.g., \cite{Jia:08:Eca}) or regulatory policy (e.g., \cite{Ciliberto/Tamer:09:Eca}) and auction markets where changes in the reserve price affect bidding outcomes. We provide details on the validity of decomposition-based predictions in these examples later.

The decomposition approach requires weaker assumptions on the parametrization of payoff functions and is simpler to implement. In this respect it contrasts with the partial identification methods of \cite{Ciliberto/Tamer:09:Eca} and \cite{Haile/Tamer:03:JPE}, which remain fully agnostic about equilibrium selection rules, and even about certain features of the game structure, but lead to relatively complex procedures for inference on counterfactual predictions. Either way, our results help empiricists by clarifying which specification details of the game are needed, and which are not, for the simpler, decomposition-based counterfactual analyses.

We perform Monte Carlo simulations and revisit several empirical applications to illustrate the implications of our results. The simulations show that the bounds can remain informative even when the policy pushes the payoff state well beyond the support of the data. We then examine their robustness to violations of the invariance condition, allowing the equilibrium selection rule to change after the policy. Such violations can push the equilibrium-based prediction outside the bounds, though only slightly for small policy changes. For within-support predictions, likewise, the gap between the decomposition-based and equilibrium-based predictions widens as the failure of invariance becomes more severe.

As an illustration of our theoretical results, we revisit the empirical application of \cite{Ciliberto/Tamer:09:Eca}. Using a model of a complete information entry game, they studied the U.S. airline market and assessed the effects on entry from a repeal of the Wright Amendment, a law that restricted entry in routes using Dallas Love Field Airport. Faced with the multiplicity of equilibria, they adopted a partial identification approach and reported maximum entry probabilities as counterfactual predictions. In our application, we produce a decomposition-based prediction and compare it with the prediction from \cite{Ciliberto/Tamer:09:Eca}. We also compare these predictions with the actual results following the repeal of the Wright Amendment in 2014. We find that the decomposition-based predictions using the pre-policy data perform well for Southwest Airlines and United Airlines out-of-sample. 

As a second empirical application, presented in the Online Appendix, we follow \cite{Goolsbee/Syverson:08:QJE} and revisit how competitors respond to a shock that lowers Southwest's cost of operating a given market. Our focus, however, is on the shock's effect on competitors' entry rather than on prices. We find that, on average, other airlines' entry behavior is consistent with deterrence. This effect is heterogeneous across markets, however, and is driven primarily by small markets. In large markets, competitors appear to ``accommodate'' Southwest, leaving their entry behavior unchanged when its cost of operating the market falls. This finding is consistent with evidence from other industries (e.g., \cite{Tenn/Wendling:14:ReStat}).\medskip

\noindent \textbf{Related Literature} The idea that we may not need to identify a full structural model to recover policy effects of interest goes back at least to \cite{Marschak:53:Cowles}, as pointed out by \cite{Heckman:10:JEL}. See \cite{Wolpin:13} for further examples and discussion. In a similar vein, \cite{Sims:80:Econometrica,Sims:82:Brookings} argued that policy analysis can be carried out using reduced-form models, without recovering the behavioral parameters of agents in the economy. Recent approaches exploring similar insights include the sufficient statistics approach (most notably in public finance, e.g., \cite{Chetty:09:ARE}; see \cite{Kleven:20:WP} for a recent overview) and policy relevant treatment effects proposed by \cite{Ichimura/Taber:00:NBER} and \cite{Heckman/Vytlacil:01:AERPP}. See \cite{Heckman:10:JEL} for a review of this literature. We convey a similar message in this paper by studying conditions for the validity of decomposition-based predictions in strategic settings.

The decomposition approach is widely used in economics, most notably in labor economics: the researcher estimates a model and then decomposes the variation in the outcome into the effects of different covariates. One example is the study of the effects of minimum wages on wage inequality (e.g., \cite{DiNardo/Fortin/Lemieux:96:Eca}): while there might be other mechanisms that affect wage inequality (e.g., unionization), once the model has been estimated, we can check how the distribution of wages would have differed if minimum wages had increased. Salient examples of the decomposition approach include the early work of \cite{Oaxaca:73:IER} and \cite{Blinder:73:JHR}, \cite{Juhn/Murphy/Pierce:93:JPE}, the nonparametric/semiparametric approach of \cite{DiNardo/Fortin/Lemieux:96:Eca}, and other extensions that have been very widely applied. See \cite{Fortin/Lemieux/Firpo:11:Handbook} for a survey. The connection between the decomposition methods and methods of program evaluation has been pointed out by \cite{Fortin/Lemieux/Firpo:11:Handbook}. \cite{Kline:11:AERPP} shows that the estimated decomposition-based prediction can be interpreted as a reweighting estimator in the program evaluation setting.

More relevant to our paper is the decomposition approach used for counterfactual policy predictions. \cite{Rothe:10:JoE} and \cite{Chernozhukov/Fernandez/Melly:13:Eca} provided inference on the full counterfactual distributions. \cite{Hsu/Lai/Lieli:22:JBES} introduced the quantile counterfactual treatment effects on a different population, and showed how we can identify and perform inference on the treatment effects using an invariance condition. While the main emphasis of this literature is on the general method of inference on various counterfactual distributions, our focus is on presenting a generic set of conditions in game-theoretic models under which such decomposition-based predictions can be accepted as valid predictions.

A common way to generate counterfactual predictions in a strategic environment is to specify and estimate a game-theoretic model, and then use the predictions from its equilibria. In many cases, point-identification of these models is not possible due to multiple equilibria. Researchers often either choose one equilibrium from the game (e.g., \cite{Jia:08:Eca}) or conduct inference on the identified set allowing for all equilibria (e.g., \cite{Ciliberto/Tamer:09:Eca}). In light of these challenges, a recent literature studies point-identification of counterfactual predictions without identifying the full details of the model. These developments have been centered around dynamic discrete choice models (see, e.g., \cite{Aguirregabiria/Suzuki:14:QME}, \cite{Norets/Tang:14:ReStud}, \cite{Arcidiacono/Miller:20:JOE}, \cite{Kalouptsidi/Scott/Souza:20:QE}). While many structural models within this class are unidentified (see the discussion in \cite{Aguirregabiria/Suzuki:14:QME}), some counterfactuals are point-identified such as those characterized by linear changes in payoffs (the so-called ``additive transfers counterfactuals'' in \cite{Kalouptsidi/Scott/Souza:17:IJIO}). See also \cite{Kalouptsidi:Kitamura:Lima:Souza:20:NBER} for partial identification of counterfactuals in a similar context. \cite{Jun/Pinkse:20:JOE} explored various approaches to produce a point-decision on partially identified counterfactual predictions from a game. Recently, \cite{Gu/Russell/Stringham:21:WP} found a useful characterization of an identified set for counterfactual predictions in a general class of discrete-outcome models. 

The message in this paper is related to that of \cite{Kocherlakota:19:JME}. Using a model of a dynamic game between the private sector and the government, he showed that to obtain an optimal policy, we can just use predictions by regressing policy payoffs on policies using historical data, without relying on a structural macroeconomic model. There are major differences between his framework and ours. First, his model is a dynamic model where the policy-maker is a player of the game, whereas ours is a static one in which the policy-maker is outside the model. Second, his model is a macroeconomic model where the analysis is based on one equilibrium that generated the data. In our setting, by contrast, we observe many independent copies of a static game and must confront the multiplicity of equilibria.

We use the solution concept of BCE due to its generality, which makes it easy to demonstrate that the decomposition method is not limited to a specific equilibrium concept. BCE has recently gained interest as a solution concept in an empirical setting due to its robustness properties and computational tractability. \cite{Magnolfi/Roncoroni:22:ReStud} adopt it in their study of entry decisions in the supermarket sector. They focus on characterizing the identified set obtained under a weak assumption on the information structure of the game, and use the set to study a policy that changes a covariate (presence of large malls).   Other empirical examples using BCE and obtaining partial identification include \cite{Gualdani/Sinha:20:WP} on discrete choice models and \cite{Syrgkanis/Tamer/Ziani:18:WP} on auctions. \cite{Han/Kaido/Magnolfi:26:WP} develop testable implications for the ordering of information structures in Bayesian games, building on the connection, established by \cite{Bergemann/Morris:16:TE}, between this ordering and BCE-based predictions. Finally, \cite{Bergemann/Brooks/Morris:22:AER} present a general framework to produce a set of counterfactual predictions in games exploiting the requirement that observations be consistent with some unknown information structure of the games. Their counterfactual policies are concerned with a change of payoff functions or the set of action profiles, whereas in our setting, the policies are mainly a change in the distribution of the payoff state vector. More importantly, our emphasis is on decomposition-based predictions and the conditions that justify their use for policy analysis in game-theoretic settings.

This paper is organized as follows. Section 2 provides an overview of our main results using a simple example. Section 3 presents the formal results for generic Bayesian games along with discussions, extensions and examples. Sections 4 and 5 contain the Monte Carlo simulations and the empirical application, respectively. Section 6 concludes. Mathematical proofs, generalizations, data details, and implementation are in the Online Appendix.

\section{Overview with a Simple Model}

In this section, we provide an overview of our main results in the context of a simple, complete information, two-player entry game model, which has been widely used in the empirical literature (e.g., \cite{Ciliberto/Tamer:09:Eca}, \cite{Jia:08:Eca}, \cite{Grieco:14:RAND} and \cite{Magnolfi/Roncoroni:22:ReStud}).
  
\subsection{Counterfactual Analysis from a Two-Player Entry Game}
\subsubsection{Two-Player Entry Game}
\label{subsubsec:simple1}
Consider a complete information game where two firms, $i=1,2$, choose a binary action $y_i \in \{0,1\}$, $y_i = 1$ representing entry in the market, and $y_i=0$ staying out of the market. The payoff for firm $i$ who faces the action profile $y = (y_1,y_2)$ is given by:
\begin{align}
	\label{utility_entry}
	u_i(y,X_i,\varepsilon_i),
\end{align}
where $(X_i,\varepsilon_i)$ is the payoff state of firm $i$, with $X_i$ and $\varepsilon_i$ denoting the characteristics of firm $i$ that are observable and unobservable to the econometrician. We let $X = (X_1,X_2)$ and $\varepsilon = (\varepsilon_1,\varepsilon_2)$, and assume for illustration purposes that $X$ and $\varepsilon$ are independent (as commonly assumed in empirical applications). We denote the set of pure strategy Nash equilibria by $\mathcal{E}$, and assume for simplicity that $\mathcal{E}$ is a finite set.

As in many empirical models used in the literature, we may consider further specifying the payoff function parametrically:
\begin{align}
	\label{util param}
	u_i(y,X_i,\varepsilon_i) = y_i(-\delta y_{-i} + X_i'\beta_i + \varepsilon_i),
\end{align}
where $\delta>0$ is a parameter measuring the effect of competition and $\beta_i$ a coefficient vector for the observed covariates. For illustration, consider a setting where $0 < X_i'\beta_i + \varepsilon_i < \delta$, so that competition in a market is detrimental to firms.\footnote{We use this constraint solely for the purpose of simplifying the exposition.} Then, there are two pure strategy Nash equilibria in this game, characterized by one firm entering, but the other staying out. More specifically, the set of pure strategy equilibria is given by
\begin{align*}
	\mathcal{E}= \{g^1, g^2\},
\end{align*}
where $w = (x,\varepsilon)$, $x = (x_1,x_2)$, $\varepsilon = (\varepsilon_1,\varepsilon_2)$, $g^1(w) = (1- 1_{A_{0,1}}(w), 1_{A_{0,1}}(w))$, and $g^2(w) = (1_{A_{1,0}}(w),1-1_{A_{1,0}}(w))$, with
\begin{align*}
	A_{0,1} &= \{(x,\varepsilon) : x_1'\beta_1 + \varepsilon_1 - \delta < 0, x_2'\beta_2 + \varepsilon_2 \ge 0\}, \text{ and }\\
	A_{1,0} &= \{(x,\varepsilon) : x_1'\beta_1 + \varepsilon_1 \ge 0, x_2'\beta_2 + \varepsilon_2 - \delta < 0\}.
\end{align*}
Hence each member $g$ of $\mathcal{E}$ is a function of $w$, and gives the prediction of the actions of the two firms when the payoff state is $w$.
\medskip

\subsubsection{Data Generation}

As often done in the empirical literature, we assume a setting with many markets where the data are generated from the entry game independently across the markets. Let $Y = (Y_1,Y_2)$ be the pair of entry indicators by firms $1$ and $2$, and define $W = (X, \varepsilon)$, the payoff state of the game. The econometrician observes $Y$ and $X$, but not $\varepsilon$. The data generation is described as follows.\medskip

Step 1: The payoff state $W$ is drawn (i.i.d.\ across the markets) from a common distribution.

Step 2: The outcome $Y$ is generated i.i.d.\ across the markets from the conditional probability:
\begin{align*}
	P\{Y = (y_1,y_2) \mid W\} = \sum_{g \in \mathcal{E}} 1\{g(W) = (y_1,y_2)\}e(g \mid W), 
\end{align*}
where $e(\csdot \mid W)$ denotes the equilibrium selection rule for given $W$, which is a conditional probability mass function on $\mathcal{E}$ given $W$.\medskip 

We introduce the equilibrium selection rule in this generic form solely to capture how a single realization of $(Y,X)$ is observed for each market, even when the game admits multiple equilibria and so predicts several possible outcomes.

\subsubsection{Counterfactual Predictions}

Our focus is on predicting the entry probability for the firms when a policy changes the payoff state $(X,\varepsilon)$ into $(f(X),\varepsilon)$, for some map $f$. Such policies have often been studied in the literature. For example, \cite{Ciliberto/Tamer:09:Eca} studied a change in an observable policy (modeled as a dummy variable in the payoff state), \cite{Jia:08:Eca} a change in the market size (i.e., increasing the population size by 10\%), \cite{Grieco:14:RAND} a change in the presence of a supercenter, and \cite{Magnolfi/Roncoroni:22:ReStud} a decrease in the presence of large malls.\footnote{While our primary focus is on policies that change the payoff state vector $X$, some situations with a policy that changes a parameter of the model fall within our framework. For example, suppose that the payoff state of firm $i$ takes the form $(X_{i,1} \beta_i + X_{i,2}'\theta_i, \varepsilon_i)$, $X_i = (X_{i,1}, X_{i,2})$, and the policy changes $\beta_i$ to be zero, or $1/2$. This policy is equivalent to the policy that fixes the value of the corresponding entry of $X_{i,1}$ to be zero, or the half of its value, with the coefficient $\beta_i$ kept the same.}

Given policy $f$, our target parameter is what we call the \bi{average equilibrium-based prediction (AEP)} of entry defined as follows:
\begin{align}
	\label{AEP def}
	\mathsf{AEP} \eqdef  \sum_{g \in \mathcal{E}_f}  \mathbf{E}\left[g(f(X),\varepsilon) e_f(g \mid f(X),\varepsilon) \right],
\end{align}
where $e_f$ denotes the equilibrium selection rule unknown to the econometrician, $\mathcal{E}_f$ the set of equilibria, both after the policy $f$, and the expectation $\mathbf{E}$ is over the joint distribution of $(X,\varepsilon)$. The AEP corresponds to the average entry probabilities of the two firms as predicted by the entry game model after the (observable) payoff state is changed from $X$ into $f(X)$.\footnote{One could consider other parameters of interest, including any (known) function of outcomes. See Section \ref{sec3}.}

It is usually difficult to estimate this prediction because we first need to recover the post-policy equilibria, $\mathcal{E}_f$, and the equilibrium selection rule, $e_f$, from data. Without any assumption about the equilibrium selection rule, the members of $\mathcal{E}_f$ are typically set-identified, since the econometrician cannot infer whether an equilibrium is played due to the payoff parameters assuming a certain value, or due to an equilibrium selection rule involved in the data generation. This further complicates the implementation of this prediction in practice.

\subsection{Overview of the Main Results}\label{subsec:main_results}
\subsubsection{The Identification of the Average Equilibrium-Based Prediction}
\label{subsubsec: suff conds for validity}
In this context of a complete information entry game, we give a general result which shows that, under an invariance condition for equilibrium selection rules and a support condition, the AEP is equal to a quantity which can be identified from data without specifying the details of the game. 

We introduce some notation. First, we define what we call the \bi{average decomposition-based prediction (ADP):} 
\begin{align}
	\label{ADP def}
	\mathsf{ADP} \eqdef \sum_{g \in \mathcal{E}} \mathbf{E}\left[ g(f(X),\varepsilon)e(g \mid f(X), \varepsilon)\right].
\end{align}
The ADP uses the pre-policy game to predict the average outcome of the policy. 
Because $X$ and $\varepsilon$ are independent, if $f(X)$ stays in the support of $X$ in the data, the ADP is point-identified as follows:
\begin{align}
	\label{ADP = E[mf(X)]}
	\mathsf{ADP} = \mathbf{E}\left[ m(f(X)) \right],
\end{align}
where $m(x) = \mathbf{E}[Y \mid X = x]$.\footnote{We assume that $f(X)$ and $X$ are mutually absolutely continuous on their common supports so that the ADP is well-defined. The assumption allows for a wide range of policies, including setting a discrete component of $X$ to zero when zero belongs to its support, or adding an amount to a continuously distributed component so that the post-policy component remains continuously distributed, as with a tax or a subsidy. However, the condition excludes policies of pinning or censoring a continuously distributed payoff state component at a point in the interior of its support.} This conditional expectation is based on the joint distribution of $(Y,X)$ from the pre-policy game. Therefore, the prediction is obtained by simply applying the pre-policy relation between $Y$ and $X$ to the counterfactual environment. The practical advantage of using the ADP is clear: we do not need to recover the set of equilibria or the equilibrium selection rule from data for the counterfactual prediction. The quantity is point-identified, and statistical inference on it is simple.

However, is the ADP a valid prediction in this setting? Our main result shows that under a set of conditions, we have the following identification:
\begin{align}
	\label{ADP = AEP}
 \mathsf{AEP} =	\mathsf{ADP}.
\end{align}
(See Section \ref{example1_detail} below for a formal result in entry games.) Two things rely only on the nonparametric causal structure of the game in (\ref{utility_entry}): the definitions of AEP and ADP, and the identification of ADP as (\ref{ADP = E[mf(X)]}). Neither requires specifying details of the game, such as the parametric payoff specification in (\ref{util param}).

For the equality (\ref{ADP = AEP}), we present two main conditions. First, the variable subject to a change due to a policy should belong to public information of all the players. This is satisfied by the complete information game considered here. Second, the equilibrium selection rules should satisfy an appropriate invariance condition. We will explain this condition below in detail. 

To see how these conditions lead to (\ref{ADP = AEP}), we focus on the policies $f$ such that $0 < f(X_i)'\beta_i + \varepsilon_i < \delta$ for simplicity. Then, the two conditions are written as follows:\medskip

(a) (Public Information Policy) It is common knowledge that the players observe $(X,\varepsilon)$.

(b) (Invariance of Sliced Equilibrium Selection Rules) If the set of equilibrium actions remains the same at a payoff state $w$ after the policy, then the selection probabilities of equilibrium actions at $w$ remain the same as well.\medskip

The public information condition (a), combined with independence between $X$ and $\varepsilon$, ensures that the posterior beliefs of the players remain invariant to the policy. As a result, for each $w$, the set of equilibrium actions at $w$ remains the same. The invariance condition (b) says that in such a case, the equilibrium selection probabilities at $w$ remain the same as well. Therefore, comparing (\ref{AEP def}) with (\ref{ADP def}) gives the equality (\ref{ADP = AEP}).

\subsubsection{Invariance of Sliced Equilibrium Selection Rules}
\label{subsubsec: invariance ESR overview}
One key condition for the equality in (\ref{ADP = AEP}) is that the equilibrium selection rule satisfies an invariance property (b). First, with $w = (x,\varepsilon)$, we let 
\begin{align*}
	\mathcal{E}(w) = \{g(w): g \in \mathcal{E}\}.
\end{align*}  
Hence, $\mathcal{E}(w)$ represents the $w$-slice of $\mathcal{E}$ before the policy, which is the set of actions supported by an equilibrium when the payoff state $W$ is fixed to be $w$ in the pre-policy game. Let $e_f(\csdot\mid w)$ be the equilibrium selection rule of the game at the payoff state $w$ after the policy $f$. Similarly, define the set of post-policy equilibria to be $\mathcal{E}_f$, and its $w$-slice $\mathcal{E}_{f}(w)$. Then, the invariance condition is stated as follows:\medskip

\noindent \textsc{Invariance Condition: } For any $w$ such that $\mathcal{E}(w) = \mathcal{E}_{f}(w)$, we have  
\begin{align}
	\label{invariance cond22}
	e(\csdot\mid w) = e_{f}(\csdot\mid w).
\end{align}
 
The invariance condition requires that, if the equilibrium action profiles remain the same after the policy \textit{at the same payoff state $w$}, their selection probability remains the same as well.

The invariance condition is already satisfied by many equilibrium selection rules used in the literature on empirical research. This includes the common assumption that the same equilibrium is played in the data and in the counterfactual (see \cite{Aguirregabiria/Mira:10:JOE}). Relatedly, many papers choose a specific equilibrium which is analyzed in both the data and the counterfactual. This includes the highest profit equilibrium for a certain firm (\cite{Jia:08:Eca}), the Pareto superior equilibrium (see \cite{DePaula:13:ARE}), or the equilibrium studied by \cite{Milgrom/Weber:82:ECMA} in common value auctions. Furthermore, this invariance condition already holds when an equilibrium selection rule is kept the same for counterfactual predictions (e.g., in \cite{Bajari/Hong/Ryan:10:Eca}, or as assumed in \cite{Berry/Haile:14:Ecma}). It is also satisfied if it is deemed constant across the equilibrium in the data and counterfactual (e.g., \cite{Aguirregabiria:12:EL}, applied in \cite{Aguirregabiria/Ho:12:JOE}). See also \cite{Aguirregabiria/Mira:13:WP} for a related result.\footnote{The modeling of the equilibrium selection rule differs slightly between \cite{Aguirregabiria/Mira:13:WP} and this paper. Their invariance condition requires the equilibrium selection rule to depend on $(w,\theta)$ only through the payoff function evaluated at $(w,\theta)$, whereas ours concerns a selection rule determined by the payoff state $w$ alone.}

Sometimes the invariance condition may be reasonable in a setting with small changes to the payoff state (e.g., small increases in minimum wages), because small changes can be well approximated by keeping the same game environment. Many of the policy experiments in entry games in empirical research are ``small policies'' which change payoff states (or the set of affected players) by only a small amount, such as 10-15\% changes in observed policy variables (as opposed to, say, doubling). For example, \cite{Jia:08:Eca} considered the increase in market size by a small amount, 10\%, and \cite{Magnolfi/Roncoroni:22:ReStud} considered a change in a market characteristic affecting only 13\% of markets. A similar reliance on small policy changes underlies \cite{Aguirregabiria:12:EL} and \cite{Aguirregabiria/Ho:12:JOE}, who approximate counterfactual outcomes locally, around the values from the pre-policy regime. 

\subsubsection{Decomposition-Based Bounds}

The identification (\ref{ADP = E[mf(X)]}) requires that 
\begin{align*}
	f(X) \in \mathbb{S}_X \eqdef \supp(X),
\end{align*}
with probability one. When this condition fails, we can employ the bound approach of \cite{Manski:89:JHR}. More specifically, define the $2 \times 1$ vector
\begin{align*}
	\mathsf{ADP}_R &\eqdef \sum_{g \in \mathcal{E}} \mathbf{E}\left[ g(f(X),\varepsilon)e(g \mid f(X), \varepsilon)1 \{f(X) \in \mathbb{S}_X \}\right].
\end{align*}
Then, with $\mathbf{1}$ denoting the $2 \times 1$ vector of ones and all inequalities holding pointwise,
\begin{align*}
	\mathsf{ADP}_R
	&\le \sum_{g \in \mathcal{E}} \mathbf{E}\left[ g(f(X),\varepsilon)e(g \mid f(X), \varepsilon)\right]\\
	&\le \mathsf{ADP}_R + P\left\{f(X) \notin \mathbb{S}_X \right\} \mathbf{1}.
\end{align*}
Thus, if we let $\Delta \eqdef P\left\{f(X) \notin \mathbb{S}_X \right\} \mathbf{1}$, we find that
\begin{align*}
	\mathsf{ADP}_{R} \le \mathsf{ADP} \le \mathsf{ADP}_{R} + \Delta.
\end{align*}
From (\ref{ADP = AEP}), this yields the following bounds for the AEP:
\begin{align}
	\label{bound2}
	\mathsf{ADP}_{R} \le  \mathsf{AEP} \le \mathsf{ADP}_R + \Delta.
\end{align}
Note that $\mathsf{ADP}_R$ is $\mathsf{ADP}$ with the restriction that $f(X) \in \mathbb{S}_X$:
\begin{align*}
	\mathsf{ADP}_R = \mathbf{E}\left[ m(f(X))1 \{f(X) \in \mathbb{S}_X \}\right].
\end{align*}
In other words, $\mathsf{ADP}_R$ in the bounds is obtained from the regression function $m(x) = \mathbf{E}[Y \mid X = x]$ without recovering the payoff functions or the equilibrium selection rule. 

\begin{figure}[t]
	\begin{center}
		\includegraphics[scale=0.42]{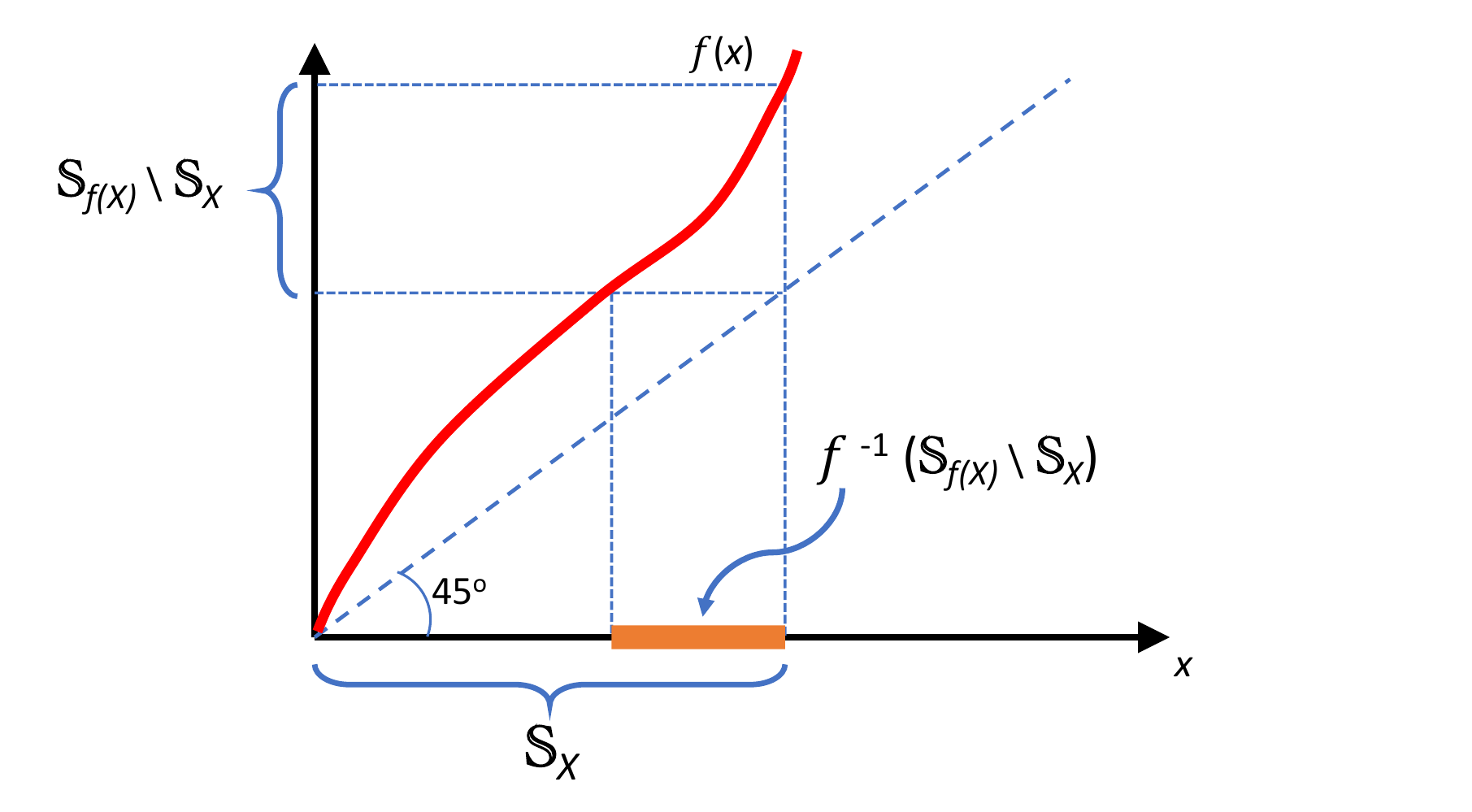}
		\caption{The Size of $\Delta$}
		\medskip
		\parbox{\textwidth}{\footnotesize \textit{Notes}: $\Delta = P\{X \in f^{-1}(\mathbb{S}_{f(X)} \setminus \mathbb{S}_X)\}$ where $\mathbb{S}_{f(X)}$ denotes the support of $f(X)$. The figure illustrates the derivation of the set $f^{-1}(\mathbb{S}_{f(X)} \setminus \mathbb{S}_X)$. When the set $\mathbb{S}_{f(X)} \setminus \mathbb{S}_X$ becomes larger, the term $\Delta$ increases.}		
		\label{fig: eeb}
	\end{center}
\end{figure}

The length of the interval in (\ref{bound2}) depends on the size of $\Delta$.\footnote{When $X$ has a known support, we can check $P\{f(X) \notin \mathbb{S}_X\} > 0$ simply by seeing whether any observations satisfy $f(X_i) \notin \mathbb{S}_X$. When $X$ has an unknown support, the support must first be estimated. Under regularity conditions, bootstrap critical values yield an asymptotically valid test. As the arguments are standard, we omit the details.} The term represents the probability that the policy sends the payoff state outside the support of the payoff state in the pre-policy game. It is identified and can be estimated because $X$ is observable. It tends to increase as the policy $f$ transforms the payoff state further away from the support of the payoff state in the pre-policy game, as illustrated in Figure \ref{fig: eeb}.

\begin{example}\label{example}
As an illustration, consider the policy which changes $X = (X_1,X_2)$ into
\begin{align*}
	f(X) = (X_1 + a_1, X_2 + a_2), \quad a_1, a_2 >0.
\end{align*}
Suppose that the pre-policy support of $X = (X_1,X_2)$ is given by the product $[x_{L,1}, x_{U,1}] \times [x_{L,2}, x_{U,2}]$, so that $f(X) \in \mathbb{S}_X$ if and only if $X_i \in [x_{L,i} - a_i, x_{U,i} - a_i]$ for each $i = 1,2$. We define
\begin{align*}
	m_i(x_1,x_2) = \mathbf{E}[Y_i \mid (X_1,X_2) = (x_1,x_2)], \quad i = 1,2.
\end{align*}
Then, we have
\begin{align}
	\mathsf{ADP}_R = \left(\begin{array}{c}
		\mathbf{E}\left[ m_1(X_1 + a_1, X_2 + a_2) 1\left\{ X_i \in [x_{L,i} - a_i, x_{U,i} - a_i], \text{ for all } i = 1,2\right\}\right]\\
		\mathbf{E}\left[ m_2(X_1 + a_1, X_2 + a_2) 1\left\{ X_i \in [x_{L,i} - a_i, x_{U,i} - a_i], \text{ for all } i = 1,2\right\}\right]
	\end{array}
	\right)\label{ADPR_example}
\end{align}
and
\begin{align*}
	\Delta &= P\left\{ X_1 \notin [x_{L,1} - a_1, x_{U,1} - a_1] \text{ or } X_2 \notin [x_{L,2} - a_2, x_{U,2} - a_2] \right\}\\ \notag
	&=P\left\{ x_{U,1} - a_1 < X_1 \text{ or } x_{U,2} - a_2 < X_2 \right\}.
\end{align*}
Since the probability in $\Delta$ is with respect to the pre-policy distribution of $X = (X_1,X_2)$, one can easily see that the term $\Delta$ increases as $a_1$ or $a_2$ increases.
\end{example}

\subsubsection{Extensions and Scope}

So far our overview presented the main result in a two-player complete information entry game where the solution concept is pure-strategy Nash equilibrium and the policy affects an observable payoff state. How far does this result extend beyond this setting?

The validity of this result extends to incomplete information games, and beyond the solution concept of pure strategy Nash equilibria. We demonstrate this in the next section by considering a generic Bayesian game which includes complete information games and private information games as special cases, and by adopting the general solution concept of Bayes Correlated Equilibria (BCE) of \cite{Bergemann/Morris:16:TE}. This solution concept is quite flexible as it includes mixed strategy Nash equilibria or, more generally, correlated equilibria as special cases.

Furthermore, the result extends to a situation where the outcome variable is not an action, but a measure of aggregate welfare or profit, as long as the latter is identified from the distribution of observed actions. It further extends to cases where $X$ and $\varepsilon$ are potentially correlated, but a control function exists.

While this paper restricts counterfactual policies to those that affect only the observed payoff state, we can extend our framework to settings where the policy changes the payoff parameters, using the notion of a \textit{policy component} in \cite{Canen/Song:25:arXiv}. For example, if the policy sets the coefficient $\beta_i$ of $X_{i,1}$ in the payoff function to zero (i.e., shutting off the effect of $X_{i,1}$ on the payoff), and if the coefficient $\beta_i$ is identified from pre-policy data, then the results of our paper carry over to this setting. As this extension is straightforward, we omit it for simplicity.

However, there are important cases that our results do not cover. They do not generally apply when the counterfactual policy alters some fundamental aspects of the environment such as changing the action space or the set of (possible) players (e.g., mergers). Examples of the latter include \cite{Hortacsu/McAdams:10:JPE}, who study how different auction formats for selling treasuries affect bidders' expected surplus, and \cite{Roberts/Sweeting:13:AER}, who study how changing a mechanism (from an auction to a sequential bidding design) affects expected revenues and payoffs in an incomplete information game. Similarly, our results do not apply to settings in which the policy changes the agents' strategy spaces. For example, \cite{Keane/Wolpin:10:IER} estimate a model in which women choose labor supply, fertility, and welfare participation, among other outcomes. In one of their counterfactual experiments, they consider a policy that eliminates a welfare program, and with it the agents' option to participate in it, to study the program's effects on labor supply across racial groups. The decomposition method does not apply here, because the probability with which agents choose the eliminated actions in the data cannot be extrapolated to the counterfactual environment. See \cite{Kalouptsidi/Scott/Souza:20:QE} for further examples and identification results for policies that change agents' choice sets.

\section{Counterfactual Analysis Using Game-Theoretic Models\label{sec3}}

\subsection{A Finite-Player Bayesian Game}
\label{subsec: a finite-player bayesian game}

We present our main results in a general set-up of strategic interactions. We first introduce a finite-player Bayesian game following \cite{Bergemann/Morris:16:TE} (BM, hereafter).\footnote{\cite{Bergemann/Morris:16:TE} considered the case where the state space and the signal space are finite sets for simplicity. In this paper, we consider more general spaces for the state and signal spaces because the researcher's empirical model often involves both discrete and continuous variables.} In our model, the Bayesian game is populated by a finite set of players indexed by $N=\{1,...,n\}$. Let $\mathbb{W}$ denote the set from which the payoff state $W$ takes values, and $\mathbb{Y} = \mathbb{Y}_1 \times ... \times \mathbb{Y}_n$ the set from which the action profile $Y = (Y_1,...,Y_n)$ takes values. Each player's action space $\mathbb{Y}_i$ can be a continuum or a countable set. We maintain the following decomposition of the payoff state throughout:
\begin{align*}
	W = (W_C,W_I) \in \mathbb{W}_C \times \mathbb{W}_I,
\end{align*}
for some sets $\mathbb{W}_C$ and $\mathbb{W}_I$, where $W_C$ is the common payoff state and $W_I = (W_{I,1},...,W_{I,n}) \in \mathbb{W}_{I,1} \times \cdots \times \mathbb{W}_{I,n}$ collects the players' private payoff states, with $W_{I,i}$ denoting player $i$'s private piece. (The subscript $C$ is mnemonic for ``common'' and $I$ for ``idiosyncratic''.)

Each player $i$ is endowed with a payoff function $u_i: \mathbb{Y} \times \mathbb{W} \rightarrow \mathbf{R}$, and chooses an action from $\mathbb{Y}_i$. Let the payoff state $W$ be drawn from a distribution $\mu_W$ on $\mathbb{W}$.\footnote{More precisely, we assume that $\mathbb{Y}$ and $\mathbb{W}$ are complete separable metric spaces, each endowed with its Borel $\sigma$-field. We suppress measure-theoretic qualifiers throughout the section, relegating details to the Online Appendix.}  Following BM, we call $B = (\mathbb{Y},\mathbb{W},u,\mu_W)$ a \bi{basic game}, where $u = (u_i)_{i \in N}$ denotes the payoff profile.

Each player $i$ observes the signal
\begin{align*}
	T_i = s_i(W), \quad s_i(w) = (w_C, w_{I,i}),
\end{align*}
taking values in $\mathbb{T}_i = \mathbb{W}_C \times \mathbb{W}_{I,i}$: the common component $W_C$ is publicly observed, and $W_{I,i}$ is player $i$'s private component. Writing $s = (s_1,...,s_n)$ and $\mathbb{T} = \mathbb{T}_1 \times...\times \mathbb{T}_n$, the signal profile $T=(T_1,...,T_n) = s(W) \in \mathbb{T}$ is a deterministic function of the payoff state.\footnote{In the general framework of BM, the signal profile is drawn from a conditional distribution $\alpha(\csdot \mid w)$ given the payoff state $w$. Our setting is the special case where $\alpha(\csdot \mid w)$ is the point mass at $s(w)$. This class still covers a wide range of information structures relevant for our examples and applications: complete information, dichotomies between public and private signals, and observable private signals, as illustrated below.} Player $i$'s belief upon observing the signal $T_i = t_i$ is the conditional distribution $\eta_i(\csdot \mid t_i)$ of the opponents' private components $W_{I,-i}$ given $T_i = t_i$. We fix a version of this conditional distribution as part of the model. We call $I = (\mathbb{W},\mathbb{T},s,\eta)$ an \bi{information structure}, where $\eta = (\eta_i)_{i=1}^n$.

A \bi{Bayesian game} $G$ consists of the basic game and information structure, i.e., $G = (B,I)$. A Bayesian game can be a complete information game or an incomplete information game depending on the information structure $I$.

We introduce strategies. Since the signal profile is determined by the payoff state, a strategy needs to condition only on the state. Let $\sigma(\csdot \mid w)$ be a conditional distribution on $\mathbb{Y}$, when the payoff state is realized to be $w \in \mathbb{W}$. Let $\Sigma$ be the collection of such conditional distributions. Following BM, we call each $\sigma \in \Sigma$ a \bi{decision rule}.

For each $i=1,...,n$, $t_i = (w_C,w_{I,i}) \in \mathbb{T}_i$ and $\sigma \in \Sigma$ and a transform $\tau_i: \mathbb{Y}_i \rightarrow \mathbb{Y}_i$, we write the expected payoff of player $i$ as
\begin{eqnarray}
	\label{utility}
    U_i(\tau_i,t_i;\sigma) = \int \int u_i(\tau_i(y_i),y_{-i},w)d\sigma(y_i,y_{-i} \mid w) d\eta_i(w_{I,-i} \mid t_i),
\end{eqnarray}
where $w = (w_C,w_{I,i},w_{I,-i})$ with the coordinates rearranged in the obvious manner. The quantity $U_i(\tau_i,t_i;\sigma)$ is player $i$'s expected payoff conditional on her signal $T_i = t_i$, when the recommended action $y_i$ is drawn from $\sigma$ and the player plays $\tau_i(y_i)$. 
 
 We say that a decision rule $\sigma \in \Sigma$ is a \bi{Bayes Correlated Equilibrium (BCE)} for $G$ if for each $i=1,...,n$, at almost every realization $t_i$ of $T_i$,
	\begin{align}
		\label{BCE}
	  U_i(\mathsf{Id},t_i;\sigma) \ge U_i(\tau_i,t_i;\sigma) \text{ for every } \tau_i: \mathbb{Y}_i \rightarrow \mathbb{Y}_i,
	\end{align}
where $\mathsf{Id}$ is the identity map on $\mathbb{Y}_i$. That is, for almost every realization of the signal, obedience holds against every deviation. Denote by $\Sigma_{\mathsf{BCE}}(G)$ the set of BCEs of game $G$. BCE generalizes other solution concepts, such as Bayes Nash equilibria. Later we show how our result carries over to these other solution concepts and provide concrete examples.

\subsection{Counterfactual Predictions from a Game}
\subsubsection{Predictions from a Bayesian Game\label{subsec:predictions}}
Predictions from a game $G$ are generated from the distribution of the observed action profile $Y$ conditional on the payoff state $W$, when such an action profile is generated by an equilibrium of the game $G$. 

The researcher observes only one action profile $Y$ from the game $G$, even when there are multiple equilibria in this game. In order to complete the description of how $Y$ is generated, we introduce a generic form of the equilibrium selection rule. For a game $G$, we define the \bi{equilibrium selection rule} (denoted by $e(\csdot;G)$) to be a distribution on $\Sigma_{\mathsf{BCE}}(G)$.  Thus, the generation of $Y$ is described as follows:\medskip

Step 1: An equilibrium $\sigma \in \Sigma_{\mathsf{BCE}}(G)$ is drawn from the distribution $e(\csdot;G)$. 

Step 2: The value of the payoff state $W=w$ is drawn from the distribution $\mu_W$.

Step 3: The action profile $Y$ is drawn from the distribution $\sigma(\csdot\mid w)$.\medskip

The three steps summarize the causal structure of the model for observed actions $Y$ and the payoff state $W$. Given a game $G$ and $w \in \mathbb{W}$, we define a \bi{(randomized) reduced form of game $G$} as\footnote{The integral with respect to the equilibrium selection rule is an integral of a real function on the space of conditional distributions which we topologize appropriately. Details can be found in the Online Appendix.}
\begin{align}
	\label{RRF}
	\rho_G(A\mid w) \eqdef \int_{\Sigma_{\mathsf{BCE}}(G)} \sigma(A\mid w) d e(\sigma;G), \enspace A \subset \mathbb{Y}.
\end{align}
The term randomized reduced form is used because the conditional distribution $\rho_G$ of $Y$ given $W = w$ reflects the causal relationship between $Y$ and $W$.

From here on, any probability statements (including expectation and conditional expectation) involving $(Y,W)$ are with respect to the joint distribution defined as follows:
\begin{align*}
	P\{Y \in A, W \in S\} = \int_{S} \rho_G(A \mid w) d\mu_W(w), \quad A \subset \mathbb{Y}, \quad S \subset \mathbb{W}.
\end{align*}
The right hand side presents the decomposition of the joint distribution of $(Y,W)$ into the conditional distribution of $Y$ given $W$ and the marginal distribution of $W$.

\subsubsection{Counterfactual Predictions}
Our counterfactual experiment involves a new policy that changes the payoff state $W$ into $f(W)$ according to some map $f:\mathbb{W} \rightarrow \mathbb{W}$. The policy changes the basic game $B$ into 
\begin{align*}
	B_f = (\mathbb{Y},\mathbb{W},u,\mu_W \circ f^{-1}),
\end{align*}
where $\mu_W \circ f^{-1}$ is the distribution of $f(W)$ when $W$ is drawn from $\mu_W$. In other words, it transforms the pre-policy game $G = (B,I)$ into a post-policy game $G_f = (B_f,I)$.\footnote{Here we regard the information structure $I$ as invariant to the policy: each player continues to observe $T_i = (W_C,W_{I,i})$ and to hold the beliefs $\eta_i$ after the policy, so that the fixed version $\eta_i$ is a version of the conditional distribution of $W_{I,-i}$ given $T_i = s_i(W)$ under $W \sim \mu_W$ both before and after the policy. This does not mean that the realized signals are unchanged: the post-policy signal is $s_i(f(W))$, which differs from $s_i(W)$ whenever the policy moves the payoff state. What remains invariant is the map from payoff states to signals and beliefs.} As we describe in Section \ref{sec:examples}, the policy can include changes to market characteristics (population size, or changes to existing market policies) or changes to observable reserve prices in auctions. It further extends to many other settings where the policy affects individual-level payoff states such as income or taxes.

A counterfactual prediction at $f(w)$ in game $G_f$ can be made from the conditional action distribution $\sigma(\csdot \mid f(w))$ of an equilibrium $\sigma \in \Sigma_{\mathsf{BCE}}(G_f)$. Let $e(\csdot;G_f)$ be the equilibrium selection rule of the game $G_f$. Then the \bi{equilibrium-based prediction of the post-policy game $G_f$} is given by
\begin{align}
	\label{rho fG}
	\rho_{G_f}(A \mid f(w)) \eqdef \int_{\Sigma_{\mathsf{BCE}}(G_f)} \sigma(A \mid f(w)) d e(\sigma;G_f), \quad A \subset \mathbb{Y}.
\end{align}
The quantity $\rho_{G_f}(A \mid f(w))$ represents the probability of the action profile from the post-policy game $G_f$ realizing in $A$ when the payoff state is $f(w)$.

An alternative way of generating a prediction is to use the following:
\begin{align}
	\label{decomp pred}
	\rho_G(A \mid f(w)) \eqdef  \int_{\Sigma_{\mathsf{BCE}}(G)} \sigma(A \mid f(w)) d e(\sigma;G).
\end{align}
We call $\rho_G(A \mid f(w))$ the \bi{decomposition-based prediction from game $G$}. When $f(w)$ is in the support of $W$, the last integral is equal to
\begin{align*}
	P\{Y \in A \mid W = f(w)\}.
\end{align*}
Hence this prediction extrapolates the relation between $Y$ and $W$ in the pre-policy game to the post-policy game. However, decomposition-based predictions may not coincide with equilibrium-based predictions in (\ref{rho fG}), when $\rho_G$ is not policy-invariant (i.e., $\rho_G \ne \rho_{G_f}$).

\subsection{Interval-Identification of Equilibrium-Based Predictions}

\subsubsection{The Main Idea}

A comparison of the two quantities in (\ref{rho fG}) and (\ref{decomp pred}) shows that when $\mathbb{S}_{f(W)} \subset \mathbb{S}_W$, the decomposition-based prediction coincides with the equilibrium-based one as long as
\begin{align}
	\label{inv RRD}
	\rho_G = \rho_{G_f},
\end{align}
$\mu_W^f$-a.e., i.e., the causal relationship between the outcome of the game and the payoff state remains the same after the policy. The equality holds, if the equilibrium selection rules remain the same before and after the policy: 
\begin{align}
	\label{eq sel rule}
	e(\csdot;G) = e(\csdot;G_f),
\end{align}
where $e(\csdot;G)$ denotes the equilibrium selection rule of the pre-policy game $G$ and $e(\csdot;G_f)$ that of the post-policy game. However, this is too strong a condition to impose in practice, because the set of equilibria can change after the policy. We therefore introduce a weaker condition that requires only the invariance of the \emph{sliced} equilibria, i.e., the set of equilibrium actions at each payoff state $w$ in the intersection of the supports of the pre-policy and post-policy state. More specifically, suppose that $W$ has a finite support for simplicity. Let $\Sigma_{\mathsf{BCE},w}(G)$ denote the set of probabilities on the equilibrium action profiles at the payoff state $w$, and let $e_w(\csdot;G)$ denote the selection probability of equilibrium actions when the payoff state is $w$. Then, we obtain the invariance $\rho_G = \rho_{G_f}$, if the following two conditions are met:\medskip

(i) $\Sigma_{\mathsf{BCE},w}(G) = \Sigma_{\mathsf{BCE},w}(G_f)$ for each $w \in \mathbb{S}_W \cap \mathbb{S}_{f(W)}$,

(ii) For each $w \in \mathbb{S}_W \cap \mathbb{S}_{f(W)}$ such that (i) holds, we have 
\begin{align*}
	e_w(\csdot;G) = e_w(\csdot;G_f).
\end{align*}
The main point of this slicing approach is that in both (i) and (ii), we focus on the \textit{same payoff state} $w$ before and after the policy. First, we show that (i) holds, when the policy transforms only the commonly observed component of the payoff state and each player's belief about the opponents' private components does not depend on that component. Then, we adopt (ii) as our invariance condition on the equilibrium selection rule, which is considerably weaker than (\ref{eq sel rule}). As discussed in Section \ref{subsubsec: invariance ESR overview}, it is already satisfied by many equilibrium selection rules used in the empirical literature, including assuming the same equilibrium is played after the policy, focusing on specific equilibria (e.g., Pareto efficient or highest-profit), or using a parametrized selection rule for counterfactual predictions. Thus, we obtain the desired validity of decomposition-based bounds.

\subsubsection{Assumptions}

Our main result gives conditions under which the decomposition-based prediction interval-identifies the equilibrium-based one.

Recall the decomposition $W = (W_C,W_I)$ of the payoff state and the signal structure $T_i = (W_C,W_{I,i})$ maintained since Section \ref{subsec: a finite-player bayesian game}: every player observes the common component $W_C$, and each player $i$ privately observes $W_{I,i}$. Throughout, for any random variable or vector $Z$, we write $\mathbb{S}_Z$ for its support.

\begin{assumption}[Information Structure Condition]
	\label{assump: information structure and policy}

	(i) The policy $f$ satisfies that
	\begin{align*}
		f(w) = (\tilde f(w_C),w_I), \quad w = (w_C, w_I) \in \mathbb{W},
	\end{align*}
	for some map $\tilde f: \mathbb{W}_C \rightarrow \mathbb{W}_C$.

	(ii) For each $i=1,...,n$, $W_{I,-i} \CI W_C \mid W_{I,i}.$

	(iii) On $\mathbb{S}_W \cap \mathbb{S}_{f(W)}$, the pre-policy and post-policy distributions of the payoff state assign probability zero to the same sets.

	(iv) $\mathbb{S}_W = \mathbb{S}_{W_C} \times \mathbb{S}_{W_I}$ and $\mathbb{S}_{f(W)} = \mathbb{S}_{\tilde f(W_C)} \times \mathbb{S}_{W_I}$.
\end{assumption}

Assumption \ref{assump: information structure and policy}(i) requires that the policy alters only a publicly observed payoff component $W_C$ of the payoff state $W = (W_C,W_I)$.\footnote{The Online Appendix provides a simple example of a private information entry game which demonstrates that when a policy changes the private information component of the payoff state, decomposition-based predictions fail. This failure occurs because changes to private components of players' payoff states alter posterior beliefs and create incentives to deviate from strategies used in the pre-policy game.} Condition (ii) requires that, given player $i$'s private payoff state, the other players' private payoff states are independent of the common payoff state. Thus, conditions (i) and (ii) ensure that players' posterior beliefs remain invariant after policy implementation. Condition (iii) requires that the policy neither create nor destroy null events on the common support: a set of payoff states there has probability zero before the policy exactly when it has probability zero after. Condition (iv) requires the supports to factor into their common and private parts, so that the common support $\mathbb{S}_W \cap \mathbb{S}_{f(W)}$ is a cylinder $(\mathbb{S}_{W_C} \cap \mathbb{S}_{\tilde f(W_C)}) \times \mathbb{S}_{W_I}$ over the common component. It holds, for example, whenever $W_C$ and $W_I$ are independent, as in the entry games of Section \ref{example1_detail}.

Assumption \ref{assump: information structure and policy} admits a broad class of information structures, as we see through examples below. In these examples, we specify the payoff state vector as $W = (X, \varepsilon)$, with $X = (X_1,...,X_n)$ and $\varepsilon = (\varepsilon_1,...,\varepsilon_n)$, so that each individual player $i$'s payoff depends on $W$ only through $(X_i, \varepsilon_i)$. The setting accommodates the information structures considered by \cite{Magnolfi/Roncoroni:22:ReStud} in the entry game model.

\begin{example}[Complete Information Game] In a complete information game, we take $\mathbb{W}_C = \mathbb{W}$ (with each $W_{I,i}$ degenerate), so that $T_i = W$. Then any policy $f$ that changes $W$ satisfies Assumption \ref{assump: information structure and policy}(i), and Assumption \ref{assump: information structure and policy}(ii) is trivially satisfied in this case.
\end{example}

\begin{example}[Public-Private Dichotomy of Signals]\label{example: public-private}
In the case of a public-private information structure where $X_i$ is publicly observed while $\varepsilon_i$ belongs to private information, we specify $T_i = (X, \varepsilon_i)$. In this case, Assumption \ref{assump: information structure and policy}(i) is satisfied whenever the policy $f$ is one that alters $X$, not $\varepsilon_i$, because we can take $W_C = X$ and $W_{I,i} = \varepsilon_i$. When $X$ is independent of $\varepsilon = (\varepsilon_1,...,\varepsilon_n)$, Assumption \ref{assump: information structure and policy}(ii) is satisfied. 
\end{example}

\begin{example}[Observable Private Signals]\label{example: observable private}
The researcher may observe signals that are part of the private information. To reflect this setting, we specify $X_i = (X_{1,i},X_{2,i})$, and take $T_i = (X_1, X_{2,i}, \varepsilon_{i})$, where $X_1 = (X_{1,i})_{i =1}^n$, so that $X_{2,i}$ belongs to private information and is observed by the researcher. In this case, as long as $f$ alters $X_1$ but not $(X_{2,i}, \varepsilon_i)$, Assumption \ref{assump: information structure and policy}(i) is satisfied once we take $W_C = X_1$ and $W_{I,i} = (X_{2,i},\varepsilon_i)$. Furthermore, if $X_1$ is conditionally independent of $(X_{2,-i},\varepsilon_{-i})$ given $(X_{2,i},\varepsilon_i)$, Assumption \ref{assump: information structure and policy}(ii) is satisfied.
\end{example}

\begin{example}[Payoff Irrelevant Signals] Part of the signals $(X_i,\varepsilon_i)$ in the previous example can be payoff irrelevant. We can reflect this by specifying the payoff $u_i$ as depending only on part of $(X_i,\varepsilon_i)$.
\end{example}

We introduce the condition about the existence of BCE before and after the policy.

\begin{assumption}
    \label{assump: BCE existence}
	$\Sigma_{\mathsf{BCE}}(G) \ne \varnothing$ and $\Sigma_{\mathsf{BCE}}(G_f) \ne \varnothing$.
\end{assumption}

We now state the invariance condition.

\begin{assumption}
	\label{assump: invariance}
     Either of the following conditions holds.\medskip

	(i) The game $G$ has a unique BCE.
    
    (ii) For each $w \in \mathbb{S}_{W} \cap \mathbb{S}_{f(W)}$, if $\Sigma_{\mathsf{BCE},w}(G) = \Sigma_{\mathsf{BCE},w}(G_f)$, then, we have
	\begin{align*}
		e_w(\csdot;G) = e_w(\csdot;G_f).
	\end{align*}
\end{assumption}

Condition (i) is common in settings spanning auctions, social interactions, and networks (see \cite{Guerre/Perrigne/Vuong:00:ECMA}, \cite{Calvo/Patacchini/Zenou:ReStud:09}, \cite{Blume/Brock/Durlauf/Jayaraman:15:JPE}, \cite{Xu:18:IER} and \cite{Eraslan/Tang:25:WP}). The invariance condition (ii) generalizes the informal condition introduced in Section \ref{subsubsec: suff conds for validity}.
	
\subsubsection{The Main Result}
In this section, we present the main result. For a real map $h$ on $\mathbb{Y} \times \mathbb{W}$, define
\begin{align}
	\label{EP DP}
	\mathsf{AEP}(h) &\eqdef \int  \int_{\mathbb{Y}} h(y,w) d \rho_{G_f}(y \mid w)d(\mu_W \circ f^{-1})(w),  \text{ and }\\ \notag
	\mathsf{ADP}_R(h) &\eqdef \int \mathbf{E}[h(Y,W) \mid W = f(w)]1\{f(w) \in \mathbb{S}_W\}d\mu_W(w).
\end{align}
The quantity $\mathsf{AEP}(h)$ is our target parameter, the equilibrium-based expectation of $h(Y,f(W))$ after the policy, while $\mathsf{ADP}_R(h)$ is its decomposition-based counterpart. Depending on the choice of $h$, which may be vector-valued, $\mathsf{AEP}(h)$ captures a variety of objects of interest: $h(y,w) = y$ yields the expected action profile, $h(y,w) = 1\{y \le t\}$ yields the CDF at $t$, and $h(y,w) = 1\{\max_{i: (w_C,w_{I,i}) \in S} y_i \le t\}$ for a given set $S$ of signal values yields the CDF of the maximum action among players with $(w_C,w_{I,i}) \in S$.

\begin{theorem}
	\label{thm: bounds1}
	Suppose that Assumptions \ref{assump: information structure and policy}-\ref{assump: invariance} hold for the pre-policy game $G = (B,I)$ and the post-policy game $G_f = (B_f,I)$. Let a map $h:\mathbb{Y} \times \mathbb{W} \rightarrow \mathbf{R}$ be such that, for all $w \in \mathbb{W}$,
	\begin{align}
		\label{h bounds}
		\underline h(w) \le \inf_{y \in \mathbb{Y}} h(y,w) \le \sup_{y \in \mathbb{Y}} h(y,w) \le \overline h(w),
	\end{align}
	for some maps $\underline h, \overline h: \mathbb{W} \rightarrow \mathbf{R}$ that are integrable with respect to $\mu_W \circ f^{-1}$.

	Then,
	\begin{align}
		\label{bounds0}
		\mathsf{ADP}_R(h) +  \Delta(\underline h)
		\le \mathsf{AEP}(h)
		\le \mathsf{ADP}_R(h) +  \Delta(\overline h),
	\end{align}
	where
	\begin{align*}
		\Delta(\overline h) \eqdef \mathbf{E}\left[\overline h(f(W))1\left\{f(W) \notin \mathbb{S}_W\right\} \right] \text{ and }
		\Delta(\underline h) \eqdef \mathbf{E}\left[\underline h(f(W))1\left\{f(W) \notin \mathbb{S}_W\right\} \right].
	\end{align*}
\end{theorem}
\medskip

 Suppose that $Y_i \in [h_L,h_U]$, $h_L < h_U$, and define the average equilibrium-based prediction for player $i$'s outcome as
\begin{align*}
	\mathsf{AEP}_i \eqdef \int  \int_{\mathbb{Y}} y_i \, d \rho_{G_f}(y \mid w)d(\mu_W \circ f^{-1})(w),
\end{align*}
and its decomposition-based counterpart as
\begin{align*}
	\mu_i(f) \eqdef \int \mathbf{E}[Y_i \mid W = f(w)]1\{f(w) \in \mathbb{S}_W\} d\mu_W(w).
\end{align*}
If we take $h(y,w) = y_i$, with $\underline h = h_L$ and $\overline h = h_U$, by Assumptions \ref{assump: information structure and policy} and \ref{assump: invariance}, we obtain the following bounds for each $i = 1,...,n$:
\begin{align}
	\label{bounds_simple}
	\mu_i(f) + h_L \csdot P\left\{f(W) \notin \mathbb{S}_W\right\} \le \mathsf{AEP}_i \le \mu_i(f) + h_U \csdot P\left\{f(W) \notin \mathbb{S}_W\right\}.
\end{align}
In an entry game where $Y_i \in \{0,1\}$, setting $h_U = 1$ and $h_L = 0$ yields bounds for the predicted entry probability of firm $i$ after the policy. When the support condition $\mathbb{S}_{f(W)} \subset \mathbb{S}_W$ holds, these bounds collapse to a point, so that $\mathsf{AEP}_i = \mu_i(f)$. We can also obtain bounds for the average treatment effect $\mathsf{ATE}_i \eqdef \mathsf{AEP}_i - \mathbf{E}[Y_i]$:
\begin{align*}
	\mu_i(f) - \mathbf{E}[Y_i] + h_L \csdot P\left\{f(W) \notin \mathbb{S}_W\right\} \le \mathsf{ATE}_i \le \mu_i(f)  - \mathbf{E}[Y_i] + h_U \csdot P\left\{f(W) \notin \mathbb{S}_W\right\}.
\end{align*}

One might ask whether the bounds in Theorem \ref{thm: bounds1} are sharp. The following proposition gives a sense in which the answer is affirmative.

\begin{proposition}
	\label{prop: sharp bounds0}
	Suppose that Assumptions \ref{assump: information structure and policy}-\ref{assump: invariance} hold for the pre-policy game $G = (B,I)$ and the post-policy game $G_f = (B_f,I)$. Suppose further that maps $h: \mathbb{Y} \times \mathbb{W} \rightarrow \mathbf{R}$, $\underline h, \overline h: \mathbb{W} \rightarrow \mathbf{R}$, are given as in Theorem \ref{thm: bounds1}, where $\mathbb{Y}$ is compact and $h(\csdot,w)$ is continuous on $\mathbb{Y}$ for each $w$, and for all $w \in \mathbb{W}$, $\overline h(w) = \max_{y \in \mathbb{Y}} h(y,w)$ and $\underline h(w) = \min_{y \in \mathbb{Y}} h(y,w)$.

	Then there are Bayesian games $G_1$ and $G_2$, with post-policy games $G_{1,f}$ and $G_{2,f}$ induced by $f$, such that
	\begin{align*}
		\mathsf{AEP}(h) &= \mathsf{ADP}_R(h) + \Delta(\overline h), \text{ under } (G_1, G_{1,f}) \text{ and }\\
		\mathsf{AEP}(h) &= \mathsf{ADP}_R(h) + \Delta(\underline h), \text{ under } (G_2, G_{2,f}).
	\end{align*}
	The games $G_1$ and $G_2$ differ only in their payoffs outside $\mathbb{S}_W$. They therefore induce the same joint distribution of $(Y,W)$ before the policy, and hence the same identified objects $\mathsf{ADP}_R(h)$, $\Delta(\overline h)$, and $\Delta(\underline h)$.
\end{proposition}

Simple shape constraints, such as $\rho_{G_f}( \csdot \mid w)$ being monotone or concave in $w$, do not necessarily improve the bounds, because a constant map satisfies such constraints trivially. The bounds arise entirely from the out-of-support payoff states, where the predictions from the game play no role and the outcome is controlled only by $\underline h$ and $\overline h$. The shape constraints may, however, help estimate $\mathsf{ADP}_R(h)$ more accurately.

We provide a brief heuristic behind Theorem \ref{thm: bounds1}. The key step in the proof is to show that, for $w \in \mathbb{S}_{W} \cap \mathbb{S}_{f(W)}$,
\begin{align}
	\label{rho eq}
	\rho_G(\csdot \mid w) = \rho_{G_f}(\csdot \mid w).
\end{align}
Once this is established, the bounds follow by splitting the integration domain into $\mathbb{S}_W \cap \mathbb{S}_{f(W)}$ (where the decomposition-based and equilibrium-based predictions coincide by (\ref{rho eq})) and its complement (where the prediction is bounded by $\underline h$ and $\overline h$). To show (\ref{rho eq}), the invariance condition in Assumption \ref{assump: invariance} reduces it to verifying that 
\begin{align*}
	\Sigma_{\mathsf{BCE},w}(G) = \Sigma_{\mathsf{BCE},w}(G_f),
\end{align*}
for $w \in \mathbb{S}_W \cap \mathbb{S}_{f(W)}$. This follows from Assumption \ref{assump: information structure and policy}: since the policy modifies only the publicly observable payoff component and $W_{I,-i} \CI W_C \mid W_{I,i}$, the $w$-slices of the BCE sets coincide on $\mathbb{S}_W \cap \mathbb{S}_{f(W)}$. We refer the reader to the Online Appendix for details.

\subsubsection{Identification of the Bounds}\label{sec:single-index-restrictions}

If we observe the actions $Y$ and the payoff state $W$ of the pre-policy game, then we can recover the bounds in Theorem \ref{thm: bounds1} from data without specifying many details of the game. However, in practice, we often do not observe the whole payoff vector $W$.

Suppose that $W = (X, \varepsilon)$, where $X$ is observed and $\varepsilon$ unobserved by the researcher. Let the policy $f$ be of the form
\begin{align}
	\label{f}
	f(W) = (f^*(X),\varepsilon), \text{ with } f^*(X) = (f_1^*(X_1),...,f_n^*(X_n)),
\end{align}
for some maps $f_i^*$, $i=1,...,n$. As for $h$, we consider $h(y,w) = h^*(y,x)$, $w = (x,\varepsilon)$, $\overline h(w) = \overline h^*(x)$, and $\underline h(w) = \underline h^*(x)$, for some maps $h^*$, $\overline h^*$, and $\underline h^*$, which depend only on observable states.

For the interval identification of $\mathsf{AEP}(h)$ through Theorem \ref{thm: bounds1}, we need to identify from data the boundary terms $\Delta(\underline h)$ and $\Delta(\overline h)$ as well as $\mathsf{ADP}_R(h)$. The identification strategy of $\mathsf{ADP}_R(h)$ depends on whether $X$ and $\varepsilon$ are independent or not. If $X$ and $\varepsilon$ are independent, we can identify $\mathsf{ADP}_R(h)$ as follows:
	\begin{align}
		\label{DP ident}
		\mathsf{ADP}_R(h) = \int \mathbf{E}\left[h^*(Y,X) \mid X = f^*(x)\right] 1\{f^*(x) \in \mathbb{S}_X\}d\mu_X(x),
	\end{align}
where $\mu_X$ denotes the distribution of $X$. The identification result allows for $\varepsilon_1,...,\varepsilon_n$ to be correlated. This correlation can come from some unobserved characteristics of the game.

The result can be extended to the case where $X$ and $\varepsilon$ are potentially correlated using a control function approach (\cite{Blundell/Powell:03:Adv} and \cite{Imbens/Newey:09:Eca}).\footnote{Game-theoretic models often involve a simultaneous system of equations. We exclude the setting where the policy variable is part of the endogenous outcomes in such equations. For example, in a two-player game with outcomes $Y = (Y_1,Y_2)$, we focus on a policy that changes the payoff state $X$ which is not one of the two outcomes. When one of the endogenous outcomes is a policy variable, the structural equations need to be transformed into a triangular system of equations to apply a control function approach. \cite{Blundell/Matzkin:14:QE} present conditions for such a transform to exist. These conditions may be implausible in some empirical applications.} More specifically, suppose that $X_i = (X_{i,a},X_{i,b})$ and 
\begin{align*}
    f_i^*(X_i) = (g_i(X_{i,a}),X_{i,b}),
\end{align*}
for some map $g_i$ so that the policy alters only $X_{i,a}$. Define 
\begin{align*}
    X_a = (X_{1,a},...,X_{n,a}) \text{ and } X_b = (X_{1,b},...,X_{n,b}),
\end{align*}
and let $g(x_a) = (g_1(x_{1,a}),...,g_n(x_{n,a}))$, $x_a = (x_{1,a},...,x_{n,a})$. We impose the following conditions on the policy variable, the control, and the unobservable.

\begin{assumption}
	\label{assump: control fn}
	(i) $X_a$ and $\varepsilon$ are conditionally independent given $X_b$.

	(ii) Whenever $(x_a,x_b,\bar\varepsilon) \in \mathbb{S}_W$ and $(x_a',x_b) \in \mathbb{S}_{X_a,X_b}$, we have $(x_a',x_b,\bar\varepsilon) \in \mathbb{S}_W$.
\end{assumption}

Assumption \ref{assump: control fn}(i) is the control-function requirement that the observed control $X_b$ absorbs the stochastic dependence between the policy variable $X_a$ and the unobservable $\varepsilon$. When $X_a$ and $\varepsilon$ are dependent due to some unobserved game-specific characteristics, we may take $X_b$ to be an observed vector including game characteristics such that conditioning on $X_b$ removes this dependence. Assumption \ref{assump: control fn}(ii) restricts the $(x_a,x_b)$-section of $\mathbb{S}_W$, namely $\{\bar\varepsilon : (x_a,x_b,\bar\varepsilon) \in \mathbb{S}_W\}$: among points $(x_a,x_b) \in \mathbb{S}_{X_a,X_b}$, this section may depend on the control $x_b$ but not on the policy variable $x_a$. Product support, $\mathbb{S}_W = \mathbb{S}_{X_a,X_b} \times \mathbb{S}_\varepsilon$, where $\mathbb{S}_\varepsilon$ is the support of $\varepsilon$, is the special case in which the section equals $\mathbb{S}_\varepsilon$ everywhere.

Under these conditions, we obtain the following identification result.
\begin{proposition}
	\label{prop: control fn}
	Suppose that Assumptions \ref{assump: information structure and policy}, \ref{assump: BCE existence}, and \ref{assump: control fn} hold. Suppose further that $h(\csdot,w) = h^*(\csdot,x)$ on $\mathbb{Y}$, $\overline h(w) = \overline h^*(x)$, and $\underline h(w) = \underline h^*(x)$, for each $w = (x,\varepsilon) \in \mathbb{W}$, for some maps $h^*$, $\overline h^*$, and $\underline h^*$. Then, the following statements hold.

	(i) $\Delta(\underline h) = \mathbf{E}\left[\underline h^*(f^*(X))1\left\{f^*(X) \notin \mathbb{S}_X\right\} \right]$ and $\Delta(\overline h) =\mathbf{E}\left[\overline h^*(f^*(X))1\left\{f^*(X) \notin \mathbb{S}_X\right\} \right]$.

	(ii) $\mathsf{ADP}_R(h) = \int \mathbf{E}\left[ h^*(Y,X) \mid (X_a,X_b) = (g(x_a),x_b)\right] 1\left\{(g(x_a),x_b) \in \mathbb{S}_{X_a,X_b}\right\} d\mu_{X_a,X_b}(x_a,x_b)$, where $\mu_{X_a,X_b}$ is the distribution of $(X_a,X_b)$.
\end{proposition}

In many examples, the payoff state takes a partial index form: $W = (X_a, Z, \varepsilon)$, where player $i$'s component of the state is $(X_{i,a}, Z_i, \varepsilon_i)$ and $Z_i = X_{i,b}'\theta_i$, with coefficient $\theta_i$, so that the covariates $X_{i,b}$ enter the game only through the indices $Z_i$. In this case, Proposition \ref{prop: control fn} applies with $(X_a,Z)$ in place of $(X_a,X_b)$. More specifically, suppose that Assumption \ref{assump: control fn} holds with $X_b$ replaced by $Z = (Z_1,...,Z_n)$. In particular, $X_a$ and $\varepsilon$ are conditionally independent given $Z$. Suppose further that $h(\csdot,w) = h^*(\csdot,x_a,z)$ on $\mathbb{Y}$ for each $w = (x_a,z,\varepsilon) \in \mathbb{W}$ for some map $h^*$. Then, Proposition \ref{prop: control fn}(ii) yields
\begin{align}
	\label{index}
	\mathsf{ADP}_R(h) =   \int \mathbf{E}\left[ h^*(Y,X_a,Z) \mid (X_a,Z) = (g(x_a),z)\right] 1\left\{(g(x_a),z) \in \mathbb{S}_{X_a,Z} \right\} d\mu_{X_a,Z}(x_a,z),
\end{align}
where $\mu_{X_a,Z}$ is the distribution of $(X_a,Z)$. We can identify and estimate $\theta_1,...,\theta_n$ (up to a scale) following the literature on multi-index models (see \cite{Ichimura/Lee:91:NSEM}, \cite{Lee:95:JOE}, \cite{Donkers/Schafghans:08:ET}, \cite{Xia:08:JASA} and \cite{Ahn/Ichimura/Powell/Ruud:18:JBES} and references therein).  The main difference here is that we do \textit{not} introduce the multi-index structure as a semiparametric restriction on a nonparametric function. It follows from the index structure of the payoff function in the game. The multi-index models are useful for dimension reduction when the game involves only a few players, and the dimension of $X_i$ is large. We provide some implementation details in the Online Appendix.

\subsection{Extension to Other Solution Concepts}\label{app:refinements}

Our results extend to other solution concepts beyond BCE. In particular, we can accommodate any refinement of BCE, which includes Bayes Nash Equilibria, among others. 

To see this, we let $\Sigma' \subset \Sigma$ be a given subcollection of decision rules $\sigma$ and consider the restricted BCE:
\begin{align}
	\label{BNE2}
	\Sigma_{\mathsf{BCE}}'(G) = \Sigma_{\mathsf{BCE}}(G) \cap \Sigma'.
\end{align}
We call this set the set of \bi{Bayes Correlated Equilibria (BCE) restricted to} $\Sigma'$.

For example, suppose that $\Sigma'$ is the collection of decision rules $\sigma$ of the following form: for any Borel $A = A_1 \times ... \times A_n \subset \mathbb{Y}$,
\begin{align}
	\label{B}
	\sigma(A \mid w) = \prod_{i=1}^n \beta_i(A_i \mid t_i), \enspace t_i = (w_C,w_{I,i}),
\end{align}
where each $\beta_i$ is a Markov kernel from $\mathbb{T}_i$ to the probability measures on $\mathbb{Y}_i$, interpreted as the conditional distribution of $Y_i$ given the signal $T_i = t_i$, so that the players' actions are drawn independently of each other given the payoff state, ruling out a correlation device. Then a BCE restricted to $\Sigma'$ is the set of \bi{Bayes Nash Equilibria (BNE)}. We can add further restrictions such as symmetry or differentiability depending on the application.

We turn to the interval-identification of equilibrium-based predictions with the BCEs restricted to $\Sigma'$. First, as before, define the equilibrium selection rules $e'(\csdot;G)$ and $e'(\csdot;G_f)$ as a probability on $\Sigma_{\mathsf{BCE}}'(G)$ and $\Sigma_{\mathsf{BCE}}'(G_f)$ respectively. Similarly, we define $\rho_G'$ and $\rho_{G_f}'$ using $\Sigma_{\mathsf{BCE}}'(G)$, $\Sigma_{\mathsf{BCE}}'(G_f)$, $e'(\csdot;G)$, and $e'(\csdot;G_f)$. Let $\mathsf{ADP}_R'(h)$ and $\mathsf{AEP}'(h)$ be the same as $\mathsf{ADP}_R(h)$ and $\mathsf{AEP}(h)$ except that we substitute $\rho_G'$ and $\rho_{G_f}'$ for $\rho_G$ and $\rho_{G_f}$ in the definition in (\ref{EP DP}). Similarly to the case of BCE, we assume that the invariance condition on $e'(\csdot;G)$ and $e'(\csdot;G_f)$ holds in terms of the BCEs restricted to $\Sigma'$. 

We also introduce a mild technical condition on $\Sigma'$. Given $B \subset \mathbb{W}$, the \bi{gluing} of $\sigma,\sigma' \in \Sigma$ \bi{across} $B$ is the decision rule $\sigma^\dagger$ that follows $\sigma$ on $B$ and $\sigma'$ off $B$, that is, $\sigma^\dagger(\csdot \mid w) = \sigma(\csdot \mid w)$ for $w \in B$, and $\sigma^\dagger(\csdot \mid w) = \sigma'(\csdot \mid w)$ for $w \notin B$. We say that $\Sigma'$ is \bi{closed under gluing across} $B$ if the gluing across $B$ of any two elements of $\Sigma'$ again belongs to $\Sigma'$. We are ready to present an analogue of Theorem \ref{thm: bounds1} as follows.
\begin{corollary}
	\label{cor: bounds}
	Suppose that Assumptions \ref{assump: information structure and policy}-\ref{assump: invariance} (in terms of the BCEs restricted to $\Sigma'$) hold for the pre-policy game $G = (B,I)$ and the post-policy game $G_f = (B_f,I)$, and that $\Sigma'$ is closed under gluing across $(\mathbb{S}_{W_C} \cap \mathbb{S}_{\tilde f(W_C)}) \times \mathbb{W}_I$.\footnote{As shown in Lemma \ref{lemm: gluing} in the Online Appendix, the condition of closure under gluing in Corollary \ref{cor: bounds} holds for the BNE class in (\ref{B}), and also for its symmetric version, in which the factors $\beta_i$ are common across the players.} Suppose further that maps $h: \mathbb{Y} \times \mathbb{W} \rightarrow \mathbf{R}$, $\underline h, \overline h: \mathbb{W} \rightarrow \mathbf{R}$, are given as in Theorem \ref{thm: bounds1}. Then,
		\begin{align}
		\label{bounds1}
		\mathsf{ADP}_R'(h) +  \Delta(\underline h) \le \mathsf{AEP}'(h) \le \mathsf{ADP}_R'(h) +  \Delta(\overline h).
	\end{align}
\end{corollary}
Hence, decomposition-based predictions can still be used for counterfactual analysis in a setting with various other solution concepts such as BNE or its further restricted versions. They coincide with the equilibrium-based predictions when $\mathbb{S}_{f(W)} \subset \mathbb{S}_W$ under the conditions of Corollary \ref{cor: bounds}.

\subsection{Examples}\label{sec:examples}

\subsubsection{Entry Games}\label{example1_detail}

Consider an entry game which received a great deal of attention in the literature (e.g., \cite{Ciliberto/Tamer:09:Eca}, \cite{Jia:08:Eca}, \cite{Grieco:14:RAND} and \cite{Magnolfi/Roncoroni:22:ReStud}). Suppose that there are $n$ firms, $i=1,...,n$, who choose a binary action $Y_i \in \{0,1\}$, $Y = (Y_1,...,Y_n)$, where $Y_i = 1$ represents entry in the market and $Y_i=0$ staying out of the market. 

We specify the payoff generically as $u_i(y, W)$, $W = (X, \varepsilon)$, where $X = (X_1,...,X_n) \in \mathbf{R}^{n d_X}$ is observed and $\varepsilon = (\varepsilon_1,...,\varepsilon_n) \in \mathbf{R}^{n d_\varepsilon}$ unobserved by the researcher. This encompasses the linear index formulation of \cite{Bresnahan/Reiss:91:JOE}, \cite{Jia:08:Eca}, and \cite{Ciliberto/Tamer:09:Eca}.

As for the information structure, we focus on two information structures that are often used in the literature on empirical research.
\begin{assumption}\label{example1_assump}
	Either of the following conditions is satisfied:\medskip
	
	(i) The game is of complete information, i.e., $T_i = (X, \varepsilon)$, for $i=1,...,n$. 
	
	(ii) The game has a public-private dichotomy of signals, i.e., $T_i = (X, \varepsilon_i)$, for $i=1,...,n$. 
\end{assumption}

For example, Assumption \ref{example1_assump}(i) is satisfied in the complete information environments of \cite{Bresnahan/Reiss:91:JOE}, \cite{Jia:08:Eca} and \cite{Ciliberto/Tamer:09:Eca}. The public-private dichotomy case is studied in depth by \cite{Grieco:14:RAND}.

As for the policy, we assume that $X$ is subject to a change by a policy whereas $\varepsilon$ is not. 
\begin{assumption}
	\label{assump: assump2} 
	(i) $f(X,\varepsilon) = (f^*(X),\varepsilon)$, for some map $f^*$.

	(ii) $X$ and $\varepsilon$ are independent.

	(iii) On $\mathbb{S}_X \cap \mathbb{S}_{f^*(X)}$, the distributions of $X$ and of $f^*(X)$ assign probability zero to the same sets.
\end{assumption}
A counterfactual policy $f$ in the above assumption was considered by all the papers cited at the beginning of this subsection. For example, \cite{Ciliberto/Tamer:09:Eca} and \cite{Grieco:14:RAND} set the values of a dummy variable in $X_i$ to 0. Meanwhile, \cite{Jia:08:Eca} changes market size, a variable in $X_i$.

We describe the invariance condition for the equilibrium selection rule in this setting. Let $G$ denote the pre-policy game and $G_f$ the post-policy game. We define $\Sigma_{\mathsf{BCE},w}(G)$ and $\Sigma_{\mathsf{BCE},w}(G_f)$ as in Assumption \ref{assump: invariance}(ii) with $w = (x,\varepsilon)$. Since these games typically have multiple equilibria, we restate the invariance condition in Assumption \ref{assump: invariance}(ii) as follows:

\begin{assumption}
	\label{assump: invar} 
	(i) BCE exists in $G$ and $G_f$.

	(ii) For each $w = (x,\varepsilon)$ in the intersection of the supports of $(X,\varepsilon)$ and $(f^*(X), \varepsilon)$ such that $\Sigma_{\mathsf{BCE},w}(G) = \Sigma_{\mathsf{BCE},w}(G_f)$, we have $e_w(\csdot; G) = e_w(\csdot; G_f)$.
\end{assumption}

Our target parameter is the conditional probability of $Y = a$, $a \in \{0,1\}^n$, after the policy, given $f^*(X) \in C$ for some set $C$, which is defined as follows:
\begin{align*}
	p_f(Y=a \mid C) \eqdef \frac{1}{P\left\{f^*(X) \in C\right\}}\int_{C \times \mathbf{R}^{n d_\varepsilon}} \int_{\Sigma_{\mathsf{BCE}}(G_f)} \sigma(a \mid w) de(\sigma;G_f) d(\mu_W \circ f^{-1})(w),
\end{align*}
where $w = (x,\varepsilon)$. For example, $p_f(Y = (1,...,1) \mid C)$ denotes the conditional probability of all firms entering the market given $f^*(X) \in C$, after the policy. The result below shows how this probability is bounded by decomposition-based predictions.

\begin{corollary}\label{example1_corollary}
	Suppose that Assumptions \ref{example1_assump}-\ref{assump: invar} hold. Then, for each $a \in \{0,1\}^n$ and $C \subset \mathbf{R}^{nd_X}$ such that $P\{f^*(X) \in C\}>0$,
	\begin{align*}
		\mathbf{E}\left[\pi\left(a \mid f^*(X)\right) \mid f^*(X) \in C\right] &\le p_f(Y=a \mid C) \\
		&\le \mathbf{E}\left[\pi\left(a \mid f^*(X)\right) \mid f^*(X) \in C\right] + P\{f^*(X) \notin \mathbb{S}_X \mid f^*(X) \in C\},
	\end{align*}
where $\pi(a \mid x) = P\{Y= a \mid X = x\}1\{x \in \mathbb{S}_X\}$.
\end{corollary}

The conditional expectation $\mathbf{E}\left[\pi(a \mid f^*(X)) \mid f^*(X) \in C\right]$ is identified using the data from the pre-policy game. We provide a step-by-step empirical implementation of this result in Section \ref{sec:empirical} and the Online Appendix.

For example, suppose that the researcher aims to obtain the predicted joint entry probability of the firms under a policy that changes $X$ into $f^*(X)$ such that $f^*(X)$ lies in the support of $X$. Then, Corollary \ref{example1_corollary} says that the counterfactual prediction is point-identified as
\begin{align*}
    \int P\{Y= (1,...,1) \mid X = f^*(x)\}d\mu_X(x),
\end{align*}
where $\mu_X$ denotes the distribution of $X$, as above. The prediction is obtained by the nonparametric regression of $1\{Y=(1,...,1)\}$ on $X$, integrating the regression function after transforming the regressors by the map $f^*$.

\subsubsection{Auctions}
\label{subsubsec:auctions}

A common approach for counterfactual analysis in the empirical auction literature is to first nonparametrically identify and estimate the distribution of valuations from the distribution of bids, and use those estimates for counterfactual analysis (see \cite{Athey/Haile:07:Handbook} for a survey). One may wonder if we can use the decomposition approach to generate counterfactual predictions without identifying the valuation distribution from data. Although a more general treatment is possible, for concreteness we focus on the first-price sealed-bid auction with independent private values studied by \cite{Guerre/Perrigne/Vuong:09:ECMA} and consider a counterfactual policy that alters the reserve price (see \citealp{Paarsch:97:JOE,Haile/Tamer:03:JPE} for two examples of such a policy).\footnote{Reserve prices are set by the seller before the auction takes place. They are the minimum value for which the seller is willing to sell the good: if no bid is higher than the reserve price during the auction, then the good remains unsold. As a result, in empirical work, they are usually considered as an auction characteristic (primitive). The reserve price is often observed by the bidders and by the researcher. If the focus is on auctions with unobserved reserve prices (as in \cite{Elyakime/Laffont/Loisel/Vuong:97:JBES}), our results do not apply. If the researchers are interested in predicting bids after setting the reserve price beyond its support in the data (which is suggested in Table 4 of \cite{Haile/Tamer:03:JPE}), they may use the bounds approach outlined above.} Because the reserve price is often a publicly observed state variable, such a policy falls within the scope of our framework.

We consider an auction model with potential bidders $i=1,...,n$. Each bidder $i$ observes a private valuation $V_i$, drawn from a distribution with common support $\mathbb{S}_{V_i} \equiv \mathbb{S}_0 \subset [0,\overline v]$ for some constant $\overline v < \infty$. The bidders commonly observe a vector of auction specific characteristics $(X,\zeta)$, where $X=(X_1,R)$, with $R$ denoting the reserve price, is observed by the researcher, while $\zeta$ represents unobserved auction heterogeneity. Each potential bidder $i$ chooses whether to enter the auction and, upon entry, which bid to submit. We encode the action of bidder $i$ as $Y_i \in \{-1\} \cup [0,\overline v]$, where $Y_i = -1$ represents staying out and otherwise $Y_i$ is the submitted bid, and the winner is the entrant who submits the highest bid above the reserve price.

The bidders share the action space and the valuation space, so their signal spaces coincide. Writing $t_i = (v_i,x,\zeta)$ for the signal of bidder $i$, we call a pair $(a,b)$ of maps of the signal, each defined at every signal value, with $a$ taking values in $\{0,1\}$ and $b$ in $[0,\overline v]$, a \bi{symmetric pure strategy profile}, the symmetry being that this one pair prescribes the behavior of every bidder: bidder $i$ enters when $a(t_i) = 1$ and then bids $b(t_i)$, and stays out otherwise. Rather than committing to a particular refinement, we take as given a class $\Sigma'$ of decision rules, each induced by a symmetric pure strategy profile, and follow Section \ref{app:refinements}, working with the BCEs restricted to $\Sigma'$. A decision rule induced by such a profile is of the product form in (\ref{B}) with factors common across the bidders, each factor being the point mass at the action that the profile prescribes at the signal. Hence the restricted BCEs are precisely the symmetric pure strategy BNEs whose induced decision rules belong to $\Sigma'$.

The policy of interest is a change in the reserve price: a change of $R$ into $f^*(R)$ for some map $f^*$. We let $G$ and $G_f$ denote the pre-policy and the post-policy auction games, and let $(a^*,b^*)$ denote a symmetric pure strategy profile whose induced decision rule belongs to $\Sigma_{\mathsf{BCE}}'(G)$. For each auction, $\widetilde{N} =\{i: a^*(V_i,X,\zeta) = 1\}$ is the set of participants, and $B_i = b^*(V_i,X,\zeta)$, $i \in \widetilde{N}$, are their bids. The researcher observes $(X,B)$, where $B = (B_i)_{i \in \widetilde{N}}$. That is, the pre-policy data are generated by this equilibrium.

The following assumption summarizes the features of this game relevant to the decomposition approach.

\begin{assumption}\label{example2_assump}
	(i) $(X,\zeta)$ is publicly known, but valuations, $V_i$, are private information. 
	
	(ii) The policy changes the reserve price $R$ into $f^*(R)$ for some map $f^*$.
	
	(iii) The class $\Sigma'$ satisfies the following three properties.

	\quad (a) For every set $D$ of values of $(X,\zeta)$, the class $\Sigma'$ is closed under gluing across the set of payoff states whose common component lies in $D$.

	\quad (b) $\Sigma_{\mathsf{BCE}}'(G)$ and $\Sigma_{\mathsf{BCE}}'(G_f)$ are both nonempty.

	\quad (c) Any two elements of $\Sigma_{\mathsf{BCE}}'(G)$ agree at almost every payoff state, and so do any two elements of $\Sigma_{\mathsf{BCE}}'(G_f)$.
\end{assumption}

Condition (iii) leaves the class $\Sigma'$ generic, so that it covers a wide range of auction settings for which an existence and uniqueness theory is available. For example, we may take $\Sigma'$ to be the class of decision rules induced by those symmetric pure strategy profiles that consist of a threshold entry rule, prescribing entry if and only if the valuation is at least a threshold value determined by $(X,\zeta)$, and a bidding strategy that is continuous and strictly increasing in the valuation on the entry region. The existence and uniqueness of such a symmetric pure strategy BNE is well studied in this auction model. (See \cite{Guerre/Perrigne/Vuong:09:ECMA}.) Property (a) holds whenever the profiles inducing $\Sigma'$ are picked out by a restriction imposed separately at each value of $(X,\zeta)$, as in this example. Since by property (c) the equilibrium of each game in $\Sigma'$ is unique up to almost sure equality, Assumption \ref{assump: invariance} is satisfied in terms of the BCEs restricted to $\Sigma'$, and we do not require a separate invariance condition for equilibrium selection rules. We denote by $(a^{*f},b^{*f})$ a symmetric pure strategy profile whose induced decision rule belongs to $\Sigma_{\mathsf{BCE}}'(G_f)$, the post-policy game being the one in which the reserve price is $f^*(R)$. Both profiles are defined at every signal value, so that each can be evaluated at the counterfactual reserve price, and we write $\mathbb{S}_{X_1,f^*(R)}$ for the support of $(X_1,f^*(R))$.

We introduce an additional assumption that is used to identify the bounds in the decomposition approach.

\begin{assumption}
	\label{example2_assump2}
	(i) $(V,\zeta)$ is conditionally independent of $R$ given $X_1$, where $V = (V_1,...,V_n)$.

	(ii) After rearranging coordinates, $\mathbb{S}_W = \mathbb{S}_{X_1,\zeta} \times \mathbb{S}_R \times \mathbb{S}_V$ and $\mathbb{S}_{f(W)} = \mathbb{S}_{X_1,\zeta} \times \mathbb{S}_{f^*(R)} \times \mathbb{S}_V$.

	(iii) For each $i=1,...,n$, $V_{-i}$ is conditionally independent of $(X,\zeta)$ given $V_{i}$.

	(iv) On $\mathbb{S}_{X_1,R} \cap \mathbb{S}_{X_1,f^*(R)}$, the distributions of $(X_1,R)$ and of $(X_1,f^*(R))$ assign probability zero to the same sets.
\end{assumption}

Assumption \ref{example2_assump2}(i) requires that the dependence between $(V,\zeta)$ and the reserve price $R$ is fully captured by the observed auction characteristics $X_1$. Assumption \ref{example2_assump2}(ii) is Assumption \ref{assump: information structure and policy}(iv) in this setting, where the common component of the payoff state is $(X_1,R,\zeta)$ and the private component is $V$: the possible values of the auction characteristics $(X_1,\zeta)$, of the reserve price, and of the valuations vary freely of one another, before and after the policy. Assumption \ref{example2_assump2}(iii) requires that, given a bidder's own private value $V_i$, the auction characteristics $(X,\zeta)$ carry no further information about the other bidders' private values $V_{-i}$. Assumption \ref{example2_assump2}(iv) is Assumption \ref{assump: information structure and policy}(iii) in this setting, applied to the observed characteristics and the reserve price.

Our object of interest is the conditional distribution of the auction revenue under the counterfactual reserve price $f^*(R)$ given $X_1 \in C$ for some set $C$ in the support of $X_1$, as predicted by the post-policy game:
\begin{align*}
	p_f(A \mid C) \eqdef P\left\{\max_{i \in \widetilde{N}^f} B_i^f \in A \mid X_1 \in C \right\}, \quad A \subset \mathbf{R}_+,
\end{align*}
where $B_i^f = b^{*f}(V_i, X_1,f^*(R),\zeta)$ and $\widetilde{N}^f = \{i: a^{*f}(V_i,X_1,f^*(R),\zeta) = 1\}$ are the bids and the set of participants generated by the post-policy equilibrium $(a^{*f},b^{*f})$.\footnote{When no bidder participates, the maximum above is taken over an empty set. We adopt the convention that the auction then results in a designated no-sale outcome, so that $\max_{i \in \widetilde{N}} B_i \notin A$ for any $A \subset \mathbf{R}_+$ whenever $\widetilde{N} = \varnothing$, and likewise for $\widetilde{N}^f$.} We define
\begin{align*}
	p(A \mid x_1,r) \eqdef P\left\{ \max_{i \in \widetilde{N}} B_i \in A \mid (X_1,R) = (x_1,r) \right\}.
\end{align*}
The conditional probability $p(A \mid x_1,r)$ is identified for all $(x_1,r)$ in the support of $(X_1,R)$, and can be estimated from the pre-policy auction data without recovering the value distribution from the data.

\begin{corollary}\label{corollary_auction}
	Suppose that Assumptions \ref{example2_assump}-\ref{example2_assump2} hold, and let $C$ be a subset of the support of $X_1$ such that $P\{X_1 \in C\}>0$. Then, the following statements hold.

	(i) For each $i = 1, \ldots, n$, almost surely on the event $\{(X_1, f^*(R)) \in \mathbb{S}_{X_1,R}\}$,
	\begin{align*}
	  a^{*f}(V_i, X_1, f^*(R), \zeta) = a^{*}(V_i, X_1, f^*(R), \zeta),
	\end{align*}
	and, almost surely on the event $\{(X_1, f^*(R)) \in \mathbb{S}_{X_1,R}, i \in \widetilde{N}^f\}$,
	\begin{align*}
	  b^{*f}(V_i, X_1, f^*(R), \zeta) = b^{*}(V_i, X_1, f^*(R), \zeta).
	\end{align*}

	(ii) For each $A \subset \mathbf{R}_+$,
	\begin{align*}
		&\mathbf{E}\left[p( A \mid X_1,f^*(R))1\{ (X_1,f^*(R)) \in \mathbb{S}_{X_1,R} \} \mid X_1 \in C \right]\\
		&\quad \quad \quad \leq p_f(A \mid C) \leq \mathbf{E}\left[p( A \mid X_1,f^*(R))1\{ (X_1,f^*(R)) \in \mathbb{S}_{X_1,R}\} \mid X_1 \in C \right] \\
		&\quad \quad \quad \quad \quad \quad \quad \quad \quad \quad \quad + P\left\{(X_1,f^*(R)) \notin \mathbb{S}_{X_1,R} \mid X_1 \in C \right\}.
	\end{align*}
\end{corollary}
The corollary says that when $\mathbb{S}_{X_1,f^*(R)} \subset \mathbb{S}_{X_1,R}$ and the conditions of the corollary are satisfied, the counterfactual quantity $p_f(A \mid C)$ is point-identified directly from data, without first recovering the valuation distribution. Hence the prediction does not rely on the conditions invoked in the literature for the nonparametric identification of the valuation distribution, and the estimation procedure is simpler.

\section{Monte Carlo Simulations}\label{sec:mc}

We conduct Monte Carlo simulations to illustrate the properties and limitations of our theoretical results. To do so, we consider the simple two-player entry game of Section \ref{subsubsec:simple1}. We adopt the policy from Example \ref{example}, which raises the payoff state $X_i$ by $a > 0$: $f(X_i) = X_i + a$. 

We use this setting to illustrate how our bounds from Theorem \ref{thm: bounds1} perform as the policy becomes more extreme (i.e., $a$ increases). We then examine the robustness of these results when the invariance assumption is violated, allowing equilibrium selection rules to change post-policy in both moderate and extreme deviations.

\subsection{The Set-Up}

Recall from (\ref{util param}) that a firm's payoff is given by $y_i(-\delta y_{-i} + X_i'\beta_i + \varepsilon_i)$, where $y_i=1$ denotes entry and $y_i = 0$ denotes exit. We set $\delta = 1$ and take $X_i$ to be uni-dimensional. For the baseline specification, we set $\beta_i = \beta = 0.5$ for both firms, and we vary $\beta$ in our simulation study.

We draw $\varepsilon_i$ i.i.d.\ from $N(0,1)$ for $i = 1, 2$. The payoff states $(X_1,X_2)'$ are drawn from a (bivariate) truncated normal distribution with mean parameter $\boldsymbol{\mu} = (\mu_1,\mu_2)' = (0,0)'$ and variance-covariance matrix 
\begin{align*}
	\boldsymbol{\Sigma} = \begin{pmatrix} 1 & 0.3 \\ 0.3 & 1 \end{pmatrix}.
\end{align*}
We take the support of each $X_i$ to be $[-2,2]$, which lets the policy push payoff states outside this support. To simulate population objects such as the AEP and ADP, we draw the game $N_{\text{sim}} = 200{,}000$ times.

As described in Section \ref{subsubsec:simple1}, the game has a region with multiple equilibria (where one firm enters and the other stays out) and regions with a unique equilibrium. For the former, our benchmark equilibrium selection rule assigns equal probability to the two pure-strategy equilibria (a 50/50 rule). In some experiments, we then vary this rule post-policy, thereby violating the invariance condition (Assumption \ref{assump: invariance}).

To calculate the AEP in (\ref{AEP def}), we first compute the Nash equilibria of the game. Then, for the multiple equilibria region, we draw the equilibrium according to the equilibrium selection rule.\footnote{For implementation, for each draw $m$, we generate a single realization $U_m \sim \text{Uniform}(0,1)$ and assign firm~1 to enter whenever $U_m < p_{1}$. By default, $p_{1} = 0.5$ (50/50 selection). As we vary $a, \beta$ and the equilibrium selection rule ($p_1$), we use the same $\{U_m\}$. This means that any variation in equilibrium outcomes across iterations is driven entirely by changes in the parameter of interest, but not by independent re-randomization of $U_m$.} Finally, we compute the vector of firm-level means of the equilibrium entry indicators over all $N_{\text{sim}}$ draws:
	\[
	\widetilde{\mathsf{AEP}} \;=\; \bigl(\widetilde{\mathsf{AEP}}_1,\; \widetilde{\mathsf{AEP}}_2\bigr)',
	\quad \text{where} \quad
	\widetilde{\mathsf{AEP}}_i \;=\; \frac{1}{N_{\text{sim}}} \sum_{m=1}^{N_{\text{sim}}} Y_{i,m}, \quad i = 1, 2.
	\]
The quantities $\widetilde{\mathsf{AEP}}_i$, $i=1,2$, are simulation-based approximations to the \emph{population} AEP. They are not finite-sample estimators and carry no sampling error beyond Monte Carlo noise.

As for the decomposition-based approach, we compute the bounds in (\ref{bound2}) for each policy. The quantity $\Delta_a \eqdef P\{f_a(X) \notin \mathbb{S}_X\}$ is computed as the fraction of post-policy observations that fall outside this support. This quantity depends on the policy $a$.  As for $\mathsf{ADP}_R$, it is given in closed form in equation (\ref{ADPR_example}). We specify the conditional expectation 
\begin{align*}
	m_i(x_1,x_2) = \mathbf{E}[Y_i \mid (X_1,X_2) = (x_1,x_2)],
\end{align*}
for each firm $i$ as a quadratic regression function and estimate it by regressing the pre-policy realized outcomes $Y_{i,m}$ on $X_{1,m}$, $X_{2,m}$, $X_{1,m}^2$, $X_{2,m}^2$ and $X_{1,m} X_{2,m}$ for all the simulated draws. The resulting OLS estimates yield fitted values $\widetilde{Y}_{i,m}$ for each draw, which we average to obtain
\[
	\widetilde{\mathsf{ADP}} \;=\; \bigl(\widetilde{\mathsf{ADP}}_1,\; \widetilde{\mathsf{ADP}}_2\bigr)',
	\qquad \text{where} \quad
	\widetilde{\mathsf{ADP}}_i \;=\; \frac{1}{N_{\text{sim}}} \sum_{m=1}^{N_{\text{sim}}} \widetilde{Y}_{i,m}, \quad i = 1, 2.
	\]
The quantity $\widetilde{\mathsf{ADP}}_R$ is computed analogously to $\widetilde{\mathsf{ADP}}$, except that each predicted outcome is included only when the post-policy payoff state falls within the pre-policy support:
\begin{align*}
	\widetilde{\mathsf{ADP}}_{R,i} = \frac{1}{N_{\text{sim}}} \sum_{m=1}^{N_{\text{sim}}} \widetilde{Y}_{i,m}\,\mathbf{1}\{f_a(X_m) \in \mathbb{S}_X\}.
\end{align*}
Because of the symmetry of this game, we present the results for firm 1 only.

\begin{figure}[t]
	\centering
			\includegraphics[width=0.45\textwidth]{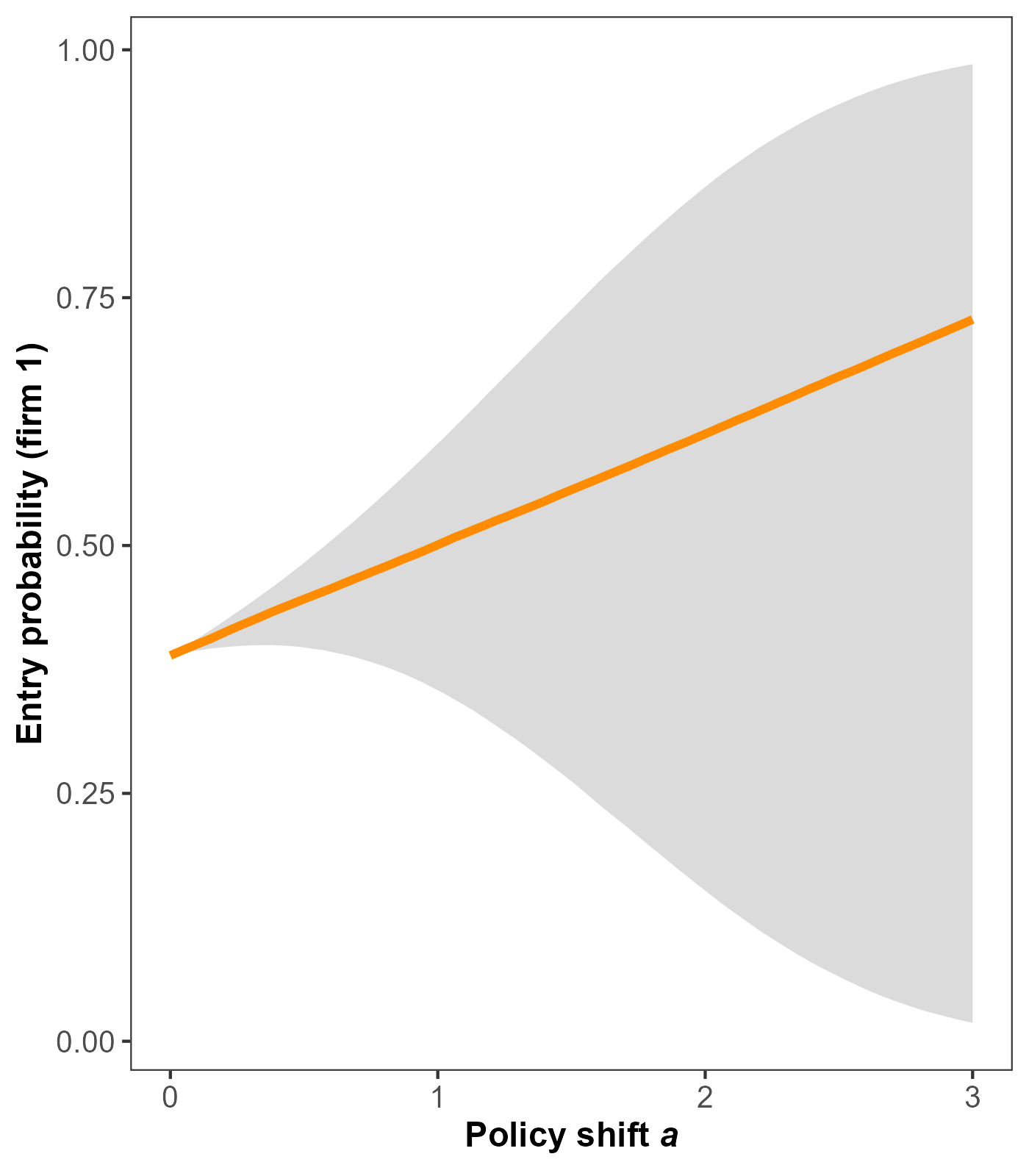}
			\caption{AEP and Bounds from Theorem \ref{thm: bounds1} When the Invariance Condition Holds}
	\medskip
	\parbox{\textwidth}{\footnotesize \textit{Notes}: The solid line is the equilibrium prediction (AEP) and the shaded region depicts the bounds from Theorem \ref{thm: bounds1}. Because the invariance condition holds, the bounds are valid: the AEP lies within the shaded region throughout. The bounds widen as the policy shift increases.}		
			\label{fig2}
			\bigskip
		\end{figure}

\subsection{Results}

Our first experiment fixes the equilibrium selection rule to be invariant and varies the policy from $a=0$, which leaves the payoff state within the pre-policy support, to $a=3$, which pushes it far outside. Figure \ref{fig2} reports the results. The equilibrium prediction (AEP) lies well inside the theoretical bounds of equation (\ref{bound2}) throughout, and, as expected, the bounds widen as $a$ grows, since $\Delta$ increases the further the payoff state moves away from the support.

\begin{figure}[t]
			\centering
			\includegraphics[width=\textwidth]{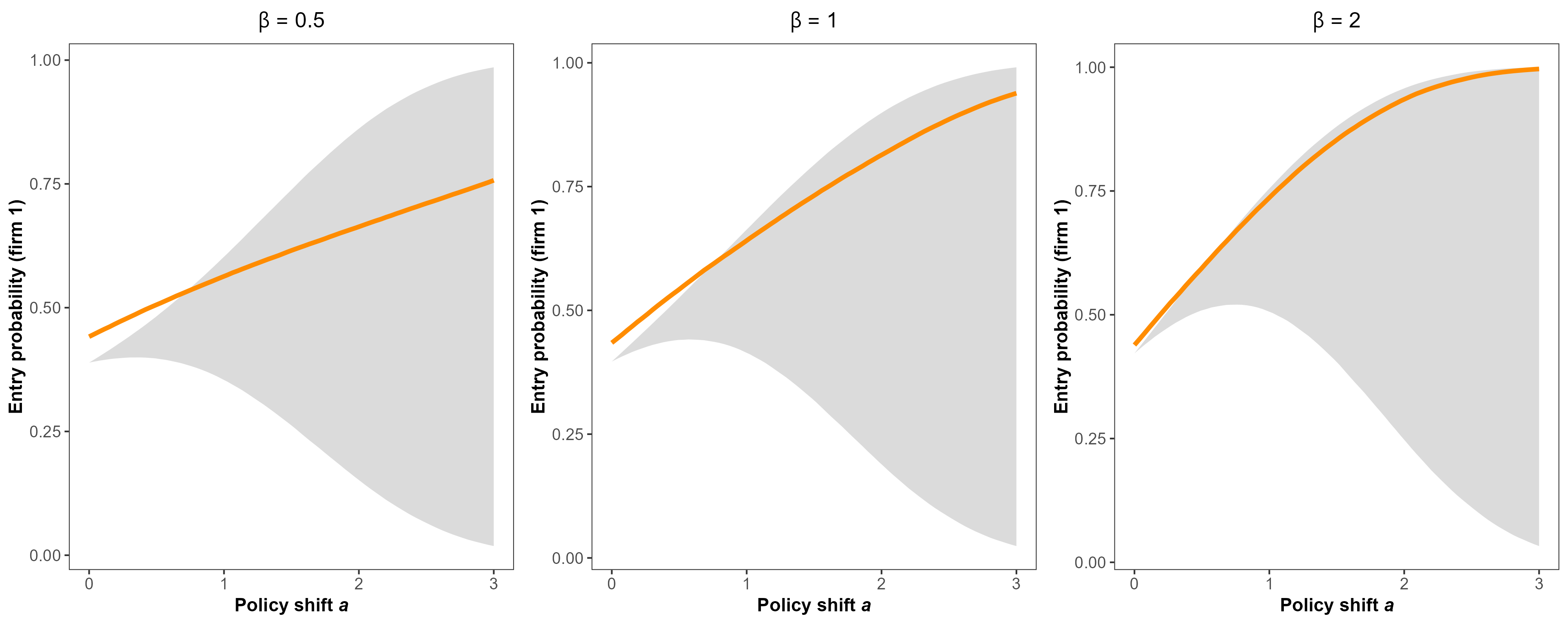}
			\caption{AEP and Bounds from Theorem \ref{thm: bounds1} When the Invariance Condition Fails}
	\medskip
	\parbox{\textwidth}{\footnotesize \textit{Notes}: The equilibrium selection rule changes from $(50,50)$ to $(100,0)$ after the policy. Thus, the invariance condition fails. The left, center, and right panels correspond to $\beta = 0.5$, $1$, and $2$, respectively. Due to the failure of the invariance condition, the AEP can fall outside the bounds.}
		\label{fig3}
		\bigskip
	\end{figure}

In the second set of exercises, we study how the bounds work when the invariance condition fails. To do so, we change the equilibrium selection rule after the policy: while the pre-policy outcomes in the multiple equilibria region ($0 < X_1'\beta+\varepsilon_1 < \delta$ and $0 < X_2'\beta+\varepsilon_2 < \delta$) are generated from a 50/50 draw, we now choose 100\% of the time the equilibrium in which firm 1 enters after the policy. The results are shown in Figure \ref{fig3}, where each panel corresponds to a different value of $\beta \in \{0.5,1,2\}$. As before, the solid line is the AEP and the shaded region depicts the bounds from our main theorem.

The figure shows that violating the invariance condition can invalidate the bounds, particularly when the policy shift is small. Under invariance, our results imply that ADP = AEP at $a=0$, so the AEP must lie within the bounds. Once invariance fails, this no longer holds, because one cannot extrapolate from the pre-policy data, and the AEP can fall outside. As the policy shift $a$ increases, however, the AEP often re-enters the bounds, since the bounds widen the further the payoff state moves beyond the support of the data.

Our final set of exercises examines how closely the ADP tracks the AEP across values of $\beta$. We also compare versions of the ADP and AEP that condition on whether the realized data fall in the single- or multiple-equilibria region. We study the case where $a=0$ (and, therefore, $\Delta = 0$): this is the setting where theory predicts that ADP and AEP coincide.\footnote{When the policy pushes the payoff state beyond its support, our theory delivers bounds rather than a single value. In such a case, we therefore neither use nor recommend the ADP as a point prediction. Our results speak only to the bounds.} Thus, we can estimate the discrepancy between the two average predictions for firm 1 as $\bigl|\widetilde{\mathsf{AEP}}_1 - \widetilde{\mathsf{ADP}}_1\bigr|$. We compute this discrepancy across specifications of the equilibrium selection rule, ranging from one in which the invariance condition holds (so the discrepancy should be 0) to specifications in which it fails by varying degrees.

	\begin{figure}[t]
		\centering
  \makebox[\textwidth][c]{
		\includegraphics[width=1.0\textwidth]{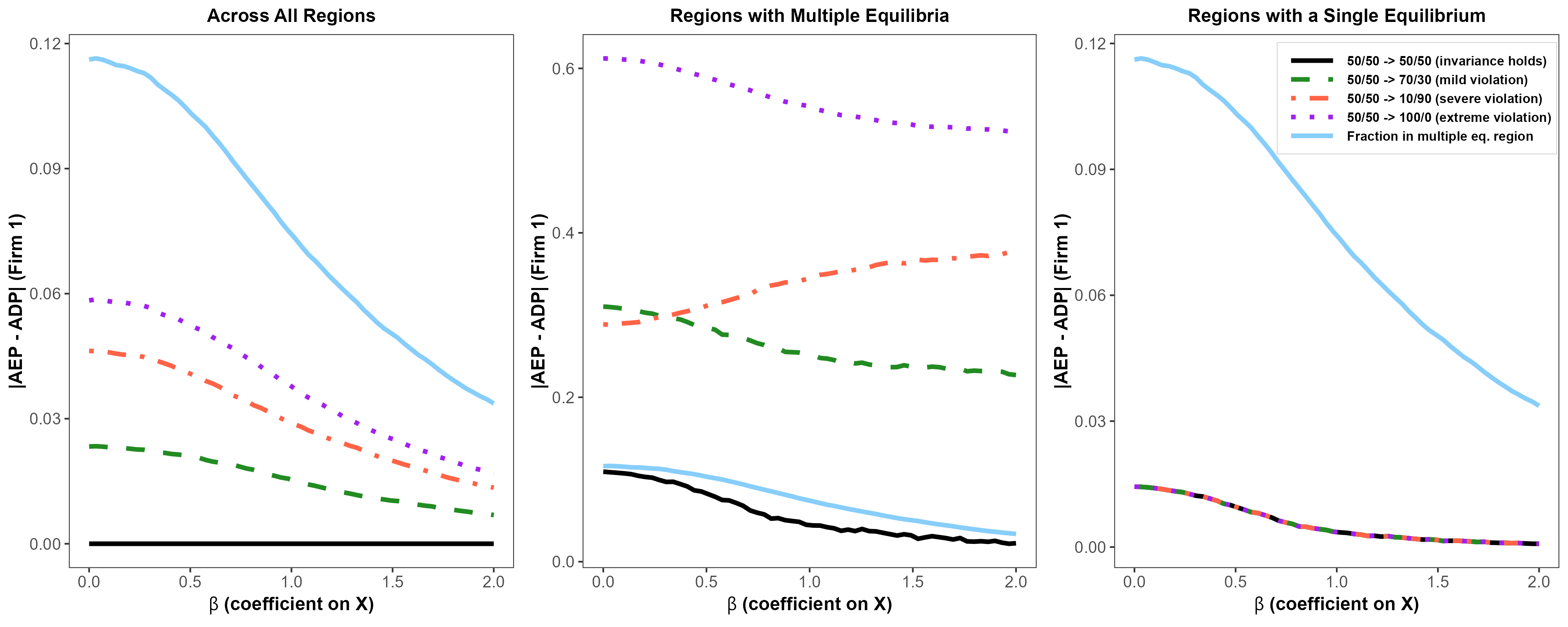}
}
		\caption{Discrepancy Between the ADP and AEP across Equilibrium Selection Rules}
	\medskip
	\parbox{\textwidth}{\footnotesize \textit{Notes}: We plot the bias $|\mathsf{AEP} - \mathsf{ADP}|$ as $\beta$ varies, fixing the policy at $a=0$ (so $\Delta = 0$). The left panel reports the baseline (unconditional) bias, while the middle and right panels report the conditional versions, conditioning on data realized in the multiple-equilibria region (middle) and the single-equilibrium region (right). We study four post-policy equilibrium selection rules: $50/50$ to $50/50$ (baseline, invariance holds), $50/50$ to $70/30$, $50/50$ to $10/90$, and $50/50$ to $100/0$. When the invariance condition holds, $\mathsf{ADP}=\mathsf{AEP}$ and the bias is 0. Since invariance holds automatically in the single-equilibrium region, all four lines coincide in the right panel. In the middle and right panels, the conditional versions of ADP and AEP show a mild gap under the invariance condition as they are conditioned on multiple-equilibria and single-equilibrium regions. The gap is reduced when $\beta$ becomes large. The middle- and right-panel biases are of opposite sign and cancel upon aggregation (left panel), confirming zero bias for the unconditional ADP, as expected.}
		\label{fig4}
		\bigskip
	\end{figure}

The results are shown in Figure \ref{fig4}. As expected, varying the equilibrium selection rule makes no difference in regions of a single equilibrium: the invariance condition is satisfied, so there is no bias from using the (conditional) ADP for counterfactual predictions. Similarly, in the region of multiple equilibria, when the invariance condition holds, there is no bias from using the decomposition-based methods. This holds across values of $\beta$. By contrast, biases arise when there are multiple equilibria and the invariance condition is violated, as shown in the middle panel of Figure \ref{fig4}, where the equilibrium selection rule changes after the policy. These conditional biases then aggregate into a bias in the unconditional ADP (left panel), which shrinks as $\beta$ grows, since the region of multiple equilibria contracts. In sum, stronger violations of the invariance condition produce larger biases, and these are most pronounced in specifications where multiple equilibria are prevalent. Even so, the biases remain small (under $0.03$ across all values of $\beta$) when violations of the invariance condition are mild and the multiple-equilibria region is small.

\section{Empirical Application: Ciliberto and Tamer (2009) Revisited}
\label{sec:empirical}

\subsection{Background}

We revisit the counterfactual analysis in \cite{Ciliberto/Tamer:09:Eca} using our results from Section \ref{example1_detail}.  They investigated the effect of the repeal of the Wright Amendment on airline entry in markets out of Dallas Love Field Airport.  The Wright Amendment had been implemented in 1979 to stimulate the use of the newer (and not as central) Dallas Fort Worth (DFW) Airport. As of the early 2000s, it restricted the flights out of the central Dallas Love Field to other cities in Texas or those from some neighboring states. In 2008, the major airlines and DFW Airport agreed to a full repeal of the amendment, to be implemented in 2014.\footnote{Most notably, the agreement reduced the number of gates at Dallas Love Field to limit its competitive impact on DFW.} The repeal could have led to significant changes in market entry, and hence in consumer welfare.

\cite{Ciliberto/Tamer:09:Eca} produced a counterfactual prediction of the outcomes after the repeal of the Wright Amendment, after estimating the identified set from a complete information entry game permitting multiple equilibria. This is the game presented in Section \ref{example1_detail}.\footnote{We provide extensive details on the empirical application, including the description of the covariates, estimators and inference procedures in the Online Appendix.} A market was defined by a route between two airports, irrespective of directions or layovers. Thus, this definition included connecting flights through a third airport. 

They modeled the Wright Amendment as an indicator covariate, $X_{i,m}^{\text{Wright}}$, equal to 1 if market $m$ was subject to the amendment (which applied to all firms in the market) and 0 otherwise. Their counterfactual experiment repeals the amendment by setting $X^{\text{Wright}}_{i,m}=0$ and studies its effect on market entry. In this setting, $\Delta = 0$: all covariates $X$ beyond the Wright Amendment remain the same after the policy, retaining the same support. Meanwhile, the support of $X^{\text{Wright}}_{i,m}$ is $\{0,1\}$ pre-policy, and $\{0\}$ post-policy: a strict subset. Hence, the probability that the policy takes $X_i$ beyond its pre-policy support is zero.\footnote{The support $\mathbb{S}_X$ in this setting is the support of $X_{i,m}$ across all pre-policy data (firms and markets). Thus, even though some markets change $X_{i,m}^{\text{Wright}}$ from 1 to 0, this change is still within the pre-policy support $\mathbb{S}_X$, which does not change post-policy.} Following \cite{Ciliberto/Tamer:09:Eca}, we assume that unobservable payoff components $\varepsilon_i$ are i.i.d.\ and independent of all covariates. Corollary \ref{example1_corollary} says that the decomposition-based prediction coincides with the equilibrium-based prediction and can be obtained from estimating all structural parameters under the equilibrium invariance condition.
 
\subsection{The Set-Up}
For this application, we follow \cite{Ciliberto/Tamer:09:Eca}, using their dataset and focusing on the four main airlines in their analysis (American, Delta, Southwest, and United). We perform a dimension reduction to resolve near multicollinearity between covariates. This reduction is useful because, in this strategic setting, the decomposition-based prediction requires including every firm-level covariate in $X_m$: the covariates $X_{j,m}$ of each rival firm $j \neq i$ shape firm $i$'s equilibrium entry decision by influencing whether firm $j$ enters.\footnote{In this context there are eight market-level covariates and two variables at the firm-market level. This generates a total of 16 covariates to be included in the analysis. (They are described in detail in the Online Appendix.) While the parametric estimator performs well with all 16 covariates, the nonparametric estimator performs better on a smaller subset.} We drop three out of eight market-level variables (market size, per capita income growth and Dallas market) that lack independent variation because of the way the data are constructed.\footnote{Both market size and per capita income growth appear well predicted by income per capita and market presence (variables that already capture economic performance at the market level and are included in the analysis). Meanwhile, the binary Dallas market variable is highly correlated with the Wright Amendment variable: by construction, any market in Dallas that does not use Dallas Love Field Airport must be using Dallas Fort Worth Airport instead. However, Dallas Fort Worth is the hub for American Airlines, and American Airlines' market presence is already included as a covariate. Thus, we assume that the dropped covariates do not affect entry conditional on the included covariates. Details are provided in the Online Appendix.} We also drop one additional firm-market-level covariate (a proxy for the firm's cost), because this variable is a deterministic function of other covariates in the analysis, such as route distance, by construction.

\subsection{Results}
We compare the results from the decomposition approach to the original results in \cite{Ciliberto/Tamer:09:Eca}. The results of this exercise are shown in Columns 1 and 2 of Table \ref{ct_decomp} using two different estimators (a linear/parametric and a nonparametric estimator), while their original results are shown in Column 3. 

In the first column, we adopt, as a benchmark, the parametric specification in which firm $i$'s expected entry in market $m$ is linear in the market characteristics: $\mathbf{E}[Y_{i,m} \mid X_m] = X_m'\gamma_i$, where $Y_{i,m}$ indicates whether firm $i$ enters market $m$. We estimate $\gamma_i$ by least squares and report the estimated policy effect: the change in entry in the post-policy game relative to the observed data, for markets previously affected by the policy. This change can be written as
\begin{align*}
	\frac{1}{|\mathcal{M}|} \sum_{m \in \mathcal{M}} f^*(X_m)'\hat{\gamma}_i - \overline{Y}_i,
\end{align*}
where $\hat \gamma_i$ is the estimator of $\gamma_i$, $\mathcal{M}$ is the set of markets previously affected by the Wright Amendment, $f^*(X_m)$ is the vector of covariate values for market $m$ after the policy, and $\overline{Y}_i$ is firm $i$'s average outcome in the data across markets in $\mathcal{M}$. In the second column, we estimate the conditional expectation $g_i(x) = \mathbf{E}[Y_{i,m} \mid X_m=x]$ nonparametrically, using a leave-one-out kernel regression estimator with a bandwidth chosen by cross-validation. (See the Online Appendix for details.) 

The linear specification in Column 1 should be viewed as an approximation to the nonparametric specification in Column 2, since the latter is the form induced by the equilibria of the entry game. Even so, for the average policy effects, the two specifications yield similar results for Southwest and United.

In the third column, we reproduce the counterfactual predictions from the main specification in \cite{Ciliberto/Tamer:09:Eca} (Table VII, Column 1). Their estimates report the maximum predicted increase in the share of Dallas Love Field markets served by each airline following the 2014 repeal of the Wright Amendment. Our predictions across specifications are broadly consistent with theirs, lying below this estimated maximum.\footnote{The nonparametric estimator in Table \ref{ct_decomp} has smaller standard errors than the parametric version, since it includes fewer variables due to the variable selection described in the previous section.}

Now we take this exercise one step further. The Wright Amendment was actually repealed in 2014. This means that we can observe how airlines entered the markets after the repeal of the Wright Amendment. We compile the 2015 data from the DB1B Market and Ticket Origin and Destination Dataset of the U.S. Department of Transportation (the same source as the original dataset), and treat it the same way as the original authors. See the Online Appendix for details. We focus on the same 81 markets from the original paper. The actual change in entry in 2015 in the data relative to the original data is shown in Column 4 of Table \ref{ct_decomp}. We then compare the counterfactual estimates from \cite{Ciliberto/Tamer:09:Eca} and our decomposition approach to the actual changes. 

\begin{table}[t]
	\begin{centering}
		\small
		\caption{\small \cite{Ciliberto/Tamer:09:Eca} Revisited: Model Predicted and Empirical Counterfactuals of the Repeal of the Wright Amendment}
		\label{ct_decomp}
		\resizebox{\columnwidth}{!}{%
			\begin{tabular}{cccccc}
				\hline 
				\hline 
				\tabularnewline
				& &\multicolumn{4}{c}{Outcome: Change in Probability of Entry in Dallas Love Markets} \\
				\\
				\cline{2-6}
				\tabularnewline
				&   & Decomposition Method & Decomposition Method & Ciliberto \& Tamer (2009) & Empirical \\
				&   &  Linear Model & Nonparametric Model  &  Maximum Predicted Entry & \\
				\tabularnewline
				\hline
				\multicolumn{1}{c}{} & &  &  &  & \\
				American Airlines & & -0.030 & 0.128 &  0.463 & -0.04   \\
				& & (0.037) & (0.036) & &\\
				\tabularnewline   
				Delta Airlines &  & -0.023 & 0.174  &  0.499 & 0.46  \\
				&		& (0.043)& (0.039) & &\\
				\tabularnewline    
				Southwest Airlines &  & 0.508 & 0.451  & 0.474  &  0.471 \\
				& & (0.037)&(0.056) & &\\
				\tabularnewline   
				United Airlines &  & -0.009 & 0.043  & 0.147  &  0 \\
				& &(0.031) &(0.016) & &\\
				\\
				\hline 
				\multicolumn{1}{c}{} & &  &  &  &\\
			\end{tabular}
		}
		\par\end{centering}
	\parbox{\textwidth}{\footnotesize
		\textit{Notes}: We report the estimated counterfactual changes to the entry of major airlines into Dallas Love Field markets following the repeal of the Wright Amendment. In Columns 1-2, we use our decomposition approach to provide point estimates of this counterfactual effect, using the same pre-2014 dataset of \cite{Ciliberto/Tamer:09:Eca}. Column 1 uses a linear model, while Column 2 reports a nonparametric estimate. Standard errors for these columns are computed by the bootstrap, following the approach in the Online Appendix with $B=999$ replications. In the third column, we reproduce the results in Table VII, Column 1 of \cite{Ciliberto/Tamer:09:Eca}, who presented the maximum change in entry of those airlines. Finally, the Wright Amendment was fully repealed in 2014, allowing all airlines to enter those markets.  The final column shows the realized values of the change in entry for those airlines in affected markets in 2015, after the repeal.}
\end{table}

A comparison of the pre-repeal predictions (Columns 1 and 2) with the realized post-repeal outcomes (Column 4) shows that both the parametric and nonparametric decomposition estimates capture the large increase in entry by Southwest and the negligible entry by American and United. This pattern is broadly consistent with multiple equilibria in an entry model: Southwest and Delta entered frequently after the repeal, while American and United stayed out. These realized values also fall within the maximum bounds reported in \cite{Ciliberto/Tamer:09:Eca}. Because the authors report only the maximum predicted entry, however, their predictions appear further from the realized values for American and United. The near-zero entry of these firms is consistent with that maximum, but it implies a predicted range of 0 to 50\% of markets entered, a wide range for policy analysis. 

However, the method does not perform well in predicting entry by Delta Airlines. This could simply be a feature of out-of-sample prediction, possibly due to changes to Delta between 2008 and 2014 (including the acquisition of Northwest Airlines, which was completed in 2010), and/or due to the definition of markets in this dataset.\footnote{Delta Airlines only operates from Dallas Love Field to Atlanta, but there are multiple connecting flights from Atlanta. Routes that include layovers are considered as separate markets by \cite{Ciliberto/Tamer:09:Eca}.} 

In the Online Appendix, we provide results for the other component in the typical aggregate decomposition (e.g., Oaxaca-Blinder): the average difference in entry across markets that are not subject to the Wright Amendment and those that are, attributable to observable characteristics alone. After all, airlines would be less likely to enter smaller markets, even if the Wright Amendment had never been in place or had been repealed. We show that a naive comparison of entry across different markets (ignoring the difference in characteristics) would overstate the effects of the policy. There, we also provide another empirical application on how the threat of entry by Southwest Airlines, originally studied by \cite{Goolsbee/Syverson:08:QJE}, affects competitor entry.

\section{Conclusion}

Decomposition methods are appealing in counterfactual analysis for their computational tractability and simplicity. However, in strategic settings, the resulting predictions may not be incentive compatible after the policy, nor account for the new coordination possibilities it induces. In this paper, we have presented a set of sufficient conditions that validate the use of the methods in strategic settings. Under the invariance condition, the agents can be viewed as playing ``the same game'' after the policy. Our result opens up a new approach to counterfactual analysis in strategic settings, bypassing the need to recover structural parameters or the set of equilibria. While the underlying intuition may seem simple, our contribution is to formalize the conditions under which the approach is valid and to delineate its scope.

There are several extensions of this paper's proposal that look promising to us. The most prominent extension is to explore a set of conditions for the decomposition-based approach in a dynamic game setting. A policy in these games generally induces a change in the agents' posteriors through a change in the future path of the payoff states, and it seems nontrivial to maintain the invariance of the posterior after the policy. \cite{Kocherlakota:19:JME} introduced independent shocks to the policy so that the posterior of the private sector for future policies remains invariant. Future work can expand on this insight and explore the validity of decomposition-based predictions in a dynamic setting. 

Our proposal can generate a wide range of intermediate approaches where the target of the prediction is of the form $h(Y,W;\theta)$, and $\theta$ is part of the structural parameters. Then, our message is that, under the conditions stated in this paper, we can perform a counterfactual analysis using the decomposition method after identifying $\theta$ alone. For example, such an intermediate approach can be used to extend this paper's framework to counterfactual analysis where the target of the prediction is welfare or profits.

\putbib[counterfactual2]
\end{bibunit}

\begin{bibunit}[econometrica]
 \newpage
 \appendix
 \numberwithin{equation}{section}

\fontsize{11pt}{13pt}\selectfont
 
 \begin{center}
 	\Large \textsc{Online Appendix to ``A Decomposition Approach to Counterfactual Analysis in Game-Theoretic Models''}
 \end{center}
 
 \date{%
 	\today%
 }
 
 \vspace*{3ex minus 1ex}
 \begin{center}
 	Nathan Canen and Kyungchul Song\\
 	\textit{University of Warwick and CEPR, and University of British Columbia}
 	\bigskip
 	\bigskip
 	\bigskip
 \end{center}

This Online Appendix consists of three appendices. Appendix \ref{App:Counterfactual Predictions in Bayesian Games} formally introduces Bayesian games in which each player's signal is a deterministic function of the payoff state, presents the decomposition-based and equilibrium-based predictions together with equilibrium selection rules with full measure-theoretic details, addresses the well-definedness and identification of these predictions, and proves the results on the validity of the decomposition approach that underlie the main text. Appendix \ref{App:Implementation of Decomposition Methods} provides details on the implementation of the decomposition method. This includes a step-by-step overview of the nonparametric estimation method used in Section \ref{sec:empirical} in the main text, as well as details about the empirical application (including data collection, empirical specifications and the implementation of the parametric and nonparametric estimators). It then provides a second empirical application based on the effects of Southwest Airlines' threat of entry on competitors' entry. Finally, Appendix \ref{App:Examples that Fail the Decomposition-Based Predictions} provides the example referenced in Section \ref{sec3}, where decomposition-based predictions do not coincide with equilibrium-based predictions, due to the failure of incentive compatibility in the counterfactual game after the policy.

\newpage

\begin{center}
	\Large \textsc{Contents}
\end{center}
\bigskip
\bigskip
\bigskip

\setcounter{section}{0}
 \titlecontents{section}[1em]{\normalfont \bfseries}
 {\contentslabel{2.3em}}
 {\hspace*{-2.3em}}
 {\titlerule*[1pc]{.}\contentspage}
 
\startcontents[sections]
\printcontents[sections]{}{0}{\setcounter{tocdepth}{2}}

 \newpage
 
 \section{Counterfactual Predictions in Bayesian Games \label{App:Counterfactual Predictions in Bayesian Games}}

 \subsection{The Set-Up: Games, Policies, and Selection Rules}
 \label{subsec:Bayesian Games}

\subsubsection{Basic Games}

We give a mathematical set-up of the Bayesian game with $n$ players introduced in the main text, where each player is indexed by $i=1,...,n$. The payoff state space has the product form
\begin{align*}
	\mathbb{W} = \mathbb{W}_C \times \mathbb{W}_I, \quad \mathbb{W}_I = \prod_{i=1}^n \mathbb{W}_{I,i},
\end{align*}
where $\mathbb{W}_C$ and $\mathbb{W}_{I,i}$, $i=1,...,n$, are complete separable metric spaces, so that $\mathbb{W}$ is endowed with the product metric and its Borel $\sigma$-field $\mathcal{B}(\mathbb{W})$, and we write the payoff state as $W = (W_C,W_I)$, $W_I = (W_{I,1},...,W_{I,n})$, with $W_C$ the common component and $W_{I,i}$ player $i$'s private component.\footnote{In the case of complete information games where there is no private payoff state, it suffices to take $\mathbb{W}_{I,i}$ to be a singleton.} Let $\mu_W$ be a probability measure on $(\mathbb{W},\mathcal{B}(\mathbb{W}))$, the prior distribution from which the payoff state $W$ is drawn. The distribution $\mu_W$ may have discrete and continuous components.

Throughout this appendix, the action spaces $\mathbb{Y}_i$, $i=1,...,n$, are complete separable metric spaces, so that $\mathbb{Y} = \mathbb{Y}_1 \times \cdots \times \mathbb{Y}_n$ is a complete separable metric space in the product metric, endowed with its Borel $\sigma$-field $\mathcal{B}(\mathbb{Y})$.

For each $i=1,...,n$, let $u_i: \mathbb{Y} \times \mathbb{W} \rightarrow \mathbf{R}$ be a measurable payoff function. We assume that each $u_i$ admits a measurable $W$-envelope $g_i: \mathbb{W} \rightarrow \mathbf{R}_+$, with $|u_i(y,w)| \le g_i(w)$ for all $(y,w) \in \mathbb{Y} \times \mathbb{W}$. Following BM, we call the quadruple $B = (\mathbb{Y},\mathbb{W},u,\mu_W)$ a \bi{basic game}, where $u = (u_i)_{i=1}^n$ denotes the profile of payoff functions. 

\subsubsection{Counterfactual Policies}
\label{subsubsec: policies}
We model the policy as a measurable map $f: \mathbb{W} \to \mathbb{W}$, which transforms the payoff state $W$ into a new payoff state $f(W)$. We define
\begin{align*}
	\mu_W^f = \mu_W \circ f^{-1}.
\end{align*}
The probability measure $\mu_W^f$ represents the distribution of the payoff state after the policy $f$ is implemented. Throughout, for any random element $Z$, we write $\mathbb{S}_Z$ for its support under $\mu_W$. Recall that on a separable metric space the support of a Borel probability measure carries full measure. In particular, $\mathbb{S}_W$ denotes the support of $\mu_W$, $\mathbb{S}_{f(W)}$ that of $\mu_W^f$, and the common support: 
\begin{align*}
	\mathbb{S}_\cap \eqdef \mathbb{S}_W \cap \mathbb{S}_{f(W)}.
\end{align*}
This appendix focuses on policies that transform only the common component of the payoff state,
\begin{align*}
	f(w) = (\tilde f(w_C), w_I), \quad w = (w_C, w_I) \in \mathbb{W},
\end{align*}
for some measurable map $\tilde f: \mathbb{W}_C \to \mathbb{W}_C$. Furthermore, we assume that $\mu_W$ and $\mu_W^f$ are mutually absolutely continuous on $\mathbb{S}_\cap$. Accordingly, for a Borel set $B \subset \mathbb{S}_\cap$, a property holding at $\mu_W$-a.e.\ $w \in B$ holds at $\mu_W^f$-a.e.\ $w \in B$, and conversely. Thus, we simply write \bi{at a.e.\ $w \in B$} without naming the measure whenever $B \subset \mathbb{S}_\cap$. 

\subsubsection{Information Structure and Bayesian Game}

We introduce the information structure of the game. As in the main text, each player $i$ observes the signal
\begin{align*}
	T_i = s_i(W), \quad s_i(w) = (w_C, w_{I,i}), \quad w = (w_C,w_I) \in \mathbb{W},
\end{align*}
taking values in $\mathbb{T}_i = \mathbb{W}_C \times \mathbb{W}_{I,i}$. We write $s = (s_1,\ldots,s_n)$ for the joint signal map and $\mathbb{T} = \mathbb{T}_1 \times \cdots \times \mathbb{T}_n$, so that the signal profile $T = (T_1,...,T_n) = s(W)$ is a deterministic function of the payoff state.
The belief map of player $i$ is a Markov kernel $\eta_i: \mathcal{B}(\mathbb{W}_{I,-i}) \times \mathbb{T}_i \to [0,1]$, which is a version of the disintegration of $\mu_W$ along its $(\mathbb{W}_C \times \mathbb{W}_{I,i})$-marginal $\mu_{W_C,W_{I,i}}$: it represents the conditional distribution of the opponents' private components $W_{I,-i}$ given $T_i = (W_C,W_{I,i})$. The disintegration is unique $\mu_{W_C,W_{I,i}}$-a.e.\ (cf. Theorem 3.4 of \cite{Kallenberg:21:Foundations}). We fix the version $\eta = (\eta_1,...,\eta_n)$ as part of the information structure. As part of the information structure, $\eta$ does not change after the policy.\footnote{Fixing a policy-invariant version of the belief map avoids the technical issues arising from null events of signals that the policy affects. A belief map specified only up to null sets of signals under the pre-policy state distribution would leave the post-policy game's obedience conditions unrestricted at signals that the pre-policy game sends with probability zero, but the post-policy game sends with positive probability. As a result, without fixing a version of the belief map, counterfactual predictions from the game may not be mathematically well-defined.} We require the fixed version $\eta$ to control the payoff envelopes uniformly across signals: for each $i=1,...,n$,
\begin{align}
	\label{eq: envelope compatibility}
	\sup_{(w_C,w_{I,i}) \in \mathbb{T}_i} \int g_i(w_C,w_{I,i},w_{I,-i}) \, d\eta_i(w_{I,-i} \mid w_C,w_{I,i}) < \infty,
\end{align}
where we rearrange the coordinates of $w = (w_C,w_{I,i},w_{I,-i})$ in the obvious manner. 

We take the tuple $I = (\mathbb{W},\mathbb{T},s,\eta)$ to be the \bi{information structure}, and, following BM, call the pair $G = (B,I)$ a \bi{Bayesian game}. 

Similarly, the post-policy Bayesian game is given by $G_f = (B_f,I)$, where 
\begin{align*}
	B_f = (\mathbb{Y},\mathbb{W},u,\mu_W^f).
\end{align*}
Comparing $G = (B,I)$ with $G_f = (B_f,I)$, it is clear that the counterfactual policy keeps the information structure $I$ and the payoff functions $u$ unchanged, while changing only the distribution of the payoff state from $\mu_W$ to $\mu_W^f$.

\subsubsection{Bayes Correlated Equilibria}

We introduce the space of decision rules which plays the role of the strategy space for the Bayesian game $G$. Since the signal profile is a deterministic injective transform $T = s(W)$ of the payoff state, a decision rule needs to condition only on the state. Let $\mathcal{K}$ be the set of Markov kernels $\kappa(\csdot \mid \csdot): \mathcal{B}(\mathbb{Y}) \times \mathbb{W} \rightarrow [0,1]$, each such kernel being Borel measurable in $w$ for every $A \in \mathcal{B}(\mathbb{Y})$ and a probability measure on $\mathcal{B}(\mathbb{Y})$ for every $w \in \mathbb{W}$. We take the space of \bi{decision rules} to be $\Sigma = \mathcal{K}$, following BM.

For each $i=1,...,n$, $t_i = (w_C,w_{I,i}) \in \mathbb{T}_i$, $\sigma \in \Sigma$, and a measurable map $\tau_i: \mathbb{Y}_i \rightarrow \mathbb{Y}_i$, we write the expected payoff of player $i$ as
\begin{align}
	\label{interim payoff}
	U_i(\tau_i,t_i;\sigma) = \int \int u_i(\tau_i(y_i),y_{-i},w)d\sigma(y_i,y_{-i} \mid w) d\eta_i(w_{I,-i} \mid w_C,w_{I,i}),
\end{align}
where $w = (w_C,w_{I,i},w_{I,-i})$ with the coordinates rearranged in the obvious manner. By the envelope bound $|u_i| \le g_i$ and (\ref{eq: envelope compatibility}), the interim expected payoff $U_i(\tau_i,t_i;\sigma)$ is well-defined and finite at every $t_i \in \mathbb{T}_i$, for every decision rule $\sigma$ and every measurable $\tau_i$.

Following BM, we say that a decision rule $\sigma \in \Sigma$ is a \bi{Bayes Correlated Equilibrium (BCE)} for $G$, if for each $i =1,...,n$, there is a $\mu_{T_i}$-null set of signals $t_i$ off which
\begin{align}
	\label{BCE obedience}
	U_i (\mathsf{Id},t_i;\sigma) \ge U_i (\tau_i,t_i;\sigma), \text{ for every measurable } \tau_i : \mathbb{Y}_i \rightarrow \mathbb{Y}_i,
\end{align}
where $\mathsf{Id}$ is the identity map $\mathbb{Y}_i \rightarrow \mathbb{Y}_i$, and $\mu_{T_i}$ denotes the $\mathbb{T}_i$-marginal of $\mu_W$. We let $\Sigma_{\mathsf{BCE}}(G) \subset \Sigma$ be the set of BCEs of the Bayesian game $G$.\footnote{Obedience in (\ref{BCE obedience}) is a property of the $\mu_W$-equivalence class of a decision rule. If $\sigma' = \sigma$ at $\mu_W$-a.e.\ $w$, then, for each $i$, at $\mu_{T_i}$-a.e.\ $t_i$ we have $U_i(\tau_i,t_i;\sigma') = U_i(\tau_i,t_i;\sigma)$ for every $\tau_i$ simultaneously. Hence $\Sigma_{\mathsf{BCE}}(G)$ is a union of $\mu_W$-equivalence classes of kernels.}

\subsubsection{Equilibrium Selection Rule}
\label{subsubsec: equivalence classes}

An equilibrium decision rule $\sigma$ summarizes the causal relationship between the outcome of the game and the payoff state. However, we face ambiguity about the relationship due to the multiplicity of BCEs. Resolving this ambiguity requires an equilibrium selection rule over equilibria, which is modeled as a probability measure on the set of decision rules. We now formulate this set as a measurable space.

For a probability measure $\mu$ on $\mathcal{B}(\mathbb{W})$, let $\widetilde\Sigma_\mu$ be the set of $\mu$-a.e.\ equivalence classes of elements of $\Sigma$, two kernels being identified when they agree $\mu$-a.e. For $\sigma \in \Sigma$, write $[\sigma]_\mu \in \widetilde\Sigma_\mu$ for its equivalence class. Writing $\mathcal{P}(\mathbb{Y})$ for the set of probability measures on $\mathbb{Y}$ with the weak topology and $\mathcal{C}_b(\mathbb{Y})$ for the space of bounded continuous real maps on $\mathbb{Y}$, we endow $\widetilde\Sigma_\mu$ with the \bi{stable topology}, the coarsest topology making the maps
\begin{align*}
	[\sigma]_\mu \mapsto \int_A \int_{\mathbb{Y}} h(y)\, d\sigma(y \mid w)\, d\mu(w)
\end{align*}
continuous for all $A \in \mathcal{B}(\mathbb{W})$ and $h \in \mathcal{C}_b(\mathbb{Y})$. These integrals depend on $\sigma$ only through $[\sigma]_\mu$. Under it, $\widetilde\Sigma_\mu$ is a complete separable metric space (Proposition 2.3.3 of \cite{Castaing/RaynaudDeFitte/Valadier:04:YoungMeasures}, p.43, applied to the base space $(\mathbb{W},\mu)$). Let $\mathcal{B}(\widetilde\Sigma_\mu)$ be its Borel $\sigma$-field. The pre-policy game uses $\mu = \mu_W$ and the post-policy game uses $\mu = \mu_W^f$. We write $\tilde\sigma$ for a generic class when the prior is clear from the context.

For the sets of equilibrium classes, we define
\begin{align*}
	\widetilde\Sigma_{\mathsf{BCE}}(G) = \{[\sigma]_{\mu_W}: \sigma \in \Sigma_{\mathsf{BCE}}(G)\} \subset \widetilde\Sigma_{\mu_W} \text{ and }  \widetilde\Sigma_{\mathsf{BCE}}(G_f) = \{[\sigma]_{\mu_W^f}: \sigma \in \Sigma_{\mathsf{BCE}}(G_f)\} \subset \widetilde\Sigma_{\mu_W^f}.
\end{align*}
Since $\Sigma_{\mathsf{BCE}}(G)$ is a union of $\mu_W$-classes, a class belongs to $\widetilde\Sigma_{\mathsf{BCE}}(G)$ if and only if each of its representatives is a BCE of $G$. We consider only those Bayesian games $G$ such that $\widetilde\Sigma_{\mathsf{BCE}}(G) \in \mathcal{B}(\widetilde\Sigma_{\mu_W})$, and analogously for $G_f$. We let $e(\csdot;G)$ be the \bi{equilibrium selection rule} for the game $G$, which is defined as a probability measure on $(\widetilde\Sigma_{\mu_W},\mathcal{B}(\widetilde\Sigma_{\mu_W}))$ such that
\begin{align*}
	e(\widetilde\Sigma_{\mathsf{BCE}}(G);G) = 1.
\end{align*}
Similarly, we define $e(\csdot;G_f)$ as the equilibrium selection rule for the post-policy game $G_f$, a probability measure on $(\widetilde\Sigma_{\mu_W^f},\mathcal{B}(\widetilde\Sigma_{\mu_W^f}))$ such that $e(\widetilde\Sigma_{\mathsf{BCE}}(G_f);G_f) = 1$.

In this framework, the uniqueness of an equilibrium is a statement about classes: the game $G$ has a \bi{unique BCE} if $\widetilde\Sigma_{\mathsf{BCE}}(G)$ is a singleton $\{\tilde\sigma_1\}$. In that case the selection rule is completely determined, $e(\csdot;G) = \delta_{\tilde\sigma_1}$, the point mass at $\tilde\sigma_1$.

\subsection{Counterfactual Predictions}
\label{subsec: well defined predictions}

\subsubsection{Randomized Reduced Form of the Game}
\label{subsubsec: reduced form}

Counterfactual predictions from the policy are generated by what we call the \textit{randomized reduced form}, which captures the causal relationship between the outcome $Y$ of the game and the payoff state $W$ in a reduced-form way, averaging over the equilibrium selection. Define the equilibrium outcome-state distribution of the game $G$ as the mixture
\begin{align}
	\label{rho_G}
	P_G = \int_{\widetilde\Sigma_{\mu_W}} (\mu_W \otimes \sigma)\, de(\tilde\sigma;G),
\end{align}
where, for each class $\tilde\sigma \in \widetilde\Sigma_{\mu_W}$, the kernel $\sigma$ denotes a representative of $\tilde\sigma$, selected by a fixed map $\tilde\sigma \mapsto \sigma$, and $\mu_W \otimes \sigma$ denotes the joint distribution on $\mathbb{Y} \times \mathbb{W}$ of $(Y,W)$ with $W \sim \mu_W$ and $Y \mid W = w \sim \sigma(\csdot \mid w)$. The measure $\mu_W \otimes \sigma$ depends on $\sigma$ only through its class $\tilde\sigma$, so $P_G$ does not depend on the choice of the selection map. The map $\tilde\sigma \mapsto (\mu_W \otimes \sigma)(E)$ is Borel for every Borel $E \subset \mathbb{Y} \times \mathbb{W}$, by the definition of the stable topology and a monotone class argument, so the mixture in (\ref{rho_G}) is well defined. The \bi{(randomized) reduced form of the game $G$} is a disintegration $\rho_G$ of $P_G$ over its $\mathbb{W}$-marginal $\mu_W$: 
\begin{align*}
	P_G(A \times B) = \int_B \rho_G(A \mid w) \, d\mu_W(w), \quad A \in \mathcal{B}(\mathbb{Y}), B \in \mathcal{B}(\mathbb{W}).
\end{align*}
Such a Markov kernel $\rho_G$ exists and is unique up to $\mu_W$-null sets (cf.\ Theorem 3.4 of \cite{Kallenberg:21:Foundations}). We fix one version throughout.

The reduced form $\rho_G$ summarizes the causal relationship between the payoff state and the outcome of the game. The outcome of the game $G$ is generated as follows.\smallskip

Step 1: The value of the payoff state $W=w$ is drawn from the distribution $\mu_W$.

Step 2: The action profile $Y$ is drawn from $\rho_G(\csdot \mid w)$.\smallskip

Thus, the joint distribution of $(Y,W)$ under $\mu_W$ is $P_G$, and the conditional distribution of $Y$ given $W \in B$ with $\mu_W(B) > 0$ is given as follows:
\begin{align}
	\label{cond prob}
	P\{Y \in A \mid W \in B\} = \frac{1}{\mu_W(B)}\int_B \rho_G(A \mid w) \, d\mu_W(w), \quad A \in \mathcal{B}(\mathbb{Y}).
\end{align}

\subsubsection{Counterfactual Predictions}
As in the original game $G$, we define the equilibrium outcome-state distribution of the post-policy game $G_f$,
\begin{align}
	\label{rho_Gf}
	P_{G_f} = \int_{\widetilde\Sigma_{\mu_W^f}} (\mu_W^f \otimes \sigma)\, de(\tilde\sigma;G_f),
\end{align}
and its reduced form $\rho_{G_f}$, a fixed disintegration of $P_{G_f}$ over $\mu_W^f$. For a real measurable map $h$ on $\mathbb{Y} \times \mathbb{W}$, the equilibrium-based and decomposition-based predictions of the main text, defined in (\ref{EP DP}), take the following form:
\begin{align}
	\label{DP}
	\mathsf{AEP}(h) &= \int_{\mathbb{W}} \int_{\mathbb{Y}} h(y,w) d\rho_{G_f}(y \mid w) d \mu_W^f(w), \text{ and }\\ \notag
	\mathsf{ADP}_R(h) &= \int_{\mathbb{S}_W} \int _{\mathbb{Y}} h(y,w) d\rho_G(y \mid w)d \mu_W^f(w).
\end{align}
The equilibrium-based prediction integrates $\rho_{G_f}$ against its own prior $\mu_W^f$, so it does not depend on the version of $\rho_{G_f}$: $\mathsf{AEP}(h) = \int h\, dP_{G_f}$. The decomposition-based prediction integrates the \emph{pre}-policy reduced form against the \emph{post}-policy prior. Since $\mu_W^f$-almost every $w \in \mathbb{S}_W$ lies in $\mathbb{S}_\cap$, and $\mu_W$ and $\mu_W^f$ are mutually absolutely continuous on $\mathbb{S}_\cap$, $\mathsf{ADP}_R(h)$ too does not depend on the version of $\rho_G$.

\subsection{Validity of Decomposition-Based Bounds}
\label{subsec: invariance}

In this section, we establish all the results stated in the main text. We first develop the invariance conditions that underpin the validity of the decomposition-based bounds, stated at the level of the sliced equilibria and their selection rules, and record them as Assumptions \ref{assump: information structure and policy2} to \ref{assump: invariance ESR}. These assumptions are the formal counterparts of Assumptions \ref{assump: information structure and policy} and \ref{assump: invariance} in the main text, making their measure-theoretic content explicit and, in part, strengthening it. Under them, each main-text result is restated as a formal version and proved: Theorem \ref{thm: bounds1} and Proposition \ref{prop: sharp bounds0} as Theorem \ref{thm: info change2} and Proposition \ref{prop: sharp bounds2}, and, in Appendix \ref{App: Proofs of the Results in the Main Text}, the identification results, Proposition \ref{prop: control fn} and Corollaries \ref{cor: bounds}, \ref{example1_corollary}, and \ref{corollary_auction}, as Proposition \ref{prop: control fn2} and Corollaries \ref{cor: bounds2}, \ref{example1_corollary2}, and \ref{corollary_auction2}. Proving each formal version thereby establishes the corresponding main-text result under those conventions.

\subsubsection{Invariance of the Sliced Equilibria}

We introduce a condition requiring the invariance of the \emph{sliced} equilibria, beginning with a counterpart of Assumption \ref{assump: information structure and policy} in the main text.

\begin{assumption}
	\label{assump: information structure and policy2}

	(i) The policy $f: \mathbb{W} \to \mathbb{W}$ satisfies that
	\begin{align*}
		f(w) = (\tilde f(w_C),w_I), \quad w = (w_C, w_I) \in \mathbb{W},
	\end{align*}
	for some measurable map $\tilde f: \mathbb{W}_C \rightarrow \mathbb{W}_C$.

	(ii) For each $i=1,...,n$ and every $w_{I,i} \in \mathbb{W}_{I,i}$, $\eta_i(\csdot \mid w_C, w_{I,i})$ is constant in $w_C \in \mathbb{W}_C$.

	(iii) For every Borel $B \subset \mathbb{S}_\cap$, $\mu_W(B) = 0$ if and only if $\mu_W^f(B) = 0$.

	(iv) $\mathbb{S}_W = \mathbb{S}_{W_C} \times \mathbb{S}_{W_I}$ and $\mathbb{S}_{f(W)} = \mathbb{S}_{\tilde f(W_C)} \times \mathbb{S}_{W_I}$, where $\mathbb{S}_{W_C}$ and $\mathbb{S}_{W_I}$ denote the supports of $W_C$ and $W_I$ under $\mu_W$, and $\mathbb{S}_{\tilde f(W_C)}$ that of $\tilde f(W_C)$ under $\mu_W$.
\end{assumption}

Condition (i) restricts attention to policies that change $w_C$ only. Condition (ii) elevates the conditional independence in Assumption \ref{assump: information structure and policy} to a pointwise restriction on the fixed belief version, removing the ambiguity of counterfactual predictions that measure-zero sets moved by the policy would create. Conditions (iii) and (iv) restate Assumption \ref{assump: information structure and policy}(iii) and (iv) of the main text: the pre- and post-policy priors share their null sets on the common support, and the common support is a cylinder over the common component, $\mathbb{S}_\cap = (\mathbb{S}_{W_C} \cap \mathbb{S}_{\tilde f(W_C)}) \times \mathbb{S}_{W_I}$.

The following assumption restates Assumption \ref{assump: BCE existence} in the main text: the sets of BCE under $G$ and $G_f$ are nonempty.

\begin{assumption}
    \label{assump: BCE existence2}
	$\Sigma_{\mathsf{BCE}}(G) \ne \varnothing$ and $\Sigma_{\mathsf{BCE}}(G_f) \ne \varnothing$.
\end{assumption}

We now formalize the invariance of sliced equilibria. For a Borel set $B \subset \mathbb{S}_\cap$, write $\sigma \simeq_B \sigma'$ if $\sigma(\csdot \mid w) = \sigma'(\csdot \mid w)$ at a.e.\ $w \in B$, an equivalence relation on $\mathcal{K}$, and write $[\sigma]_B$ for the class of $\sigma$, which depends on $\sigma$ only through its $\mu_W$-class, and equally only through its $\mu_W^f$-class. We define the \bi{$B$-sliced equilibrium class sets}
\begin{align}
	\label{B slice}
	\widetilde\Sigma_{\mathsf{BCE},B}(G) = \left\{ [\sigma]_B : \sigma \in \Sigma_{\mathsf{BCE}}(G) \right\} \text{ and } \widetilde\Sigma_{\mathsf{BCE},B}(G_f) = \left\{ [\sigma]_B : \sigma \in \Sigma_{\mathsf{BCE}}(G_f) \right\}.
\end{align}

Write $\mu_{T_i}$ and $\mu_{T_i}^f$ for the distributions of the signal $T_i = (W_C, W_{I,i})$ under $\mu_W$ and $\mu_W^f$, with supports $\mathbb{S}_{W_C,W_{I,i}}$ and $\mathbb{S}_{\tilde f(W_C),W_{I,i}}$. For a subcollection $\Sigma' \subset \Sigma$, the restricted BCEs are $\Sigma_{\mathsf{BCE}}'(G) = \Sigma_{\mathsf{BCE}}(G) \cap \Sigma'$ and $\Sigma_{\mathsf{BCE}}'(G_f) = \Sigma_{\mathsf{BCE}}(G_f) \cap \Sigma'$, with the $B$-sliced sets $\widetilde\Sigma_{\mathsf{BCE},B}'(G)$ and $\widetilde\Sigma_{\mathsf{BCE},B}'(G_f)$ formed as in (\ref{B slice}). Given $B \in \mathcal{B}(\mathbb{W})$, $\Sigma'$ is \bi{closed under gluing across $B$} if, for any $\sigma, \sigma' \in \Sigma'$, the decision rule equal to $\sigma$ on $B$ and to $\sigma'$ off $B$ belongs to $\Sigma'$. Under Assumption \ref{assump: information structure and policy2} and other technical conditions, the lemma below shows that the two games offer the same menu of equilibrium behaviors over every region of the common support.

\begin{lemma}[Invariance of the Sliced Equilibria]
	\label{lemm: master lemma}
	Suppose that a game-policy pair $(G,f)$ satisfies Assumption \ref{assump: information structure and policy2}, and let $\Sigma' \subset \Sigma$ be such that $\Sigma_{\mathsf{BCE}}'(G) \neq \varnothing$ and $\Sigma_{\mathsf{BCE}}'(G_f) \neq \varnothing$, and $\Sigma'$ is closed under gluing across $(\mathbb{S}_{W_C} \cap \mathbb{S}_{\tilde f(W_C)}) \times \mathbb{W}_I$. Then, for every Borel $B \subset \mathbb{S}_\cap$,
	\begin{align*}
		\widetilde\Sigma_{\mathsf{BCE},B}'(G) = \widetilde\Sigma_{\mathsf{BCE},B}'(G_f).
	\end{align*}
\end{lemma}
\medskip

\subsubsection{Invariance of Sliced Equilibrium Selection Rules}

We introduce an invariance condition for the equilibrium selection rules. First, we define the joint distribution of the selected equilibrium, the outcome, and the payoff state as a measure $M_G$ on $\widetilde\Sigma_{\mu_W} \times \mathbb{Y} \times \mathbb{W}$:
\begin{align*}
	M_G = \int_{\widetilde\Sigma_{\mu_W}} \delta_{\tilde\sigma} \otimes (\mu_W \otimes \sigma)\, de(\tilde\sigma;G),
\end{align*}
where $\delta_{\tilde\sigma}$ is the point mass at $\tilde\sigma$. Let $q_G$ be a disintegration of $M_G$ over its $(\widetilde\Sigma_{\mu_W} \times \mathbb{W})$-marginal $e(\csdot;G) \otimes \mu_W$: a Markov kernel with
\begin{align*}
	M_G(D \times A \times B) = \int_D \int_B q_G(A \mid \tilde\sigma, w)\, d\mu_W(w)\, de(\tilde\sigma;G),
\end{align*}
for Borel $D \subset \widetilde\Sigma_{\mu_W}$, $A \subset \mathbb{Y}$, and $B \subset \mathbb{W}$. The kernel $q_G(\csdot \mid \tilde\sigma, w)$ is the action distribution prescribed by the selected equilibrium at $w$: for $e(\csdot;G)$-a.e.\ $\tilde\sigma$,
\begin{align*}
	q_G(\csdot \mid \tilde\sigma, w) = \sigma(\csdot \mid w), \enspace \mu_W\text{-a.e.\ } w, 
\end{align*}
by the uniqueness of the disintegration of $\mu_W \otimes \sigma$. For each $w \in \mathbb{W}$, we define the \bi{$w$-slice} of the equilibrium selection rule, a probability measure on $\mathcal{P}(\mathbb{Y})$, by
\begin{align}
	\label{e_w}
	e_w(D;G) = e\big(\{\tilde\sigma \in \widetilde\Sigma_{\mu_W} : q_G(\csdot \mid \tilde\sigma, w) \in D\};G\big), \quad D \in \mathcal{B}(\mathcal{P}(\mathbb{Y})),
\end{align}
the formal counterpart of the $w$-slice in Assumption \ref{assump: invariance}(ii) of the main text. Since any two versions of $q_G$ agree $e(\csdot;G) \otimes \mu_W$-a.e., the slice is unambiguous at $\mu_W$-a.e.\ $w$. We define $M_{G_f}$, $q_{G_f}$, and $e_w(\csdot;G_f)$ analogously, with $G_f$ and $\mu_W^f$ in place of $G$ and $\mu_W$, so that $e_w(\csdot;G_f)$ is likewise unambiguous at $\mu_W^f$-a.e.\ $w$. Averaging a slice returns the reduced form at that state: for each $A \in \mathcal{B}(\mathbb{Y})$,
\begin{align}
	\label{eq: barycenter}
	\rho_G(A \mid w) &= \int_{\mathcal{P}(\mathbb{Y})} \pi(A)\, de_w(\pi;G), \enspace \text{$\mu_W$-a.e.\ } w, \text{ and }\\ \notag
	\rho_{G_f}(A \mid w) &= \int_{\mathcal{P}(\mathbb{Y})} \pi(A)\, de_w(\pi;G_f), \enspace \text{$\mu_W^f$-a.e.\ } w.
\end{align}
On $\mathbb{S}_\cap$, both slices are thus unambiguous and both identities in (\ref{eq: barycenter}) hold at a.e.\ $w$. Now, we state the invariance condition in terms of these slices.

\begin{assumption}
	\label{assump: invariance ESR}
	For every Borel set $B \subset \mathbb{S}_\cap$ such that $\widetilde\Sigma_{\mathsf{BCE},B}(G) = \widetilde\Sigma_{\mathsf{BCE},B}(G_f)$, we have
	\begin{align}
		\label{eq: invariance sliced eq}
		e_w(\csdot;G) = e_w(\csdot;G_f), \enspace \text{a.e.\ } w \in B.
	\end{align}
\end{assumption}

The assumption is the formal counterpart of Assumption \ref{assump: invariance}(ii) in the main text. The following corollary establishes that when the equilibrium behaviors stay the same before and after the policy, Assumption \ref{assump: invariance ESR} holds. In particular, suppose that the game $G$ has a unique BCE class, $\widetilde\Sigma_{\mathsf{BCE}}(G) = \{[\sigma_1]_{\mu_W}\}$, as in Assumption \ref{assump: invariance}(i) of the main text. Then, under Assumptions \ref{assump: information structure and policy2} and \ref{assump: BCE existence2}, Lemma \ref{lemm: master lemma} with $\Sigma' = \Sigma$ and $B = \mathbb{S}_\cap$ gives $\widetilde\Sigma_{\mathsf{BCE},\mathbb{S}_\cap}(G_f) = \widetilde\Sigma_{\mathsf{BCE},\mathbb{S}_\cap}(G) = \{[\sigma_1]_{\mathbb{S}_\cap}\}$, so that every BCE of $G$ and every BCE of $G_f$ coincide with $\sigma_1$ at a.e.\ $w$ of the common support, whether or not $G_f$ has a unique BCE class. The hypothesis of the corollary then holds, and the equilibrium selection plays no role.

\begin{corollary}
	\label{cor: cylinder slice}
	Suppose that Assumptions \ref{assump: information structure and policy2}(iii) and \ref{assump: BCE existence2} hold, and that every $\sigma \in \Sigma_{\mathsf{BCE}}(G)$ and every $\sigma_f \in \Sigma_{\mathsf{BCE}}(G_f)$ satisfy 
	\begin{align*}
		\sigma(\csdot \mid w) = \sigma_f(\csdot \mid w), \enspace \text{a.e.\ } w \in \mathbb{S}_\cap.
	\end{align*}

    Then, every pair of equilibrium selection rules $e(\csdot;G)$ and $e(\csdot;G_f)$ satisfies Assumption \ref{assump: invariance ESR}. Moreover, for any $\sigma_1 \in \Sigma_{\mathsf{BCE}}(G)$, 
	\begin{align}
		\label{equal RRF}
	   \rho_G(\csdot \mid w) = \rho_{G_f}(\csdot \mid w) = \sigma_1(\csdot \mid w), \enspace \text{a.e.\ } w \in \mathbb{S}_\cap.
	\end{align}
\end{corollary}

The equality (\ref{equal RRF}) is exactly the reduced-form invariance on which the validity of the decomposition-based bounds rests. When the game $G$ has a unique BCE class, the corollary therefore already delivers the bounds in Theorem \ref{thm: info change2} below under Assumptions \ref{assump: information structure and policy2} and \ref{assump: BCE existence2} alone.

\subsubsection{Validity of Decomposition-Based Bounds: the General Result}
\label{subsec: main result}

The following theorem is the formal version of Theorem \ref{thm: bounds1}, established under Assumptions \ref{assump: information structure and policy2} to \ref{assump: invariance ESR}.

\begin{theorem}
	\label{thm: info change2}
	Suppose that Assumptions \ref{assump: information structure and policy2} to \ref{assump: invariance ESR} hold for the pre-policy game $G = (B,I)$ and the post-policy game $G_f = (B_f,I)$. Let $h:\mathbb{Y} \times \mathbb{W} \rightarrow \mathbf{R}$ be a measurable map such that, for all $w \in \mathbb{W}$,
	\begin{align}
		\label{h bounds2}
		\underline h(w) \le \inf_{y \in \mathbb{Y}} h(y,w) \le \sup_{y \in \mathbb{Y}} h(y,w) \le \overline h(w),
	\end{align}
	for some measurable maps $\underline h, \overline h: \mathbb{W} \rightarrow \mathbf{R}$ that are $\mu_W^f$-integrable on $\mathbb{W}$.

	Then,
    \begin{align}
		\label{bounds master}
		\mathsf{ADP}_R(h) +  \Delta(\underline h) \le \mathsf{AEP}(h) \le \mathsf{ADP}_R(h) + \Delta(\overline h),
	\end{align}
	where
	\begin{align*}
		\Delta(\overline h) = \int_{\mathbb{W} \setminus \mathbb{S}_W} \overline h(w) d \mu_W^f(w),
	\end{align*}
	and $\Delta(\underline h)$ is defined similarly.
\end{theorem}

We turn to the problem of sharpness of the decomposition-based bounds. The following proposition is a restatement of Proposition \ref{prop: sharp bounds0}.

\begin{proposition}
	\label{prop: sharp bounds2}
	Suppose that $\mathbb{Y}$ is compact. Let $h: \mathbb{Y} \times \mathbb{W} \rightarrow \mathbf{R}$ and its envelopes $\underline h, \overline h: \mathbb{W} \rightarrow \mathbf{R}$ be as in Theorem \ref{thm: info change2}, such that
	\begin{align*}
		\overline h(w) = \max_{y \in \mathbb{Y}} h(y,w) \text{ and } \underline h(w) = \min_{y \in \mathbb{Y}} h(y,w),
	\end{align*}
	where $h(\csdot,w)$ is continuous on $\mathbb{Y}$ for each $w$. Suppose further that the pair $(\mu_W, f)$ satisfies Assumption \ref{assump: information structure and policy2}(iii).

	Then there are Bayesian games $G_1$ and $G_2$, with post-policy games $G_{1,f}$ and $G_{2,f}$ induced by $f$, that satisfy Assumptions \ref{assump: information structure and policy2} to \ref{assump: invariance ESR} and
	\begin{align*}
		\mathsf{AEP}(h) &= \mathsf{ADP}_R(h) + \Delta(\overline h), \text{ under } (G_1, G_{1,f}) \text{ and }\\
		\mathsf{AEP}(h) &= \mathsf{ADP}_R(h) + \Delta(\underline h), \text{ under } (G_2, G_{2,f}).
	\end{align*}
	The games $G_1$ and $G_2$ differ only in their payoffs outside $\mathbb{S}_W$. They therefore induce the same joint distribution of $(Y,W)$ before the policy, and hence the same identified objects $\mathsf{ADP}_R(h)$, $\Delta(\overline h)$, and $\Delta(\underline h)$.
\end{proposition}

The final claim gives the sharpness its identification content: the two extremal structures $(G_1, G_{1,f})$ and $(G_2, G_{2,f})$ are observationally equivalent in the pre-policy environment, so that, given the distribution of the pre-policy observables, neither endpoint of the interval in Theorem \ref{thm: info change2} can be excluded.
 
\subsubsection{Identification of the Decomposition-Based Bounds}
\label{App: Proofs of the Results in the Main Text}

We consider the identification of the decomposition-based bounds in the partially observed setting of Section \ref{sec:single-index-restrictions} of the main text: 
\begin{align*}
	W = (X_a,X_b,\varepsilon),
\end{align*}
with $X = (X_a,X_b)$ observed and $\varepsilon$ unobserved, and the policy acting only on the observed component, $f(w) = (f_a(x_a),x_b,\varepsilon)$ with $f^*(x) = (f_a(x_a),x_b)$, for a measurable map $f_a$. Throughout this subsection, the common component is the observable part and the private components are the unobservables: $W_C = X$ and $W_I = \varepsilon$, with $W_{I,i} = \varepsilon_i$. 

The policy takes the form $f(w) = (\tilde f(w_C), w_I)$ with $\tilde f = f^*$, and Assumption \ref{assump: information structure and policy2}(iv) reads $\mathbb{S}_W = \mathbb{S}_X \times \mathbb{S}_\varepsilon$ and $\mathbb{S}_{f(W)} = \mathbb{S}_{f^*(X)} \times \mathbb{S}_\varepsilon$, where $\mathbb{S}_X = \mathbb{S}_{X_a,X_b}$, $\mathbb{S}_{f^*(X)}$, and $\mathbb{S}_\varepsilon$ denote the supports of $X = (X_a,X_b)$, $f^*(X)$, and $\varepsilon$. The target function $h$ depends on the state only through the observables: we write $h^*(y,x) = h(y,w)$, with envelopes $\overline h^*(x) = \overline h(w)$ and $\underline h^*(x) = \underline h(w)$. Here $h$ is as in Theorem \ref{thm: info change2}, satisfying (\ref{h bounds2}), and $\overline h, \underline h$ are both $\mu_W$-integrable and $\mu_W^f$-integrable.

We restate Assumption \ref{assump: control fn} of the main text.

\begin{assumption}
\label{assump: control fn2}
(i) $X_a$ and $\varepsilon$ are conditionally independent given $X_b$.

(ii) Whenever $(x_a,x_b,\bar\varepsilon) \in \mathbb{S}_W$ and $(x_a',x_b) \in \mathbb{S}_{X_a,X_b}$, we have $(x_a',x_b,\bar\varepsilon) \in \mathbb{S}_W$.
\end{assumption}

By part (i) of this assumption, the conditional distribution of $\varepsilon$ given $(X_a,X_b)$ has a version that does not depend on $x_a$, which we fix and denote $\mu_{\varepsilon\mid X_b}(\csdot \mid x_b)$. Under the split $W_C = X$ and $W_I = \varepsilon$, part (ii) is implied by the product form: if $(x_a,x_b,\bar\varepsilon) \in \mathbb{S}_W = \mathbb{S}_X \times \mathbb{S}_\varepsilon$, then $\bar\varepsilon \in \mathbb{S}_\varepsilon$, so $(x_a',x_b,\bar\varepsilon) \in \mathbb{S}_W$ for every $(x_a',x_b) \in \mathbb{S}_X$. The following is the formal version of Proposition \ref{prop: control fn}.

\begin{proposition}
\label{prop: control fn2}
Suppose that Assumptions \ref{assump: information structure and policy2}, \ref{assump: BCE existence2}, and \ref{assump: control fn2} hold. Then the following statements hold.\medskip

(i) $\Delta(\underline h) = \mathbf{E}\left[\underline h^*(f^*(X))1\left\{f^*(X) \notin \mathbb{S}_X\right\}\right] \text{ and } \Delta(\overline h) = \mathbf{E}\left[\overline h^*(f^*(X))1\left\{f^*(X) \notin \mathbb{S}_X\right\}\right].$

(ii) With $\mu_{X_a,X_b}$ denoting the distribution of $(X_a,X_b)$,
\begin{align*}
	\mathsf{ADP}_R(h) = \int \mathbf{E}\left[ h^*(Y,X) \mid (X_a,X_b) = (f_a(x_a),x_b)\right] 1\left\{(f_a(x_a),x_b) \in \mathbb{S}_{X_a,X_b}\right\} d\mu_{X_a,X_b}(x_a,x_b).
\end{align*}
\end{proposition}

Hence the integrand of Proposition \ref{prop: control fn2}(ii), evaluated at the policy-shifted points $(f_a(x_a),x_b)$, is identified from the distribution of the observables.

\subsubsection{Restriction to Subclasses of Equilibria}
We consider the setting of Section \ref{app:refinements}, where we focus on a restricted class of Bayes correlated equilibria such as BNE. For a subcollection $\Sigma' \subset \Sigma$, with the restricted BCEs $\Sigma_{\mathsf{BCE}}'(G)$ and $\Sigma_{\mathsf{BCE}}'(G_f)$ as defined before Lemma \ref{lemm: master lemma}, we define the associated restricted selection rules, reduced forms, and decomposition-based predictions as $e'(\csdot;G)$, $\rho_G'$, and $\mathsf{ADP}_R'(h)$, respectively, with the restricted equilibrium-based prediction $\mathsf{AEP}'(h)$ defined as $\mathsf{AEP}(h)$ with $\rho_{G_f}'$ in place of $\rho_{G_f}$. As with $\mathsf{ADP}_R(h)$, the value of $\mathsf{ADP}_R'(h)$ does not depend on the version of $\rho_G'$: $\mu_W^f$-almost every $w \in \mathbb{S}_W$ lies in $\mathbb{S}_\cap$, where $\mu_W^f$ is absolutely continuous with respect to $\mu_W$ by Assumption \ref{assump: information structure and policy2}(iii).

For the restricted equilibrium concept in Section \ref{app:refinements} of the main text, we adopt throughout the same measurability convention used for the unrestricted BCEs in Appendix \ref{subsubsec: equivalence classes}: we consider only subcollections $\Sigma' \subset \Sigma$ and games $G$ for which the restricted equilibrium-class sets $\widetilde\Sigma_{\mathsf{BCE}}'(G) = \{[\sigma]_{\mu_W}: \sigma \in \Sigma_{\mathsf{BCE}}'(G)\}$ and $\widetilde\Sigma_{\mathsf{BCE}}'(G_f) = \{[\sigma]_{\mu_W^f}: \sigma \in \Sigma_{\mathsf{BCE}}'(G_f)\}$ belong to $\mathcal{B}(\widetilde\Sigma_{\mu_W})$ and $\mathcal{B}(\widetilde\Sigma_{\mu_W^f})$, respectively. This guarantees that the restricted selection rules $e'(\csdot;G)$ and $e'(\csdot;G_f)$ and the associated reduced forms $\rho_G'$ and $\rho_{G_f}'$ are well-defined, exactly as in the unrestricted case.

The formal version of Corollary \ref{cor: bounds} is presented below. Assumption \ref{assump: invariance ESR}, read in terms of the restricted BCEs, replaces the sliced sets $\widetilde\Sigma_{\mathsf{BCE},B}$ and the slices $e_w$ by their primed counterparts $\widetilde\Sigma_{\mathsf{BCE},B}'$ and $e_w'$, the latter constructed from $e'$ exactly as $e_w$ from $e$ in (\ref{e_w}).\footnote{More explicitly, define 
\begin{align*}
	M_G' = \int \delta_{\tilde\sigma} \otimes (\mu_W \otimes \sigma)\, de'(\tilde\sigma;G),
\end{align*}
on $\widetilde\Sigma_{\mu_W} \times \mathbb{Y} \times \mathbb{W}$. Let $q_G'$ be a disintegration of $M_G'$ over $e'(\csdot;G) \otimes \mu_W$, and set, with $e_w'(\csdot;G_f)$ defined analogously from $\mu_W^f$, 
\begin{align*}
	e_w'(D;G) = e'(\{\tilde\sigma : q_G'(\csdot \mid \tilde\sigma, w) \in D\};G).
\end{align*}
}
  
The following is the formal version of Corollary \ref{cor: bounds}, in the manner of Theorem \ref{thm: info change2}.

\begin{corollary}
\label{cor: bounds2}
Let $\Sigma' \subset \Sigma$ be a subcollection which has nonempty restricted equilibrium-class sets $\widetilde\Sigma_{\mathsf{BCE}}'(G)$ and $\widetilde\Sigma_{\mathsf{BCE}}'(G_f)$ that are Borel. Suppose that Assumptions \ref{assump: information structure and policy2} to \ref{assump: invariance ESR}, read in terms of the restricted BCEs $\Sigma_{\mathsf{BCE}}'(G)$ and $\Sigma_{\mathsf{BCE}}'(G_f)$, hold, that $h$, $\underline h$, and $\overline h$ are as in Theorem \ref{thm: info change2}, and that either of the following holds:\medskip

(a) $\Sigma'$ is closed under gluing across $(\mathbb{S}_{W_C} \cap \mathbb{S}_{\tilde f(W_C)}) \times \mathbb{W}_I$, or 

(b) every $\sigma \in \Sigma_{\mathsf{BCE}}'(G)$ and every $\sigma_f \in \Sigma_{\mathsf{BCE}}'(G_f)$ satisfy $\sigma(\csdot \mid w) = \sigma_f(\csdot \mid w)$ at a.e.\ $w \in \mathbb{S}_\cap$.\medskip 

Then
\begin{align*}
	\mathsf{ADP}_R'(h) + \Delta(\underline h) \le \mathsf{AEP}'(h) \le \mathsf{ADP}_R'(h) + \Delta(\overline h).
\end{align*}
\end{corollary}

Condition (a) requires the closure of $\Sigma'$ under gluing, in the sense defined before Lemma \ref{lemm: master lemma}, across the cylinder $(\mathbb{S}_{W_C} \cap \mathbb{S}_{\tilde f(W_C)}) \times \mathbb{W}_I$. Lemma \ref{lemm: gluing} below verifies this closure, across every cylinder $D \times \mathbb{W}_I$ with $D \subset \mathbb{W}_C$ Borel, for the product-form class in (\ref{B}) and for its symmetric refinement.\footnote{By the \bi{symmetric refinement} of the product-form class in (\ref{B}) we mean the following. Suppose that the players share a common action space and a common private-component space, $\mathbb{Y}_1 = \cdots = \mathbb{Y}_n$ and $\mathbb{W}_{I,1} = \cdots = \mathbb{W}_{I,n}$, so that $\mathbb{T}_1 = \cdots = \mathbb{T}_n = \mathbb{T}_*$, say. The symmetric refinement of (\ref{B}) is the subcollection of those $\sigma \in \Sigma$ of the form (\ref{B}) whose factors are common across the players: there is a single conditional distribution $\beta(\csdot \mid t)$ of an action given a signal $t \in \mathbb{T}_*$ such that $\beta_i = \beta$ for all $i = 1,\ldots,n$, i.e., $\sigma(A \mid w) = \prod_{i=1}^n \beta(A_i \mid t_i)$, with $t_i = (w_C,w_{I,i})$, for any $A = A_1 \times \cdots \times A_n \subset \mathbb{Y}$.} Condition (b) arises, for example, when each game has a unique restricted BCE class and the two classes coincide at a.e.\ $w$ of the common support, as Lemma \ref{lemm: auction invariance} establishes for the auction; the equilibrium selection then plays no role.

\begin{lemma}
	\label{lemm: gluing}
	Let $B = D \times \mathbb{W}_I$ with $D \subset \mathbb{W}_C$ Borel. Then the following statements hold.

	(i) Let $\sigma, \sigma' \in \Sigma$ have the product form in (\ref{B}), with factors $(\beta_i)_{i=1}^n$ and $(\beta_i')_{i=1}^n$. Then the gluing of $\sigma$ and $\sigma'$ across $B$ has the product form in (\ref{B}), with the factors
	\begin{align}
		\label{glued beta}
		\beta_i^\dagger(\csdot \mid t_i) = 1\{w_C \in D\}\, \beta_i(\csdot \mid t_i) + 1\{w_C \notin D\}\, \beta_i'(\csdot \mid t_i), \quad t_i = (w_C,w_{I,i}).
	\end{align}

	(ii) The product-form class in (\ref{B}) is closed under gluing across $B$. If the players share a common action space and a common private-component space, $\mathbb{Y}_1 = \cdots = \mathbb{Y}_n$ and $\mathbb{W}_{I,1} = \cdots = \mathbb{W}_{I,n}$, then the symmetric refinement of (\ref{B}) is likewise closed under gluing across $B$.
\end{lemma}

\subsection{Examples}

\subsubsection{Entry Games}
We consider the entry-game setting of Section \ref{example1_detail} in the main text, where the payoff state is $W = (X,\varepsilon)$, with $X = (X_1,\ldots,X_n)$ observed and $\varepsilon = (\varepsilon_1,\ldots,\varepsilon_n)$ unobserved, each $X_i$ taking values in $\mathbf{R}^{d_X}$ and each $\varepsilon_i$ in $\mathbf{R}^{d_\varepsilon}$, and the policy acts as $f(X,\varepsilon) = (f^*(X),\varepsilon)$, where $f^*(X) = (f_a(X_a),X_b)$, for a measurable map $f_a: \mathbf{R} \to \mathbf{R}$ with $X_a \in \mathbf{R}$. We impose the following counterpart of Assumptions \ref{example1_assump} to \ref{assump: invar} in the main text, with the measure-theoretic conventions of this appendix made explicit. Throughout, the payoffs of the entry game are maintained to satisfy the envelope condition (\ref{eq: envelope compatibility}).

\begin{assumption}
\label{assump: entry2}
(i) The information structure satisfies one of the following: 

\quad (a) the game is of complete information, i.e., $T_i = (X,\varepsilon)$ for $i = 1,\ldots,n$; or 

\quad (b) the game has a public-private dichotomy of signals, i.e., $T_i = (X,\varepsilon_i)$ for $i = 1,\ldots,n$.

(ii) For every Borel $B \subset \mathbb{S}_X \cap \mathbb{S}_{f^*(X)}$, $P\{X \in B\} = 0$ if and only if $P\{f^*(X) \in B\} = 0$.

(iii) $X$ and $\varepsilon$ are independent.

(iv) A BCE exists in $G$ and $G_f$.

(v) For the given pair of equilibrium selection rules, $e_w(\csdot;G) = e_w(\csdot;G_f)$ at a.e.\ $w$ in the intersection of the supports of $(X,\varepsilon)$ and $(f^*(X),\varepsilon)$.

(vi) Under the public-private dichotomy, each player $i$'s belief map $(x,\varepsilon_i) \mapsto P\{\varepsilon_{-i} \in \csdot \mid X = x, \varepsilon_i\}$ is constant in $x$.
\end{assumption}

Corollary \ref{example1_corollary2} below is the formal version of Corollary \ref{example1_corollary}, in the manner of Theorem \ref{thm: info change2}. In its statement, the prediction $p_f(Y = a \mid C)$ is defined through the post-policy reduced form, in accordance with (\ref{DP}):
\begin{align*}
	p_f(Y = a \mid C)\, P\{f^*(X) \in C\} = \int 1\{x \in C\}\, \rho_{G_f}(\{a\} \mid w)\, d\mu_W^f(w), \quad w = (x,\varepsilon).
\end{align*}
The conditional choice probability $\pi(a \mid x) = P\{Y = a \mid X = x\}1\{x \in \mathbb{S}_X\}$ constituting the bounds is obtained by averaging the randomized reduced form over the unobserved component,
\begin{align*}
	\pi(a \mid x) = 1\{x \in \mathbb{S}_X\} \int \rho_G(\{a\} \mid x,\varepsilon)\, d\mu_\varepsilon(\varepsilon),
\end{align*}
where the averaging is over the marginal $\mu_\varepsilon$: by the independence of $X$ and $\varepsilon$ (Assumption \ref{assump: entry2}(iii)), $\mu_\varepsilon$ is a version of the conditional distribution of $\varepsilon$ given $X$, and we take the fixed version $\mu_{\varepsilon \mid X_b}$ of Assumption \ref{assump: control fn2}(i) to be $\mu_\varepsilon$ in this setting.

\begin{corollary}
\label{example1_corollary2}
Suppose that Assumption \ref{assump: entry2} holds. Then, for each $a \in \{0,1\}^n$ and each Borel set $C \subset \mathbf{R}^{n d_X}$ with $P\{f^*(X) \in C\} > 0$,
\begin{align*}
	\mathbf{E}\left[\pi\left(a \mid f^*(X)\right) \mid f^*(X) \in C\right] &\le p_f(Y=a \mid C) \\
	&\le \mathbf{E}\left[\pi\left(a \mid f^*(X)\right) \mid f^*(X) \in C\right] + P\{f^*(X) \notin \mathbb{S}_X \mid f^*(X) \in C\},
\end{align*}
where $\pi(a \mid x) = P\{Y= a \mid X = x\}1\{x \in \mathbb{S}_X\}$.
\end{corollary}

\subsubsection{Auctions}
\label{subsubsec: auctions appendix}

We consider the auction setting of Section \ref{subsubsec:auctions}, specializing the maintained framework of Appendix \ref{App:Counterfactual Predictions in Bayesian Games} in addition to Assumptions \ref{assump: auction2a} and \ref{assump: auction2b} below. Throughout this subsection, $\overline v < \infty$ denotes the upper bound on valuations and bids fixed in Section \ref{subsubsec:auctions}, and, as there, $X = (X_1,R)$. The auction is placed in that framework as follows. The action of bidder $i$ is $Y_i \in \{-1\} \cup [0,\overline v]$, the payoff state is $W = (X_1,R,V,\zeta)$, the common component is $W_C = (X_1,R,\zeta)$, and the private component of bidder $i$ is $W_{I,i} = V_i$, so that the signal of bidder $i$ is $T_i = (X_1,R,\zeta,V_i)$. The policy replaces $R$ by $f^*(R)$ and is therefore of the form $f(w) = (\tilde f(w_C),w_I)$, with
\begin{align*}
	\tilde f(x_1,r,\zeta) = (x_1,f^*(r),\zeta).
\end{align*}
Each bidder's belief map $v_i \mapsto P\{V_{-i} \in \csdot \mid V_i = v_i\}$ is a fixed version. Valuations and bids lie in $[0,\overline v]$, so the payoffs are bounded and the envelope conditions, including (\ref{eq: envelope compatibility}), hold with a constant envelope.

The bidders share the action space and the valuation space, so their signal spaces coincide. Write $t_i = (v_i,x_1,r,\zeta)$ for the signal of bidder $i$, and $t$ for a generic element of the common signal space. A \bi{symmetric pure strategy profile} is a pair $(a',b')$ of Borel maps of $t$, defined at every point, with the entry rule $a'$ taking values in $\{0,1\}$ and the bidding strategy $b'$ in $[0,\overline v]$, the symmetry being that this one pair prescribes the behavior of every bidder: bidder $i$ enters when $a'(t_i) = 1$ and then bids $b'(t_i)$. The profile induces the decision rule that, at each state $w = (x_1,r,v,\zeta)$, sets $Y_i = b'(t_i)$ when $a'(t_i) = 1$ and $Y_i = -1$ otherwise. We take the induced rule to be of the product form in (\ref{B}) with factors common across the bidders, each $\beta_i(\csdot \mid t_i)$ being the point mass at the prescription at $t_i$, so it is a BNE of a game exactly when it is a BCE of that game.

We impose the following counterpart of Assumptions \ref{example2_assump} to \ref{example2_assump2} in the main text, with the measure-theoretic conventions of this appendix made explicit.

\begin{assumption}
\label{assump: auction2a}
(i) $(X_1,R,\zeta)$ is publicly known, and the valuations $V_i$ are private information.

(ii) The policy changes the reserve price $R$ into $f^*(R)$ for some measurable map $f^*$.

(iii) There is a class $\Sigma'$ of decision rules, each induced by a symmetric pure strategy profile, with the following three properties.

\quad (a) For every Borel set $D$ of values of $(x_1,r,\zeta)$, the class $\Sigma'$ is closed under gluing across the set of states whose common component lies in $D$.

\quad (b) $\Sigma_{\mathsf{BCE}}'(G) \ne \varnothing$ and $\Sigma_{\mathsf{BCE}}'(G_f) \ne \varnothing$.

\quad (c) Any two elements of $\Sigma_{\mathsf{BCE}}'(G)$ agree at $\mu_W$-a.e.\ $w$, and any two elements of $\Sigma_{\mathsf{BCE}}'(G_f)$ agree at $\mu_W^f$-a.e.\ $w$.
\end{assumption}

Condition (iii) leaves the class $\Sigma'$ generic, in the manner of the restricted classes of Corollary \ref{cor: bounds2}, so that it covers a wide range of auction settings for which an existence and uniqueness theory is available. Property (c) requires each game to have a single equilibrium in $\Sigma'$ up to almost sure equality. Property (a) holds for any class $\Sigma'$ of the following form. For each value $(x_1,r,\zeta)$ of the common component, let $\mathcal{Q}_{x_1,r,\zeta}$ be a set of pairs of maps of the valuation, the first map taking values in $\{0,1\}$ and the second in $[0,\overline v]$, and let $\Sigma'$ be the class of decision rules induced by those symmetric pure strategy profiles $(a',b')$ such that $(a'(\csdot,x_1,r,\zeta),b'(\csdot,x_1,r,\zeta)) \in \mathcal{Q}_{x_1,r,\zeta}$ for every $(x_1,r,\zeta)$. To see that property (a) holds for such a class, let $D$ be a Borel set of values of $(x_1,r,\zeta)$, let $(a_1,b_1)$ and $(a_2,b_2)$ be two profiles inducing members of $\Sigma'$, and consider the pair of maps that coincides with $(a_1,b_1)$ at every signal with $(x_1,r,\zeta) \in D$ and with $(a_2,b_2)$ at every other signal. This pair is a symmetric pure strategy profile, because its two maps are Borel when $D$ is a Borel set. At each $(x_1,r,\zeta)$, the pair of maps of the valuation given by this third profile is that of $(a_1,b_1)$ or that of $(a_2,b_2)$, and hence belongs to $\mathcal{Q}_{x_1,r,\zeta}$. The decision rule induced by the third profile is the gluing of the decision rules induced by $(a_1,b_1)$ and $(a_2,b_2)$ across the set of states whose common component lies in $D$, and therefore the gluing belongs to $\Sigma'$.

\begin{assumption}
\label{assump: auction2b}
(i) $(V,\zeta)$ is conditionally independent of $R$ given $X_1$, where $V = (V_1,\ldots,V_n)$.

(ii) After rearranging coordinates, $\mathbb{S}_W = \mathbb{S}_{X_1,\zeta} \times \mathbb{S}_R \times \mathbb{S}_V$ and $\mathbb{S}_{f(W)} = \mathbb{S}_{X_1,\zeta} \times \mathbb{S}_{f^*(R)} \times \mathbb{S}_V$, where $\mathbb{S}_{X_1,\zeta}$, $\mathbb{S}_{f^*(R)}$, and $\mathbb{S}_V$ denote the supports of $(X_1,\zeta)$, $f^*(R)$, and $V$.

(iii) For each $i = 1,\ldots,n$, $V_{-i}$ is conditionally independent of $(X_1,R,\zeta)$ given $V_i$.

(iv) For every Borel $B \subset \mathbb{S}_{X_1,R} \cap \mathbb{S}_{X_1,f^*(R)}$, $P\{(X_1,R) \in B\} = 0$ if and only if $P\{(X_1,f^*(R)) \in B\} = 0$, where $\mathbb{S}_{X_1,R}$ and $\mathbb{S}_{X_1,f^*(R)}$ denote the supports of $(X_1,R)$ and $(X_1,f^*(R))$.
\end{assumption}

By Assumption \ref{assump: auction2b}(i), the conditional distribution of $(V,\zeta)$ given $(X_1,R) = (x_1,r)$ has a version not depending on $r$, which we fix from here on and denote by $\mu_{V,\zeta \mid X_1}(\csdot \mid x_1)$.

Fix symmetric pure strategy profiles $(a^*,b^*)$ and $(a^{*f},b^{*f})$ whose induced decision rules $\sigma^*$ and $\sigma^{*f}$ belong to $\Sigma_{\mathsf{BCE}}'(G)$ and $\Sigma_{\mathsf{BCE}}'(G_f)$, respectively. Define the entry decisions and bids of each bidder $i$, before and after the policy, as
\begin{align*}
	E_i &= a^*(V_i,X_1,R,\zeta), \quad B_i = b^*(V_i,X_1,R,\zeta), \text{ and }\\
	E_i^f &= a^{*f}(V_i,X_1,f^*(R),\zeta), \quad B_i^f = b^{*f}(V_i,X_1,f^*(R),\zeta),
\end{align*}
and the sets of entrants as $\widetilde N = \{i : E_i = 1\}$ and $\widetilde N^f = \{i : E_i^f = 1\}$. All almost-sure statements are under the distribution of $(V,X_1,R,\zeta)$, and a maximum over an empty set of entrants is set to $-1 \notin \mathbf{R}_+$, so no-sale events contribute zero below. The post-policy prediction of focus is 
\begin{align*}
	p_f(A \mid C) = P\left\{\max_{i \in \widetilde N^f} B_i^f \in A \mid X_1 \in C\right\}, \enspace A \subset \mathbf{R}_+, \ C \subset \mathbf{R}^{d_{X_1}},
\end{align*}
with $P\{X_1 \in C\} > 0$. We define
\begin{align*}
	p(A \mid x_1,r) = P\Big\{\max_{i \in \widetilde N} B_i \in A \ \mid (X_1,R) = (x_1,r)\Big\},
\end{align*}
a conditional probability fixed up to null sets of the distribution of $(X_1,R)$. The map $p(A \mid \csdot)$ is identified $\mu_{X_1,R}$-a.e.\ from the distribution of the observables. The following is the formal version of Corollary \ref{corollary_auction}, in the manner of Theorem \ref{thm: info change2}.

\begin{corollary}
\label{corollary_auction2}
Suppose that Assumptions \ref{assump: auction2a} and \ref{assump: auction2b} hold. Let $A \subset \mathbf{R}_+$ be a Borel set and $C$ a Borel subset of the support of $X_1$ with $P\{X_1 \in C\} > 0$. Then the following statements hold.

(i) For each $i = 1, \ldots, n$,
\begin{align*}
	a^{*f}(V_i, X_1, f^*(R), \zeta) &= a^{*}(V_i, X_1, f^*(R), \zeta), \text{ a.e.\ on the event } \{(X_1, f^*(R)) \in \mathbb{S}_{X_1,R}\} \text{ and }\\
    b^{*f}(V_i, X_1, f^*(R), \zeta) &= b^{*}(V_i, X_1, f^*(R), \zeta), \text{ a.e.\ on the event } \{(X_1, f^*(R)) \in \mathbb{S}_{X_1,R}, \ i \in \widetilde{N}^f\}.
\end{align*}

(ii) Let $S = \{ (X_1,f^*(R)) \in \mathbb{S}_{X_1,R} \}$. Then,
\begin{align*}
	&\mathbf{E}\left[p( A \mid X_1,f^*(R))1_S \mid X_1 \in C \right]\\
	&\quad \quad \quad \leq p_f(A \mid C) \leq \mathbf{E}\left[p( A \mid X_1,f^*(R))1_S \mid X_1 \in C \right] + P\left\{S^c \mid X_1 \in C \right\}.
\end{align*}
\end{corollary}

\subsection{Mathematical Proofs}\label{subsec: preliminaries}

In this section, we present the proofs of the results in Section \ref{subsec: invariance}. We begin with two preliminary lemmas.

\begin{lemma}
	\label{lemm: image support}
	Let $\mu$ be a Borel probability measure on a separable metric space $\mathbb{X}$, and let $g: \mathbb{X} \to \mathbb{X}'$ be a continuous map into a metric space $\mathbb{X}'$. Then
	\begin{align*}
		\mathrm{supp}(\mu \circ g^{-1}) = \mathrm{cl}\big(g(\mathrm{supp}\,\mu)\big),
	\end{align*}
	where $\mathrm{supp}$ denotes the support of a measure and $\mathrm{cl}$ the closure of a set.
\end{lemma}

\begin{proof}
Write $\nu = \mu \circ g^{-1}$, and recall that $\mathrm{supp}(\nu) = \{y: \nu(U) >0 \text{ for every open } U \text{ such that } y \in U\}$, which is closed. Let $y = g(x)$ with $x \in \mathrm{supp}\,\mu$, and let $U$ be an open set containing $y$. By continuity, $g^{-1}(U)$ is an open set containing $x$, so $\nu(U) = \mu(g^{-1}(U)) > 0$. Hence $g(\mathrm{supp}\,\mu) \subset \mathrm{supp}\,\nu$ and, the support being closed, $\mathrm{cl}(g(\mathrm{supp}\,\mu)) \subset \mathrm{supp}\,\nu$.

Conversely, let $y \notin \mathrm{cl}(g(\mathrm{supp}\,\mu))$. Then some open set $U$ containing $y$ satisfies $U \cap g(\mathrm{supp}\,\mu) = \varnothing$, so that $g^{-1}(U) \cap \mathrm{supp}\,\mu = \varnothing$ and $\nu(U) = \mu(g^{-1}(U)) \le \mu(\mathbb{X} \setminus \mathrm{supp}\,\mu) = 0$. Here $\mu(\mathbb{X} \setminus \mathrm{supp}\,\mu) = 0$, because $\mathbb{X} \setminus \mathrm{supp}\,\mu$ is the union of the open $\mu$-null sets and, $\mathbb{X}$ being a separable metric space, hence second countable, countably many of them cover it. Hence $y \notin \mathrm{supp}\,\nu$.
\end{proof}

\begin{lemma}
	\label{lemm: null transfer}
	Suppose that a game-policy pair $(G,f)$ satisfies Assumption \ref{assump: information structure and policy2}, and fix $i \in \{1,\ldots,n\}$. Then the following statements hold.
	
	(i) $\mathbb{S}_{W_C,W_{I,i}} = \mathbb{S}_{W_C} \times \mathbb{S}_{W_{I,i}}$ and $\mathbb{S}_{\tilde f(W_C),W_{I,i}} = \mathbb{S}_{\tilde f(W_C)} \times \mathbb{S}_{W_{I,i}}$.
	
	(ii) For every Borel set $N \subset \mathbb{S}_{W_C,W_{I,i}} \cap \mathbb{S}_{\tilde f(W_C),W_{I,i}}$, 
	\begin{align*}
		\mu_{T_i}(N) = 0 \text{ if and only if } \mu_{T_i}^f(N) = 0.
		\end{align*}
\end{lemma}

\begin{proof}
(i) Applying Lemma \ref{lemm: image support} to the coordinate projections of $\mathbb{S}_W = \mathbb{S}_{W_C} \times \mathbb{S}_{W_I}$ and $\mathbb{S}_{f(W)} = \mathbb{S}_{\tilde f(W_C)} \times \mathbb{S}_{W_I}$ in Assumption \ref{assump: information structure and policy2}(iv) gives the two product identities, the support $\mathbb{S}_{W_{I,i}}$ being common to the two distributions, as the policy leaves $W_I$ unchanged.

(ii) Let $N$ be as in the statement, and suppose that $\mu_{T_i}(N) = 0$. By (i), every signal $(w_C,w_{I,i}) \in N$ has $w_C \in \mathbb{S}_{W_C} \cap \mathbb{S}_{\tilde f(W_C)}$. Under $\mu_W^f$, almost every state has $w_I \in \mathbb{S}_{W_I}$. Hence, by Assumption \ref{assump: information structure and policy2}(iv), $\mu_W^f$-almost every state with signal in $N$ lies in
\begin{align*}
	\mathbb{S}_\cap = (\mathbb{S}_{W_C} \cap \mathbb{S}_{\tilde f(W_C)}) \times \mathbb{S}_{W_I},
\end{align*}
so that $\mu_{T_i}^f(N) = \mu_W^f(s_i^{-1}(N) \cap \mathbb{S}_\cap)$, which vanishes by Assumption \ref{assump: information structure and policy2}(iii), since $\mu_W(s_i^{-1}(N)) = \mu_{T_i}(N) = 0$. The converse implication, with the roles of the two distributions exchanged, holds by the same argument.
\end{proof}

\begin{proof}[Proof of Lemma \ref{lemm: master lemma}]
Since $G$ and $G_f$ share the information structure $I$, the interim payoff $U_i(\tau_i, t_i; \sigma)$ in (\ref{interim payoff}) is the \emph{same} function in both games, and at $t_i = (w_C, w_{I,i})$ it depends on $\sigma$ only through the values $\sigma(\csdot \mid w')$ at states $w' = (w_C, w_{I,i}, w_{I,-i})$ with $w_{I,-i}$ drawn from $\eta_i(\csdot \mid t_i)$.
The two games differ only in where obedience is imposed. A decision rule $\sigma \in \Sigma$ belongs to $\Sigma_{\mathsf{BCE}}(G)$ (resp.\ $\Sigma_{\mathsf{BCE}}(G_f)$) if and only if, for each $i$, off a $\mu_{T_i}$-null (resp.\ $\mu_{T_i}^f$-null) set of signals, $U_i(\mathsf{Id}, t_i; \sigma) \ge U_i(\tau_i, t_i; \sigma)$ for every $\tau_i$.

Now fix $\sigma \in \Sigma_{\mathsf{BCE}}'(G)$ and $\sigma_f \in \Sigma_{\mathsf{BCE}}'(G_f)$, and write $\widetilde B \eqdef (\mathbb{S}_{W_C} \cap \mathbb{S}_{\tilde f(W_C)}) \times \mathbb{W}_I$, so that, by Assumption \ref{assump: information structure and policy2}(iv), $\mathbb{S}_\cap = (\mathbb{S}_{W_C} \cap \mathbb{S}_{\tilde f(W_C)}) \times \mathbb{S}_{W_I} \subset \widetilde B$. Let $\sigma^\dagger$ be the gluing of $\sigma$ and $\sigma_f$ across $\widetilde B$: 
\begin{align*}
  \sigma^\dagger(\csdot \mid w) = \sigma(\csdot \mid w) 1\{	w \in \widetilde B\} + \sigma_f(\csdot \mid w) 1\{w \notin \widetilde B\}.
\end{align*}
By the closure under gluing, $\sigma^\dagger \in \Sigma'$. We now prove that $\sigma^\dagger \in \Sigma_{\mathsf{BCE}}'(G_f)$.

Fix $i$, and choose a Borel $\mu_{T_i}$-null set $N_0$ such that, at every $t_i \notin N_0$, the obedience inequalities $U_i(\mathsf{Id},t_i;\sigma) \ge U_i(\tau_i,t_i;\sigma)$ hold for every deviation $\tau_i$, and a Borel $\mu_{T_i}^f$-null set $N_1$ with the same property for $\sigma_f$. The set $N_2 \eqdef N_0 \cap \mathbb{S}_{W_C,W_{I,i}} \cap \mathbb{S}_{\tilde f(W_C),W_{I,i}}$ is a $\mu_{T_i}$-null Borel subset of $\mathbb{S}_{W_C,W_{I,i}} \cap \mathbb{S}_{\tilde f(W_C),W_{I,i}}$. By Lemma \ref{lemm: null transfer}, $\mu_{T_i}^f(N_2) = 0$. As $\mu_{T_i}^f$ is the distribution of $(\tilde f(W_C), W_{I,i})$, with its support $\mathbb{S}_{\tilde f(W_C),W_{I,i}}$, the set
\begin{align*}
	N \eqdef N_1 \cup N_2 \cup \big(\mathbb{T}_i \setminus \mathbb{S}_{\tilde f(W_C),W_{I,i}}\big)
\end{align*}
is a $\mu_{T_i}^f$-null Borel set.

To show $\sigma^\dagger \in \Sigma_{\mathsf{BCE}}'(G_f)$, it suffices to check the obedience inequalities for $\sigma^\dagger$ at every signal $t_i = (w_C,w_{I,i}) \notin N$. Such a signal lies in $\mathbb{S}_{\tilde f(W_C),W_{I,i}}$, which equals $\mathbb{S}_{\tilde f(W_C)} \times \mathbb{S}_{W_{I,i}}$ by Lemma \ref{lemm: null transfer}, so $w_C \in \mathbb{S}_{\tilde f(W_C)}$ and $w_{I,i} \in \mathbb{S}_{W_{I,i}}$. Fix $t_i \notin N$ and consider two cases, according to whether $w_C \in \mathbb{S}_{W_C}$.

Suppose first that $w_C \in \mathbb{S}_{W_C}$. Then $w_C \in \mathbb{S}_{W_C} \cap \mathbb{S}_{\tilde f(W_C)}$, so every state whose common component is $w_C$ lies in $\widetilde B$, where $\sigma^\dagger$ agrees with $\sigma$. Hence $U_i(\tau_i,t_i;\sigma^\dagger) = U_i(\tau_i,t_i;\sigma)$ for every $\tau_i$. Moreover, $t_i \in \mathbb{S}_{W_C} \times \mathbb{S}_{W_{I,i}}$, which equals $\mathbb{S}_{W_C,W_{I,i}}$ by Lemma \ref{lemm: null transfer}, and $t_i \in \mathbb{S}_{\tilde f(W_C),W_{I,i}}$ as noted, so if $t_i$ belonged to $N_0$ it would belong to $N_2 \subset N$. Hence $t_i \notin N_0$, the obedience inequalities hold at $t_i$ for $\sigma$, and therefore for $\sigma^\dagger$. 

Suppose next that $w_C \notin \mathbb{S}_{W_C}$. Then no state with common component $w_C$ lies in $\widetilde B$, so $\sigma^\dagger$ agrees with $\sigma_f$ at every such state and $U_i(\tau_i,t_i;\sigma^\dagger) = U_i(\tau_i,t_i;\sigma_f)$ for every $\tau_i$. Recall that $t_i$ was chosen outside $N$. Hence, $t_i \notin N_1$. The obedience inequalities hold at $t_i$ for $\sigma_f$, and therefore for $\sigma^\dagger$. Thus, we conclude that $\sigma^\dagger \in \Sigma_{\mathsf{BCE}}'(G_f)$.

Since $\mathbb{S}_\cap \subset \widetilde B$ and $\sigma^\dagger = \sigma$ everywhere on $\widetilde B$, we get 
\begin{align*}
	[\sigma]_{\mathbb{S}_\cap} = [\sigma^\dagger]_{\mathbb{S}_\cap} \in \widetilde\Sigma_{\mathsf{BCE},\mathbb{S}_\cap}'(G_f),
\end{align*}
and, $\sigma$ being arbitrary, $\widetilde\Sigma_{\mathsf{BCE},\mathbb{S}_\cap}'(G) \subset \widetilde\Sigma_{\mathsf{BCE},\mathbb{S}_\cap}'(G_f)$. The reverse inclusion follows by interchanging $G$ and $G_f$: the cylinder $\widetilde B$ is symmetric in the two games, and Lemma \ref{lemm: null transfer} transfers null sets in both directions. Finally, fix a Borel $B \subset \mathbb{S}_\cap$. Since $\sigma_1 \simeq_{\mathbb{S}_\cap} \sigma_2$ implies $\sigma_1 \simeq_B \sigma_2$, the equality of the $\mathbb{S}_\cap$-sliced sets implies that of the $B$-sliced ones.
\end{proof}

\begin{proof}[Proof of Corollary \ref{cor: cylinder slice}]
	Fix a pair of equilibrium selection rules $e(\csdot;G)$ and $e(\csdot;G_f)$, and fix $\sigma_1 \in \Sigma_{\mathsf{BCE}}(G)$. The hypothesis, applied to the pair $(\sigma_1, \sigma_f)$, gives
	\begin{align}
		\label{chain}
		\sigma_f(\csdot \mid w) = \sigma_1(\csdot \mid w), \enspace \text{a.e. } w \in \mathbb{S}_\cap,
	\end{align}
	for every $\sigma_f \in \Sigma_{\mathsf{BCE}}(G_f)$. Applying it once more to the pairs $(\sigma, \sigma_f)$, with $\sigma_f \in \Sigma_{\mathsf{BCE}}(G_f)$ fixed, and combining with (\ref{chain}), gives
	\begin{align*}
		\sigma(\csdot \mid w) = \sigma_f(\csdot \mid w) = \sigma_1(\csdot \mid w), \enspace \text{a.e. } w \in \mathbb{S}_\cap,
	\end{align*}
	for every $\sigma \in \Sigma_{\mathsf{BCE}}(G)$.

	Fix $D \in \mathcal{B}(\mathcal{P}(\mathbb{Y}))$ and a Borel $B \subset \mathbb{S}_\cap$. Since $q_G(\csdot \mid \tilde\sigma, w) = \sigma(\csdot \mid w)$ at $\mu_W$-a.e.\ $w$ for $e(\csdot;G)$-a.e.\ $\tilde\sigma$, and hence $q_G(\csdot \mid \tilde\sigma, w) = \sigma_1(\csdot \mid w)$ at a.e.\ $w \in \mathbb{S}_\cap$, by (\ref{e_w}) and Fubini's theorem,
	\begin{align*}
		\int_B e_w(D;G)\, d\mu_W(w) &= \int_{\widetilde\Sigma_{\mu_W}} \int_B 1\{q_G(\csdot \mid \tilde\sigma, w) \in D\}\, d\mu_W(w)\, de(\tilde\sigma;G) \\
		&= \int_B \delta_{\sigma_1(\csdot \mid w)}(D)\, d\mu_W(w).
	\end{align*}
	As the Borel set $B \subset \mathbb{S}_\cap$ is arbitrary, $e_w(D;G) = \delta_{\sigma_1(\csdot \mid w)}(D)$ at a.e.\ $w \in \mathbb{S}_\cap$. Taking $D$ in a countable generating $\pi$-system of $\mathcal{B}(\mathcal{P}(\mathbb{Y}))$ and discarding the countably many null sets gives $e_w(\csdot;G) = \delta_{\sigma_1(\csdot \mid w)}$ at a.e.\ $w \in \mathbb{S}_\cap$. The same argument, with $q_{G_f}$ and the prior $\mu_W^f$ in place of $q_G$ and $\mu_W$, gives $e_w(\csdot;G_f) = \delta_{\sigma_1(\csdot \mid w)}$ at a.e.\ $w \in \mathbb{S}_\cap$. In particular, the equality (\ref{eq: invariance sliced eq}) of Assumption \ref{assump: invariance ESR} holds for $B = \mathbb{S}_\cap$ and hence for every Borel $B \subset \mathbb{S}_\cap$, whatever the sliced sets. This proves Assumption \ref{assump: invariance ESR}. 
	
	Finally, by (\ref{eq: barycenter}), at a.e.\ $w \in \mathbb{S}_\cap$, for each fixed $A \in \mathcal{B}(\mathbb{Y})$, we have 
	\begin{align*}
		\rho_G(A \mid w) = \int \pi(A)\, de_w(\pi;G) = \sigma_1(A \mid w), \enspace \text{a.e. } w \in \mathbb{S}_\cap.
	\end{align*}
	Likewise, $\rho_{G_f}(A \mid w) = \sigma_1(A \mid w)$. Taking $A$ in a countable generating $\pi$-system of $\mathcal{B}(\mathbb{Y})$ and discarding the countably many null sets, we obtain 
	\begin{align*}
		\rho_G(\csdot \mid w) = \sigma_1(\csdot \mid w) = \rho_{G_f}(\csdot \mid w), \enspace \text{a.e.\ } w \in \mathbb{S}_\cap.
	\end{align*}
\end{proof}

\begin{proof}[Proof of Theorem \ref{thm: info change2}]
First, we let
\begin{align*}
	m_G(h; w) = \int_{\mathbb{Y}} h(y,w)d\rho_{G}(y\mid w),
\end{align*}
and define $m_{G_f}(h;w)$ analogously with $G$ replaced by $G_f$. Since $\rho_G(\csdot \mid w)$ and $\rho_{G_f}(\csdot \mid w)$ are Markov kernels and (\ref{h bounds2}) holds, we have $m_G(h;w) \in [\underline h(w), \overline h(w)]$ and $m_{G_f}(h;w) \in [\underline h(w), \overline h(w)]$ at every $w$, so both are finite. Both maps are measurable in $w$.\footnote{A monotone class argument from $h = 1_{A \times B}$ gives the measurability for bounded $h$, and the truncations $h_n = \max(\min(h,n),-n)$ converge to $h$ pointwise, dominated at each $w$ by the finite envelope bound, so $m_G(h_n;w) \rightarrow m_G(h;w)$, as $n \to \infty$, at every $w$, and likewise for $m_{G_f}$.} Both are also $\mu_W^f$-integrable by the integrability of the envelopes $\underline h, \overline h$. Thus, we obtain the following inequalities:
\begin{align}
	\label{bounds22}
	&\int_{\mathbb{S}_W} m_{G_f}(h; w) d \mu_W^f(w) + \int_{\mathbb{W} \setminus \mathbb{S}_W} \underline h(w) d \mu_W^f(w)\\ \notag
	&\le \int_{\mathbb{W}} m_{G_f}(h; w) d \mu_W^f(w) \le \int_{\mathbb{S}_W} m_{G_f}(h; w) d \mu_W^f(w) + \int_{\mathbb{W} \setminus \mathbb{S}_W} \overline h(w) d \mu_W^f(w),
\end{align}
and the middle term equals $\mathsf{AEP}(h)$ by (\ref{DP}). Since $\mu_W^f$ is concentrated on $\mathbb{S}_{f(W)}$, it suffices to show that, at a.e.\ $w \in \mathbb{S}_\cap$,
\begin{align}
	\label{eq323}
	m_{G}(h; w) = m_{G_f}(h; w),
\end{align}
because this implies
\begin{align*}
	\int_{\mathbb{S}_W} m_{G_f}(h;w) d \mu_W^f(w) = \int_{\mathbb{S}_W} m_{G}(h;w) d \mu_W^f(w) = \mathsf{ADP}_R(h),
\end{align*}
which, combined with (\ref{bounds22}), yields the desired bounds.

By Lemma \ref{lemm: master lemma}, with $\Sigma' = \Sigma$, we have $\widetilde\Sigma_{\mathsf{BCE},\mathbb{S}_\cap}(G) = \widetilde\Sigma_{\mathsf{BCE},\mathbb{S}_\cap}(G_f)$. Hence, by Assumption \ref{assump: invariance ESR}, 
\begin{align}
	\label{slice equality}
	e_w(\csdot;G) = e_w(\csdot;G_f), \enspace \text{a.e.\ } w \in \mathbb{S}_\cap.
\end{align}
Fix $A \in \mathcal{B}(\mathbb{Y})$. Both identities in (\ref{eq: barycenter}) hold at a.e.\ $w \in \mathbb{S}_\cap$, so combining them with (\ref{slice equality}) gives $\rho_G(A \mid w) = \rho_{G_f}(A \mid w)$ at a.e.\ $w \in \mathbb{S}_\cap$, and the $\pi$-system argument in the proof of Corollary \ref{cor: cylinder slice} gives $\rho_G(\csdot \mid w) = \rho_{G_f}(\csdot \mid w)$ at a.e.\ $w \in \mathbb{S}_\cap$. Then (\ref{eq323}) follows, $h(\csdot,w)$ being bounded between $\underline h(w)$ and $\overline h(w)$.
\end{proof}

\begin{proof}[Proof of Proposition \ref{prop: sharp bounds2}]
Take $I$ to be the complete information structure. We take each $\mathbb{W}_{I,i}$ to be a singleton and identify $\mathbb{W}$ with $\mathbb{W}_C \times \{\bar w_I\}$, with $\mathbb{W}_I = \{\bar w_I\}$. Then $T_i = s_i(W) = W$ and each belief map $\eta_i$ is the point mass at the degenerate opponents' private component. The condition (\ref{eq: envelope compatibility}) and Assumption \ref{assump: information structure and policy2}(ii) then hold trivially, with $g_i \equiv 1$ an envelope of the payoffs constructed below. Assumption \ref{assump: information structure and policy2}(i) holds with $\tilde f = f$, the private components being degenerate. Assumption \ref{assump: information structure and policy2}(iv) holds automatically: $\mu_W = \mu_{W_C} \otimes \delta_{\bar w_I}$ and $\mu_W^f = (\mu_{W_C} \circ \tilde f^{-1}) \otimes \delta_{\bar w_I}$, and the support of a product measure is the product of the supports of its factors. Assumption \ref{assump: information structure and policy2}(iii) holds by hypothesis.

Since $\mathbb{Y}$ is compact and $h$ is measurable with $h(\csdot,w)$ continuous on $\mathbb{Y}$ for each $w$, the measurable maximum theorem (cf.\ Theorem 18.19 of \cite{Aliprantis/Border:06:InfiniteDim}) yields measurable selectors $\overline a, \underline a: \mathbb{W} \rightarrow \mathbb{Y}$ with $h(\overline a(w),w) = \overline h(w)$ and $h(\underline a(w),w) = \underline h(w)$ at every $w$. Fix a measurable map $b: \mathbb{W} \to \mathbb{Y}$, let $c = \overline a$ on $\mathbb{S}_W$ and $c = b$ off $\mathbb{S}_W$, and set
\begin{align*}
	u_i(a,w) = 1\{a_i = c_i(w)\} \text{ and } \sigma(\csdot \mid w) = \delta_{c(w)},
\end{align*}
where $\delta_a$ is the point mass at $a \in \mathbb{Y}$. The set $\{(a,w) : a_i = c_i(w)\}$ is the preimage of the closed diagonal of $\mathbb{Y}_i \times \mathbb{Y}_i$ under the Borel map $(a,w) \mapsto (a_i, c_i(w))$, so $u_i$ is jointly measurable, and $\delta_{c(w)}(A) = 1\{w \in c^{-1}(A)\}$, so $\sigma \in \mathcal{K} = \Sigma$. Playing $c_i(w)$ is strictly dominant for player $i$ at every $w$, and $\sigma$ plays $c(w)$. Let $G = (B,I)$ and $G_f = (B_f,I)$, where $B = (\mathbb{Y},\mathbb{W},u,\mu_W)$ and $B_f = (\mathbb{Y},\mathbb{W},u,\mu_W^f)$. Both games depend on the choice of map $b$ only through the payoffs at the payoff state outside $\mathbb{S}_W$. We carry a general $b$ through the argument and specialize to the two choices $b = \overline a$ and $b = \underline a$, yielding the pairs $(G_1,G_{1,f})$ and $(G_2,G_{2,f})$, only at the end.

As $\sigma$ recommends the strictly dominant profile at every state, it is a BCE of both games and Assumption \ref{assump: BCE existence2} holds. Under complete information the signal is the state, so $\mu_{T_i} = \mu_W$ and $\mu_{T_i}^f = \mu_W^f$. Let $\sigma' \in \Sigma_{\mathsf{BCE}}(G)$, fix $i$, and discard the Borel $\mu_W$-null set off which the obedience inequalities hold against every deviation. At every remaining $w$, the constant deviation $\tau_i \equiv c_i(w)$ gives
\begin{align*}
	\sigma'(\{y : y_i = c_i(w)\} \mid w) = U_i(\mathsf{Id},w;\sigma') \ge U_i(\tau_i,w;\sigma') = 1.
\end{align*}
Therefore, $\sigma'(\csdot \mid w) = \delta_{c(w)} = \sigma(\csdot \mid w)$ at $\mu_W$-a.e.\ $w$. Conversely, any decision rule equal to $\sigma$ at $\mu_W$-a.e.\ $w$ is obedient at $\mu_W$-a.e.\ signal. Hence, the same argument applying under $\mu_W^f$,
\begin{align*}
	\Sigma_{\mathsf{BCE}}(G) &= \{\sigma' \in \Sigma : \sigma'(\csdot \mid w) = \sigma(\csdot \mid w) \text{ at $\mu_W$-a.e.\ } w\}, \text{ and}\\
	\Sigma_{\mathsf{BCE}}(G_f) &= \{\sigma' \in \Sigma : \sigma'(\csdot \mid w) = \sigma(\csdot \mid w) \text{ at $\mu_W^f$-a.e.\ } w\},
\end{align*}
that is, $\widetilde\Sigma_{\mathsf{BCE}}(G) = \{[\sigma]_{\mu_W}\}$ and $\widetilde\Sigma_{\mathsf{BCE}}(G_f) = \{[\sigma]_{\mu_W^f}\}$, singletons, closed. The equilibrium selection rules are the point masses at these classes, and $\rho_G$, $\rho_{G_f}$, $\mathsf{ADP}_R(h)$, and $\mathsf{AEP}(h)$ are well-defined.

Since $\mu_W(\mathbb{S}_W)=1$ and $\sigma = \delta_{\overline a(\csdot)}$ on $\mathbb{S}_W$, every BCE of $G$ agrees with $\delta_{\overline a(\csdot)}$ at $\mu_W$-a.e.\ $w$, so $P_G = \mu_W \otimes \delta_{\overline a(\csdot)}$ and $\rho_G = \delta_{\overline a(\csdot)}$ at $\mu_W$-a.e.\ $w$, and hence also at a.e.\ $w \in \mathbb{S}_\cap$. Since $\mu_W^f(\mathbb{S}_{f(W)}) = 1$, $\mu_W^f$-almost every $w \in \mathbb{S}_W$ lies in $\mathbb{S}_\cap$. Since the selector $\overline a$ satisfies $h(\overline a(w),w) = \overline h(w)$ at every $w$,
\begin{align*}
	\mathsf{ADP}_R(h) = \int_{\mathbb{S}_W} h(\overline a(w),w)\, d\mu_W^f(w) = \int_{\mathbb{S}_W} \overline h(w)\, d\mu_W^f(w) = \int_{\mathbb{S}_\cap} \overline h(w)\, d\mu_W^f(w),
\end{align*}
independent of $c$.

On the post-policy side, every BCE of $G_f$ agrees with $\sigma$ at $\mu_W^f$-a.e.\ $w$, so $P_{G_f} = \mu_W^f \otimes \sigma$ and $\rho_{G_f}(\csdot \mid w) = \sigma(\csdot \mid w)$ at $\mu_W^f$-a.e.\ $w$. Since $\sigma = \delta_{\overline a(\csdot)}$ on $\mathbb{S}_W$ and $\sigma = \delta_{c(\csdot)}$ off $\mathbb{S}_W$,
\begin{align}
	\label{decomp32}
	\mathsf{AEP}(h) = \int_{\mathbb{S}_{f(W)}} h(c(w),w)\, d\mu_W^f(w) = \mathsf{ADP}_R(h) + \int_{\mathbb{S}_{f(W)} \setminus \mathbb{S}_W} h(c(w),w)\, d\mu_W^f(w).
\end{align}

Since every BCE of $G$ equals $\sigma$ at a.e.\ $w \in \mathbb{S}_\cap$, and every BCE of $G_f$ equals $\sigma$ at a.e.\ $w \in \mathbb{S}_\cap$, any pair coincides at a.e.\ $w \in \mathbb{S}_\cap$, and, by Corollary \ref{cor: cylinder slice}, Assumption \ref{assump: invariance ESR} holds for every pair of equilibrium selection rules.

The bounds in (\ref{bounds master}) are achieved by the two choices of $b$, giving two pairs of games $(G_1,G_{1,f})$ and $(G_2,G_{2,f})$. Taking $b = \overline a$ gives $c = \overline a$ throughout, so $h(c(w),w) = \overline h(w)$ off $\mathbb{S}_W$, the last integral in (\ref{decomp32}) is $\Delta(\overline h)$, the domains $\mathbb{S}_{f(W)} \setminus \mathbb{S}_W$ and $\mathbb{W} \setminus \mathbb{S}_W$ differing by the $\mu_W^f$-null set $\mathbb{W} \setminus \mathbb{S}_{f(W)}$, and $\mathsf{AEP}(h) = \mathsf{ADP}_R(h) + \Delta(\overline h)$ under $(G_1,G_{1,f})$. Taking $b = \underline a$ gives $h(c(w),w) = \underline h(w)$ off $\mathbb{S}_W$, so the integral is $\Delta(\underline h)$, and $\mathsf{AEP}(h) = \mathsf{ADP}_R(h) + \Delta(\underline h)$ under $(G_2,G_{2,f})$. The two pre-policy games $G_1$ and $G_2$ differ only in payoffs off $\mathbb{S}_W$, so they share the reduced form $\rho_G = \delta_{\overline a(\csdot)}$ at $\mu_W$-a.e.\ $w$, and hence the same joint distribution $P_G = \mu_W \otimes \rho_G$ of $(Y,W)$ before the policy. The identified objects $\mathsf{ADP}_R(h)$, $\Delta(\overline h)$, and $\Delta(\underline h)$ are common to the two pairs of games.
\end{proof}

\begin{proof}[Proof of Proposition \ref{prop: control fn2}]
We first establish the following support identity:
\begin{align}
	\label{support identity}
	1\{(f_a(x_a),x_b,\bar\varepsilon) \in \mathbb{S}_W\} = 1\{(f_a(x_a),x_b) \in \mathbb{S}_{X_a,X_b}\}, \enspace \mu_{X_a,X_b,\varepsilon}\text{-a.e.},
\end{align}
where $\mu_{X_a,X_b,\varepsilon}$ denotes the joint distribution of $(X_a,X_b,\varepsilon)$. The inequality $\le$ holds pointwise: if every neighborhood of $(f_a(x_a),x_b,\bar\varepsilon)$ carries positive $\mu_W$-mass, then, the coordinate projection being continuous, every neighborhood of $(f_a(x_a),x_b)$ carries positive $\mu_{X_a,X_b}$-mass, so that $(f_a(x_a),x_b) \in \mathbb{S}_{X_a,X_b}$. The inequality $\ge$ follows from the cylinder condition in Assumption \ref{assump: control fn2}(ii): for $\mu_{X_a,X_b,\varepsilon}$-a.e.\ $(x_a,x_b,\bar\varepsilon)$, we have $(x_a,x_b,\bar\varepsilon) \in \mathbb{S}_W$, so whenever $(f_a(x_a),x_b) \in \mathbb{S}_{X_a,X_b}$, applying the cylinder condition with $x_a' = f_a(x_a)$ gives $(f_a(x_a),x_b,\bar\varepsilon) \in \mathbb{S}_W$. This proves (\ref{support identity}).

We also note a null-set transfer: for every Borel $N \subset \mathbb{S}_{X_a,X_b}$ with $\mu_{X_a,X_b}(N) = 0$,
\begin{align}
	\label{obs null transfer}
	P\{f^*(X) \in N\} = 0.
\end{align}
Indeed, $P\{f^*(X) \in N\} = \mu_W^f(N \times \mathbb{W}_I) = \mu_W^f(B)$, where $B \eqdef (N \cap \mathbb{S}_{f^*(X)}) \times \mathbb{S}_\varepsilon$ is Borel, by Assumption \ref{assump: information structure and policy2}(iv). Since $\mu_W(B) \le \mu_{X_a,X_b}(N) = 0$, Assumption \ref{assump: information structure and policy2}(iii) gives $\mu_W^f(B) = 0$.

(i) Since $f(W) = (f^*(X),\varepsilon)$ and $\overline h(w) = \overline h^*(x)$, the definition of $\Delta(\overline h)$ in Theorem \ref{thm: info change2} and the support identity (\ref{support identity}) give
\begin{align*}
	\Delta(\overline h) = \mathbf{E}\left[\overline h^*(f^*(X))1\left\{(f_a(X_a),X_b,\varepsilon) \notin \mathbb{S}_W\right\}\right]
	= \mathbf{E}\left[\overline h^*(f^*(X))1\left\{f^*(X) \notin \mathbb{S}_X\right\}\right],
\end{align*}
where the expectations are finite by the $\mu_W^f$-integrability of $\overline h$. The identification of $\Delta(\underline h)$ follows by the same argument with $\underline h^*$ in place of $\overline h^*$.

(ii) Define, at every $(x_a,x_b,\bar\varepsilon) \in \mathbb{W}$,
\begin{align*}
	m_h(x_a,x_b,\bar\varepsilon) = \int h^*(y,(x_a,x_b))\, d\rho_G(y \mid x_a,x_b,\bar\varepsilon),
\end{align*}
a measurable map satisfying $\underline h^*(x_a,x_b) \le m_h(x_a,x_b,\bar\varepsilon) \le \overline h^*(x_a,x_b)$ at every point. By (\ref{cond prob}), $m_h$ is a version of the conditional expectation $\mathbf{E}[h^*(Y,X) \mid (X_a,X_b,\varepsilon) = \csdot]$. Since the map $(x_a,x_b) \mapsto \mu_{\varepsilon | X_b}(\csdot \mid x_b)$ is the version of the conditional distribution of $\varepsilon$ given $(X_a,X_b)$ used throughout, we have 
\begin{align}
	\label{eq445}
	\mathbf{E}[h^*(Y,X) \mid (X_a,X_b) = (x_a,x_b)] = \int m_h(x_a,x_b,\bar\varepsilon)\, d\mu_{\varepsilon | X_b}(\bar\varepsilon \mid x_b), \enspace \mu_{X_a,X_b}\text{-a.e.\ } (x_a,x_b),
\end{align}
where the latter map is measurable because $\mu_{\varepsilon|X_b}$ is a Markov kernel and $m_h$ is measurable and bounded by the envelopes. The equality above holds off a Borel $\mu_{X_a,X_b}$-null set $N$. Furthermore, the null-set transfer (\ref{obs null transfer}), applied to $N \cap \mathbb{S}_{X_a,X_b}$, gives $P\{f^*(X) \in N \cap \mathbb{S}_{X_a,X_b}\} = 0$. Thus, for $\mu_{X_a,X_b}$-a.e.\ $(x_a,x_b)$ with $(f_a(x_a),x_b) \in \mathbb{S}_{X_a,X_b}$,
\begin{align}
	\label{eq3}
	\mathbf{E}\left[ h^*(Y,X) \mid (X_a,X_b) = (f_a(x_a),x_b)\right] = \int m_h(f_a(x_a),x_b,\bar \varepsilon) d\mu_{\varepsilon | X_b}(\bar \varepsilon \mid x_b).
\end{align}
By the definition of $\mathsf{ADP}_R(h)$ in (\ref{DP}) with the inner integral written as $m_h$, the change of variables under which $\mu_W^f$ is the distribution of $(f^*(X),\varepsilon)$, and the support identity (\ref{support identity}), we write
\begin{align*}
	\mathsf{ADP}_R(h) &= \int m_h(f_a(x_a),x_b,\bar \varepsilon)  1\left\{(f_a(x_a),x_b) \in \mathbb{S}_{X_a,X_b} \right\} d\mu_{X_a,X_b,\varepsilon}(x_a,x_b,\bar \varepsilon)\\
	&= \int \int m_h(f_a(x_a),x_b,\bar \varepsilon)  d\mu_{\varepsilon | X_b}(\bar \varepsilon \mid x_b) 1\left\{(f_a(x_a),x_b) \in \mathbb{S}_{X_a,X_b} \right\} d\mu_{X_a,X_b}(x_a,x_b).
\end{align*}
The second equality invokes disintegration of $\mu_{X_a,X_b,\varepsilon}$ along $(X_a,X_b)$, using the fixed kernel $(x_a,x_b) \mapsto \mu_{\varepsilon | X_b}(\csdot \mid x_b)$ as the version of the conditional distribution of $\varepsilon$ given $(X_a,X_b)$. By (\ref{eq3}), the last double integral is equal to
\begin{align*}
	\int \mathbf{E}\left[ h^*(Y,X) \mid (X_a,X_b) = (f_a(x_a),x_b)\right]  1\left\{(f_a(x_a),x_b) \in \mathbb{S}_{X_a,X_b} \right\}  d\mu_{X_a,X_b}(x_a,x_b),
\end{align*}
which is the equality asserted in part (ii).
\end{proof}

\begin{proof}[Proof of Corollary \ref{cor: bounds2}]
The restricted selection rules $e'(\csdot;G)$ and $e'(\csdot;G_f)$ are concentrated on the restricted equilibrium-class sets $\widetilde\Sigma_{\mathsf{BCE}}'(G)$ and $\widetilde\Sigma_{\mathsf{BCE}}'(G_f)$, which are Borel by hypothesis, and the reduced forms $\rho_G'$ and $\rho_{G_f}'$ are fixed disintegrations of the corresponding mixtures $P_G'$ and $P_{G_f}'$, formed as in (\ref{rho_G}) and (\ref{rho_Gf}) with $e'$ in place of $e$.

We establish the invariance
\begin{align}
	\label{invar}
	\rho_G'(\csdot \mid w) = \rho_{G_f}'(\csdot \mid w) \quad \text{at a.e.\ } w \in \mathbb{S}_\cap.
\end{align}
Under hypothesis (a), Lemma \ref{lemm: master lemma} applies to $\Sigma'$: the restricted BCE sets are nonempty by the assumption of the corollary, and the closure under gluing across $(\mathbb{S}_{W_C} \cap \mathbb{S}_{\tilde f(W_C)}) \times \mathbb{W}_I$ is hypothesis (a). The lemma gives $\widetilde\Sigma_{\mathsf{BCE},B}'(G) = \widetilde\Sigma_{\mathsf{BCE},B}'(G_f)$ for every Borel $B \subset \mathbb{S}_\cap$. By Assumption \ref{assump: invariance ESR} in terms of the BCEs, at $B = \mathbb{S}_\cap$, we have 
\begin{align*}
  e_w'(\csdot;G) = e_w'(\csdot;G_f) \quad \text{at a.e.\ } w \in \mathbb{S}_\cap.	
\end{align*}
The invariance in (\ref{invar}) then follows exactly as in the final paragraph of the proof of Theorem \ref{thm: info change2}, run with the restricted objects, the identities (\ref{eq: barycenter}) holding for them by the same construction. Under hypothesis (b), the restricted form of Corollary \ref{cor: cylinder slice} yields the invariance (\ref{invar}) directly.

By (\ref{invar}), the identity (\ref{eq323}) holds for the primed objects, $h(\csdot,w)$ being bounded between $\underline h(w)$ and $\overline h(w)$. The proof of Theorem \ref{thm: info change2} derives the bounds from (\ref{eq323}) through the inequalities (\ref{bounds22}), an argument that applies verbatim here.
\end{proof}

\begin{proof}[Proof of Lemma \ref{lemm: gluing}]
(i) Let $\sigma^\dagger$ be the gluing of $\sigma$ and $\sigma'$ across $B$: $\sigma^\dagger(\csdot \mid w) = \sigma(\csdot \mid w)$ for $w \in B$ and $\sigma^\dagger(\csdot \mid w) = \sigma'(\csdot \mid w)$ otherwise. It belongs to $\mathcal{K} = \Sigma$: it is a probability measure at each $w$, and $w \mapsto \sigma^\dagger(A \mid w)$ is Borel for each Borel $A \subset \mathbb{Y}$, since $B$ is Borel and $\sigma, \sigma' \in \mathcal{K}$. Each $\beta_i^\dagger$ is a Markov kernel from $\mathbb{T}_i$ to the probability measures on $\mathbb{Y}_i$. Since $\mathbb{W} = \mathbb{W}_C \times \mathbb{W}_I$, we have $1\{w \in B\} = 1\{w_C \in D\}$, a function of each signal $t_i = (w_C,w_{I,i})$. Fix a Borel $A = A_1 \times \cdots \times A_n \subset \mathbb{Y}$ and $w \in \mathbb{W}$, and write $x = 1\{w_C \in D\}$ and $y = 1\{w_C \notin D\}$, so that $xy = 0$, $x^n = x$, and $y^n = y$. Expanding the product, the cross terms vanish and
\begin{align*}
	\prod_{i=1}^n \beta_i^\dagger(A_i \mid t_i) = x \prod_{i=1}^n \beta_i(A_i \mid t_i) + y \prod_{i=1}^n \beta_i'(A_i \mid t_i) = x\, \sigma(A \mid w) + y\, \sigma'(A \mid w) = \sigma^\dagger(A \mid w).
\end{align*}
Hence $\sigma^\dagger$ satisfies (\ref{B}) with the factors $\beta_i^\dagger$.

(ii) For the class in (\ref{B}), part (i) exhibits the gluing of any two members as a member. For the symmetric refinement, choose factorizations with common factors, $\beta_i = \beta$ and $\beta_i' = \beta'$ for all $i$, which exist by the definition of the refinement. The signal spaces coincide, $\mathbb{T}_1 = \cdots = \mathbb{T}_n$, and the switch $1\{w_C \in D\}$ does not depend on $i$, so the factors $\beta_i^\dagger$ in (\ref{glued beta}) are common across the players, and the gluing lies in the symmetric refinement.
\end{proof}

\begin{proof}[Proof of Corollary \ref{example1_corollary2}]
We first place the entry game within the scope of Theorem \ref{thm: info change2} and Proposition \ref{prop: control fn2}. By the independence of $X$ and $\varepsilon$ in Assumption \ref{assump: entry2}(iii), $\mu_W = \mu_X \otimes \mu_\varepsilon$ and $\mu_W^f = (\mu_X \circ f^{*-1}) \otimes \mu_\varepsilon$, where $\mu_X$ denotes the distribution of $X$ under $\mu_W$, so that
\begin{align}
	\label{entry supports}
	\mathbb{S}_W = \mathbb{S}_X \times \mathbb{S}_\varepsilon, \quad \mathbb{S}_{f(W)} = \mathbb{S}_{f^*(X)} \times \mathbb{S}_\varepsilon, \quad \mathbb{S}_\cap = (\mathbb{S}_X \cap \mathbb{S}_{f^*(X)}) \times \mathbb{S}_\varepsilon.
\end{align}
Assumption \ref{assump: information structure and policy2}(i) holds with $\tilde f = f^*$ under the public-private dichotomy, where $W_C = X$ and $W_{I,i} = \varepsilon_i$, and with $\tilde f = f$ under complete information, where $W_C = (X,\varepsilon)$ and each private component is a singleton. Assumption \ref{assump: information structure and policy2}(ii) holds in the latter case because each belief map is a point mass, and in the former because the fixed belief version of Assumption \ref{assump: entry2}(vi) is constant in $x$, that is, in the entire common component. Assumption \ref{assump: entry2}(ii) supplies Assumption \ref{assump: information structure and policy2}(iii): by Fubini's theorem, a Borel subset of the common support in (\ref{entry supports}) is $\mu_W$-null if and only if $\mu_\varepsilon$-almost every $\varepsilon$-section of it, a subset of $\mathbb{S}_X \cap \mathbb{S}_{f^*(X)}$, is $\mu_X$-null, which by Assumption \ref{assump: entry2}(ii) holds if and only if the set is $\mu_W^f$-null. Assumption \ref{assump: information structure and policy2}(iv) is the first two identities in (\ref{entry supports}) under the public-private dichotomy, and holds trivially under complete information, the private components being degenerate. Assumption \ref{assump: entry2}(iv) supplies Assumption \ref{assump: BCE existence2}, and Assumption \ref{assump: entry2}(v) supplies the conclusion of Assumption \ref{assump: invariance ESR} at $B = \mathbb{S}_\cap$, hence at every Borel subset. As for Proposition \ref{prop: control fn2}, Assumption \ref{assump: control fn2}(i) follows from Assumption \ref{assump: entry2}(iii), with the fixed version $\mu_{\varepsilon \mid X_b} = \mu_\varepsilon$, and Assumption \ref{assump: control fn2}(ii) from the product form of $\mathbb{S}_W$ in (\ref{entry supports}). Proposition \ref{prop: control fn2} is stated for the split $W_C = X$ and $W_I = \varepsilon$, which is the split at hand under the public-private dichotomy but not under complete information, where $W_C = (X,\varepsilon)$. This is immaterial here: its proof uses that split only through the product support forms of Assumption \ref{assump: information structure and policy2}(iv), and (\ref{entry supports}) supplies those forms under either information structure.

Take $h(y,w) = h^*(y,x)$ with $h^*(y,x) = 1\{y=a\}1\{x \in C\}$, so that $\overline h^*(x) = 1\{x \in C\}$ and $\underline h^*(x) = 0$. By the definition of $p_f$ above,
\begin{align*}
	\mathsf{AEP}(h) =(\mu_W \circ f^{-1})(C \times \mathbf{R}^{n d_\varepsilon}) \times  p_f(Y = a \mid C),
\end{align*}
and, by Proposition \ref{prop: control fn2}(i) with these envelopes,
\begin{align*}
	\Delta(\overline h) = P\left\{ f^*(X) \notin \mathbb{S}_X, f^*(X) \in C \right\} \text{ and } \Delta(\underline h) = 0.
\end{align*}

Turning to the decomposition-based prediction, Proposition \ref{prop: control fn2}(ii) gives
\begin{align}
	\mathsf{ADP}_R(h) &= \int \mathbf{E}\left[h^*(Y,X) \mid X = f^*(x)\right] 1\{ f^*(x) \in \mathbb{S}_X\} d\mu_X(x)\\ \notag
	&= \int P\left\{Y = a \mid X = f^*(x) \right\} 1\{f^*(x) \in C \cap \mathbb{S}_X \} d\mu_X(x)\\ \notag
	&= \mathbf{E}\left[ \pi(a \mid f^*(X)) \mid f^*(X) \in C \right] P\left\{f^*(X) \in C\right\}.
\end{align}
The second equality evaluates the conditioning identity 
\begin{align*}
	\mathbf{E}[h^*(Y,X) \mid X = x'] = P\{Y = a \mid X = x'\}1\{x' \in C\},
\end{align*}
valid at $\mu_X$-a.e.\ $x'$, at the shifted point $x' = f^*(x)$.\footnote{The substitution replaces $x'$ by $f^*(x)$ on both sides of the identity, so that a statement holding at $\mu_X$-a.e.\ $x'$ is invoked at $\mu_X$-a.e.\ $x$. It is not automatic, because the evaluation points $f^*(x)$ are distributed as $f^*(X)$, whose distribution need not be absolutely continuous with respect to $\mu_X$, so that the exceptional set $N$ of the identity, though $\mu_X$-null, could in principle have $\mu_X \circ f^{*-1}(N)>0$. To see how such a substitution is valid, fix Borel versions of the two conditional expectations and let $N$ be a Borel $\mu_X$-null set off which the identity holds. Then $N \cap \mathbb{S}_X \cap \mathbb{S}_{f^*(X)}$ is a Borel $\mu_X$-null subset of $\mathbb{S}_X \cap \mathbb{S}_{f^*(X)}$, so Assumption \ref{assump: entry2}(ii) gives $P\{f^*(X) \in N \cap \mathbb{S}_X \cap \mathbb{S}_{f^*(X)}\} = 0$, and since $f^*(X) \in \mathbb{S}_{f^*(X)}$ almost surely, $P\{f^*(X) \in N, f^*(X) \in \mathbb{S}_X\} = 0$. The identity therefore holds at $\mu_X$-a.e.\ $x$ at which the indicator $1\{f^*(x) \in \mathbb{S}_X\}$ does not vanish. The same argument shows that $\pi(a \mid f^*(X))$ does not depend on the version of $P\{Y = a \mid X = \csdot\}$ chosen, so that the bounds of the corollary are well defined.} The third equality uses the definition of $\pi$.

By Theorem \ref{thm: info change2},
\begin{align*}
	\mathsf{ADP}_R(h) \le \mathsf{AEP}(h) \le \mathsf{ADP}_R(h) + \Delta(\overline h).
\end{align*}
Dividing through by $P \{f^*(X) \in C\}$, we obtain the desired result.
\end{proof}

We prepare the proof of Corollary \ref{corollary_auction2} with three lemmas. The first places the auction within the scope of the results of this appendix, the second isolates the invariance content of the corollary, and the third collects the hypotheses of the two results from which the corollary is assembled, Corollary \ref{cor: bounds2} and Proposition \ref{prop: control fn2}.

\begin{lemma}
	\label{lemm: auction framework}
	Suppose that Assumptions \ref{assump: auction2a}(ii) and \ref{assump: auction2b} hold. Then the auction, as placed in the framework of Appendix \ref{App:Counterfactual Predictions in Bayesian Games} in Section \ref{subsubsec: auctions appendix}, satisfies Assumption \ref{assump: information structure and policy2}. Moreover, after the rearrangement of coordinates in Assumption \ref{assump: auction2b}(ii), $\mathbb{S}_{W_C} = \mathbb{S}_{X_1,\zeta} \times \mathbb{S}_R$ and $\mathbb{S}_{X_1,R} = \mathbb{S}_{X_1} \times \mathbb{S}_R$.
\end{lemma}

\begin{proof}
Parts (i), (ii), and (iv) of Assumption \ref{assump: information structure and policy2} are read off from the setup. Part (i) holds with the map $\tilde f$ fixed in Section \ref{subsubsec: auctions appendix}, measurable by Assumption \ref{assump: auction2a}(ii), the policy altering only $R$, a coordinate of $W_C$. For part (ii), the fixed belief version depends on the signal only through $V_i$, consistently with Assumption \ref{assump: auction2b}(iii). Hence, it is constant in the entire common component and invariant along the policy as that part requires, common to $G$ and $G_f$. For part (iv), Assumption \ref{assump: auction2b}(ii) supplies the product forms of $\mathbb{S}_W$ and $\mathbb{S}_{f(W)}$, and projecting them onto the coordinates of $W_C$ by Lemma \ref{lemm: image support} gives $\mathbb{S}_{W_C} = \mathbb{S}_{X_1,\zeta} \times \mathbb{S}_R$ and $\mathbb{S}_{\tilde f(W_C)} = \mathbb{S}_{X_1,\zeta} \times \mathbb{S}_{f^*(R)}$, so that $\mathbb{S}_W = \mathbb{S}_{W_C} \times \mathbb{S}_V$ and $\mathbb{S}_{f(W)} = \mathbb{S}_{\tilde f(W_C)} \times \mathbb{S}_V$, as that part requires. Projecting further onto the coordinates $(x_1,r)$ gives $\mathbb{S}_{X_1,R} = \mathbb{S}_{X_1} \times \mathbb{S}_R$. These are the two identities of the lemma.

Assumption \ref{assump: information structure and policy2}(iii) requires that $\mu_W$ and $\mu_W^f$ share their null sets among the Borel subsets of $\mathbb{S}_\cap$, while Assumption \ref{assump: auction2b}(iv) grants the corresponding property for the distributions of $(X_1,R)$ and $(X_1,f^*(R))$. Since $f^*(R)$ is a function of $R$, Assumption \ref{assump: auction2b}(i) makes $\mu_{V,\zeta \mid X_1}(\csdot \mid x_1)$ the conditional distribution of $(V,\zeta)$ given $(X_1,f^*(R)) = (x_1,r)$ as well. Hence, writing $B_{x_1,r} = \{(v,\zeta) : (x_1,r,v,\zeta) \in B\}$ for the $(x_1,r)$-section of a Borel set $B \subset \mathbb{W}$ and $g(x_1,r) = \mu_{V,\zeta \mid X_1}(B_{x_1,r} \mid x_1)$, a Borel map, we find that
\begin{align*}
	\mu_W(B) = \int g \, d\mu_{X_1,R} \text{ and }  \mu_W^f(B) = \int g \, d\mu_{X_1,f^*(R)},
\end{align*}
where $\mu_{X_1,R}$ and $\mu_{X_1,f^*(R)}$ denote the distributions of $(X_1,R)$ and $(X_1,f^*(R))$. As $g$ is nonnegative, $\mu_W(B) = 0$ if and only if $\mu_{X_1,R}(\{g > 0\}) = 0$, and $\mu_W^f(B) = 0$ if and only if $\mu_{X_1,f^*(R)}(\{g > 0\}) = 0$. Suppose now that $B \subset \mathbb{S}_\cap$. If the section $B_{x_1,r}$ is nonempty, there is a point $(x_1,r,v,\zeta) \in B$. Since $B \subset \mathbb{S}_\cap$, this point lies in both $\mathbb{S}_W$ and $\mathbb{S}_{f(W)}$, so that $(x_1,r) \in \mathbb{S}_{X_1,R} \cap \mathbb{S}_{X_1,f^*(R)}$ by Lemma \ref{lemm: image support}. Since $g(x_1,r) = 0$ at every $(x_1,r)$ whose section is empty, the Borel set $\{g > 0\}$ is contained in $\mathbb{S}_{X_1,R} \cap \mathbb{S}_{X_1,f^*(R)}$, and Assumption \ref{assump: auction2b}(iv) delivers part (iii).
\end{proof}

\begin{lemma}
	\label{lemm: auction invariance}
	Under Assumptions \ref{assump: auction2a} and \ref{assump: auction2b}, the following statements hold.

	(i) For each $i$, at $\mu_{T_i}^f$-a.e.\ signal $(v_i,x_1,r,\zeta)$ with $(x_1,r) \in \mathbb{S}_{X_1,R}$, the profiles $(a^*,b^*)$ and $(a^{*f},b^{*f})$ prescribe the same entry decision and, where entry is prescribed, the same bid.

	(ii) For any $\sigma \in \Sigma_{\mathsf{BCE}}'(G)$ and $\sigma' \in \Sigma_{\mathsf{BCE}}'(G_f)$, 
	\begin{align*}
		\sigma(\csdot \mid w) = \sigma'(\csdot \mid w), \enspace \text{a.e.\ } w \in \mathbb{S}_\cap.
	\end{align*}

	(iii) For every pair of restricted equilibrium selection rules, 
	\begin{align*}
		\rho_G'(\csdot \mid w) = \sigma^*(\csdot \mid w), \enspace \mu_W\text{-a.e.\ } w \text{ and }
		\rho_{G_f}'(\csdot \mid w) = \sigma^{*f}(\csdot \mid w), \enspace \mu_W^f\text{-a.e.\ } w.
	\end{align*}
\end{lemma}

\begin{proof}
	By Lemma \ref{lemm: auction framework}, the auction satisfies Assumption \ref{assump: information structure and policy2}. Let $\sigma^*$ and $\sigma^{*f}$ be the decision rules induced by $(a^*,b^*)$ and $(a^{*f},b^{*f})$, fixed in Section \ref{subsubsec: auctions appendix}, so that $\sigma^* \in \Sigma_{\mathsf{BCE}}'(G)$ and $\sigma^{*f} \in \Sigma_{\mathsf{BCE}}'(G_f)$. By the uniqueness property (c) of Assumption \ref{assump: auction2a}(iii), every element of $\Sigma_{\mathsf{BCE}}'(G)$ is equal to $\sigma^*$ at $\mu_W$-a.e.\ $w$ and every element of $\Sigma_{\mathsf{BCE}}'(G_f)$ is equal to $\sigma^{*f}$ at $\mu_W^f$-a.e.\ $w$. The restricted equilibrium-class sets are therefore the singletons $\{[\sigma^*]_{\mu_W}\}$ and $\{[\sigma^{*f}]_{\mu_W^f}\}$.

	(i) The two profiles agree on the common support:
	\begin{align*}
		\sigma^*(\csdot \mid w) = \sigma^{*f}(\csdot \mid w), \enspace \text{a.e.\ } w \in \mathbb{S}_\cap.
	\end{align*}
	Indeed, by property (a) of Assumption \ref{assump: auction2a}(iii) with the Borel set $\mathbb{S}_{W_C} \cap \mathbb{S}_{\tilde f(W_C)}$ of common-component values, $\Sigma'$ is closed under gluing across $(\mathbb{S}_{W_C} \cap \mathbb{S}_{\tilde f(W_C)}) \times \mathbb{W}_I$. We apply Lemma \ref{lemm: master lemma} to obtain the middle equality in
	\begin{align}
		\label{eqs}
		\{[\sigma^*]_{\mathbb{S}_\cap}\} = \widetilde\Sigma_{\mathsf{BCE},\mathbb{S}_\cap}'(G) = \widetilde\Sigma_{\mathsf{BCE},\mathbb{S}_\cap}'(G_f) = \{[\sigma^{*f}]_{\mathbb{S}_\cap}\}.
	\end{align}
	By property (c) of Assumption \ref{assump: auction2a}(iii), every element of $\Sigma_{\mathsf{BCE}}'(G)$ is $\simeq_{\mathbb{S}_\cap}$-equivalent to $\sigma^*$, and every element of $\Sigma_{\mathsf{BCE}}'(G_f)$ to $\sigma^{*f}$, which gives the first and the last equalities.

	Now fix $i$. Under $\mu_W^f$, almost every state lies in $\mathbb{S}_{f(W)} = \mathbb{S}_{X_1,\zeta} \times \mathbb{S}_{f^*(R)} \times \mathbb{S}_V$, so such a state with $(x_1,r) \in \mathbb{S}_{X_1,R} = \mathbb{S}_{X_1} \times \mathbb{S}_R$ lies in $\mathbb{S}_W$ as well, hence in $\mathbb{S}_\cap$. At a.e.\ such state, the kernels $\sigma^*(\csdot \mid w)$ and $\sigma^{*f}(\csdot \mid w)$ coincide by (\ref{eqs}), and, being point masses at the two profiles' prescriptions as they are from pure strategies, they coincide only if the prescribed actions agree. Bids lie in $[0,\overline v]$ and staying out is coded $-1$, so agreement of the actions is agreement of the entry decisions together with agreement of the bids where entry is prescribed. The states at which bidder $i$'s prescriptions disagree and $(x_1,r) \in \mathbb{S}_{X_1,R}$ form the inverse image under $s_i$ of a Borel set of signals, which is therefore $\mu_{T_i}^f$-null.

	(ii) By the proof of (i), every element of $\Sigma_{\mathsf{BCE}}'(G)$ is $\simeq_{\mathbb{S}_\cap}$-equal to $\sigma^*$ and every element of $\Sigma_{\mathsf{BCE}}'(G_f)$ to $\sigma^{*f}$. With the agreement of $\sigma^*$ and $\sigma^{*f}$ in (\ref{eqs}), this is (ii).

	(iii) As $e'(\csdot;G)$ is the point mass at $[\sigma^*]_{\mu_W}$, the representative $\sigma$ selected for that class agrees with $\sigma^*$ at $\mu_W$-a.e.\ $w$, so $\mu_W \otimes \sigma = \mu_W \otimes \sigma^*$ and the probability in (\ref{rho_G}), formed with $e'$ in place of $e$, reduces to $P_G' = \mu_W \otimes \sigma^*$. Thus $\rho_G'$ and $\sigma^*$ are both disintegrations of $P_G'$ over $\mu_W$, whence 
	\begin{align*}
		\rho_G'(\csdot \mid w) = \sigma^*(\csdot \mid w), \enspace \mu_W\text{-a.e. } w.
	\end{align*}
	The same argument, with $\mu_W^f$ and $\sigma^{*f}$ in place of $\mu_W$ and $\sigma^*$, gives the second identity.
\end{proof}

\begin{lemma}
	\label{lemm: auction applicability}
	Suppose that Assumptions \ref{assump: auction2a} and \ref{assump: auction2b} hold, and let $\Sigma'$ be the class of decision rules in Assumption \ref{assump: auction2a}(iii). Then the following statements hold.

	(i) The hypotheses of Corollary \ref{cor: bounds2} hold for the class $\Sigma'$ and for any $h$, $\underline h$, and $\overline h$ as in Theorem \ref{thm: info change2}.

	(ii) Under the relabeling
	\begin{align}
		\label{auction relabeling}
		x_a = r, \quad x_b = (x_1,\zeta), \quad \varepsilon_i = v_i, \text{ and } f_a = f^*,
	\end{align}
	the hypotheses of Proposition \ref{prop: control fn2} hold for any $h$, $\underline h$, and $\overline h$ as in Theorem \ref{thm: info change2} that depend on the state only through the common component: $h(y,w) = h^*(y,x_1,r,\zeta)$, $\overline h(w) = \overline h^*(x_1,r,\zeta)$, and $\underline h(w) = \underline h^*(x_1,r,\zeta)$ for some maps $h^*$, $\overline h^*$, and $\underline h^*$, with $\underline h$ and $\overline h$ also $\mu_W$-integrable.
\end{lemma}

\begin{proof}
(i) Assumption \ref{assump: information structure and policy2} holds by Lemma \ref{lemm: auction framework}, and Assumption \ref{assump: BCE existence2} in restricted terms is the existence property (b) of Assumption \ref{assump: auction2a}(iii). Assumption \ref{assump: invariance ESR} in restricted terms follows from Lemma \ref{lemm: auction invariance}(ii) through the restricted form of Corollary \ref{cor: cylinder slice}, and alternative (b) of Corollary \ref{cor: bounds2} is Lemma \ref{lemm: auction invariance}(ii) itself.

(ii) Under the relabeling (\ref{auction relabeling}), the split $W_C = X$ and $W_I = \varepsilon$ of Proposition \ref{prop: control fn2} is the split in the auction setting, the map written $f^*$ in that proposition is $\tilde f$ here, and the requirement of that proposition's setting that $h$ and its envelopes depend on the state only through $x = (x_a,x_b)$ is the stated restriction on $h$, $\overline h$, and $\underline h$. Assumption \ref{assump: control fn2}(i) requires $R$ and $V$ to be conditionally independent given $(X_1,\zeta)$, which follows from Assumption \ref{assump: auction2b}(i), and we take $\mu_{V \mid X_1,\zeta}$ as its fixed version. Assumption \ref{assump: control fn2}(ii) follows from Assumption \ref{assump: information structure and policy2}(iv), established in Lemma \ref{lemm: auction framework}: a state of $\mathbb{S}_W = \mathbb{S}_{W_C} \times \mathbb{S}_V$ remains in $\mathbb{S}_W$ when $r$ is replaced by any $r'$ with $(x_1,r',\zeta) \in \mathbb{S}_{W_C}$. The remaining hypotheses of the proposition, Assumptions \ref{assump: information structure and policy2} and \ref{assump: BCE existence2}, were verified in (i).
\end{proof}

\begin{proof}[Proof of Corollary \ref{corollary_auction2}]
Write $S = \{(X_1, f^*(R)) \in \mathbb{S}_{X_1,R}\}$ and let $1_S$ denote its indicator.

(i) Fix $i$. The signal $(V_i,X_1,f^*(R),\zeta)$ has distribution $\mu_{T_i}^f$, so it almost surely avoids the exceptional null set of Lemma \ref{lemm: auction invariance}(i), and on $S$ its $(x_1,r)$-coordinates lie in $\mathbb{S}_{X_1,R}$. The result (i) follows from the lemma.

(ii) The bounds come from Corollary \ref{cor: bounds2}, applied to the restricted class $\Sigma'$, and their three terms are then evaluated by Lemma \ref{lemm: auction invariance} and Proposition \ref{prop: control fn2}. The restricted selection rules $e'(\csdot;G)$ and $e'(\csdot;G_f)$ and the reduced forms $\rho_G'$ and $\rho_{G_f}'$ are those of Appendix \ref{App:Counterfactual Predictions in Bayesian Games}, formed with $\Sigma'$ in place of $\Sigma$. Take
\begin{align*}
	h(y,w) = 1\Big\{\max_{i : y_i \neq -1} y_i \in A\Big\} 1\{x_1 \in C\}, \quad \overline h(w) = 1\{x_1 \in C\}, \quad \underline h = 0.
\end{align*}
Bids lie in $[0,\overline v]$ and staying out is coded $-1$. This means that $\max_{i : y_i \neq -1} y_i = \max_i y_i$ at every $y$, including the no-sale case. Hence $h$ is Borel and $\underline h \le h \le \overline h$ everywhere, and, $A$ being a subset of $\mathbf{R}_+$, no-sale events contribute zero.

By Lemma \ref{lemm: auction applicability}(i), the hypotheses of Corollary \ref{cor: bounds2} hold for $\Sigma'$, and that corollary gives
\begin{align*}
	\mathsf{ADP}_R'(h) + \Delta(\underline h) \le \mathsf{AEP}'(h) \le \mathsf{ADP}_R'(h) + \Delta(\overline h).
\end{align*}

It remains to evaluate the three terms. By Lemma \ref{lemm: auction invariance}(iii), $\rho_{G_f}'(\csdot \mid w) = \sigma^{*f}(\csdot \mid w)$ at $\mu_W^f$-a.e.\ $w$ and $\rho_G'(\csdot \mid w) = \sigma^*(\csdot \mid w)$ at $\mu_W$-a.e.\ $w$. The first identity, with the definition of $p_f$ through $(a^{*f},b^{*f})$, gives
\begin{align*}
	\mathsf{AEP}'(h) = P\Big\{\max_{i \in \widetilde N^f} B_i^f \in A,\ X_1 \in C\Big\} = p_f(A \mid C)\, P\{X_1 \in C\}.
\end{align*}

The remaining two terms come from Proposition \ref{prop: control fn2}, whose hypotheses hold under the relabeling (\ref{auction relabeling}) by Lemma \ref{lemm: auction applicability}(ii). Since $\rho_G$ enters the proposition only through the definition (\ref{DP}) of $\mathsf{ADP}_R$ and the identification (\ref{cond prob}), its proof applies with $\rho_G'$ in place of $\rho_G$, the conditional expectations being read under $P_G' = \mu_W \otimes \rho_G' = \mu_W \otimes \sigma^*$. It gives $\Delta(\underline h) = 0$ and
\begin{align*}
	\Delta(\overline h) &= \mathbf{E}\big[1\{X_1 \in C\}1\{\tilde f(W_C) \notin \mathbb{S}_{W_C}\}\big],\\
	\mathsf{ADP}_R'(h) &= \mathbf{E}\big[m(X_1,f^*(R),\zeta)\, 1\{\tilde f(W_C) \in \mathbb{S}_{W_C}\}\, 1\{X_1 \in C\}\big],
\end{align*}
where $m(x_1,r,\zeta) = P\{\max_{i \in \widetilde N} B_i \in A \mid W_C = (x_1,r,\zeta)\}$. Indeed, under $P_G'$ the winning bid is $\max_{i \in \widetilde N} B_i$ and the indicator $1\{x_1 \in C\}$ is a function of $W_C$, so the conditional expectation in part (ii) of the proposition is $1\{x_1 \in C\}\, m(x_1,r,\zeta)$ at $\mu_{W_C}$-a.e.\ $(x_1,r,\zeta)$. By the null-set transfer (\ref{obs null transfer}) in the proof of that proposition, the value of $\mathsf{ADP}_R'(h)$ does not depend on the version chosen for $m$.

The corollary states its bounds in terms of the observables, and two reductions bring these terms to that form. First, note that $\mathbb{S}_{W_C} = \mathbb{S}_{X_1,\zeta} \times \mathbb{S}_R$ and $\mathbb{S}_{X_1,R} = \mathbb{S}_{X_1} \times \mathbb{S}_R$ by Lemma \ref{lemm: auction framework}, while $(X_1,\zeta) \in \mathbb{S}_{X_1,\zeta}$ almost surely, so both $\{\tilde f(W_C) \in \mathbb{S}_{W_C}\}$ and $S$ agree with $\{f^*(R) \in \mathbb{S}_R\}$ up to a null event. Hence $\Delta(\overline h) = P(S^c \cap \{X_1 \in C\})$, and the indicator in $\mathsf{ADP}_R'(h)$ may be replaced by $1_S$. Second, we integrate out $\zeta$. Let $\mu_{\zeta \mid X_1}(\csdot \mid x_1)$ be the $\zeta$-marginal of the fixed kernel $\mu_{V,\zeta \mid X_1}(\csdot \mid x_1)$, a version of the conditional distribution of $\zeta$ given $(X_1,R) = (x_1,r)$ that does not depend on $r$, by Assumption \ref{assump: auction2b}(i). As $1_S$ and $1\{X_1 \in C\}$ are functions of $(X_1,R)$, the law of iterated conditional expectations gives
\begin{align*}
	\mathsf{ADP}_R'(h) = \mathbf{E}\big[\tilde p(X_1,f^*(R))\, 1_S\, 1\{X_1 \in C\}\big], \text{ with } \tilde p(x_1,r) = \int m(x_1,r,z)\, d\mu_{\zeta \mid X_1}(z \mid x_1),
\end{align*}
and, again by the law of iterated conditional expectations, $\tilde p$ is a version of $p(A \mid \csdot)$. Replacing it by any other version leaves the expectation unchanged. Indeed, two versions differ only on a Borel $\mu_{X_1,R}$-null set $N$, the set $N \cap \mathbb{S}_{X_1,R} \cap \mathbb{S}_{X_1,f^*(R)}$ is null under the distribution of $(X_1,f^*(R))$ by Assumption \ref{assump: auction2b}(iv), and $(X_1,f^*(R)) \in \mathbb{S}_{X_1,f^*(R)}$ almost surely, so that $P(\{(X_1,f^*(R)) \in N\} \cap S) = 0$. Dividing the bounds by $P\{X_1 \in C\}$ gives the inequalities in (ii).
\end{proof}

\section{Implementation of Decomposition-Based Predictions\label{App:Implementation of Decomposition Methods}}
 
 In this section, we explain the implementation details for estimating the decomposition-based prediction $\mathsf{ADP}_R$ defined in (\ref{EP DP}) in the main text. Throughout this section, we assume that $X$ and $\varepsilon$ are independent. As in Section \ref{sec:single-index-restrictions}, we consider the policy of the following form: $f(W) = (f^*(X),\varepsilon)$ for some map $f^*$. If we define
 \begin{align}
 	\label{cond exp}
 	\mu(x) = \mathbf{E}[Y \mid X =x],
 \end{align}
 the prediction $\mathsf{ADP}_R$ is identified as follows:
 \begin{align}
 	\label{ADP_R}
 	\mathsf{ADP}_R = \mathbf{E}\left[\mu(f^*(X))1\{f^*(X) \in \mathbb{S}_X\}\right],
 \end{align}
where we recall that $\mathbb{S}_X$ denotes the support of $X$ (in the data from the pre-policy game). Our explanation is divided into two settings: the setting with a fully nonparametric method and the setting that makes use of a multi-index structure.
 
We introduce a sample unit notation $m$ (mnemonic for `market'), so we assume we observe i.i.d.\ draws $(Y_m,X_m)$, $m=1,...,M$, from the joint distribution of $(Y,X)$ in an equilibrium of the Bayesian game. Hence the sample size is $M$, which we let go to infinity in the asymptotic inference, while the number of players $n$ stays fixed. For each market $m=1,...,M$, we have $Y_m = (Y_{1,m},...,Y_{n,m})$, the profile of actions in the $m$-th market, where $Y_{i,m}$ represents the observed action of player $i$ in market $m$. Similarly we have $X_m = (X_{1,m},...,X_{n,m})$, where $X_{i,m}$ denotes the $d$-dimensional vector of observed payoff characteristics for player $i$ in market $m$. 
  
\subsection{Implementation Using a Nonparametric Decomposition Method}

\subsubsection{Implementation}

 Decomposition-based predictions can be generated using various nonparametric estimation methods. For illustration, we explain a way to construct such predictions using a kernel estimation method. Suppose that we observe $M$ markets which we view as i.i.d.\ draws from an $n$-player representative game. For each player $i$, let $X_{i,m} = [X_{i,m,1}',X_{i,m,2}']'$, where $X_{i,m,1}$ takes values from a finite subset of $\mathbf{R}^{q_1}$ and $X_{i,m,2} \in \mathbf{R}^{q_2}$ is a vector of continuous variables in market $m=1,...,M$. We also assume that $Y_i \ge 0$ for all $i = 1,...,n$. In this environment, we focus on estimating $\mathsf{ADP}_R$ defined in (\ref{ADP_R}). We first construct a leave-one-out kernel estimator of the $i$-th component $\mu_i(x) = \mathbf{E}[Y_i \mid X =x]$ of $\mu(x) = \mathbf{E}[Y \mid X =x]$ for each $m = 1,...,M$ as
 \begin{align}
 	\label{hat mu}
    \hat \mu_{i,m,b}(x) = \frac{\displaystyle \sum_{k=1, k \ne m}^M Y_{i,k} 1\{X_{k,1} = x_1\}K_b(X_{k,2} - x_2)}{\displaystyle \sum_{k = 1, k \ne m}^M 1\{X_{k,1} = x_1\}K_b(X_{k,2} - x_2)}, \quad  x = (x_1,x_2),
 \end{align}
 where $Y_{i,k}$ is the $i$-th component of $Y_k$, $X_{k,1} = [X_{1,k,1}',...,X_{n,k,1}']'$, $X_{k,2} = [X_{1,k,2}',...,X_{n,k,2}']'$ and $K_b(x_2) = K(x_{2,1}/b,...,x_{2,nq_2}/b)/b^{nq_2}$, $x_2 = (x_{2,1},...,x_{2,nq_2})'$, and $K$ is a multivariate kernel and $b$ is a bandwidth. For an estimator of $\mathsf{ADP}_R$, we consider the following:
 \begin{align}
    \widehat{\mathsf{ADP}}_R = \frac{1}{M}\sum_{m=1}^M \hat \mu_{m,b}\left(f^*(X_m)\right)1\{f^*(X_m) \in \mathbb{S}_X\},
 \end{align}
where $\hat \mu_{m,b}(x) = [\hat \mu_{1,m,b}(x),...,\hat \mu_{n,m,b}(x)]'$. The estimated interval of the average policy effect on player $i$'s action is given by
\begin{align*}
	\left[ \widehat{\mathsf{ADP}}_{R,i} - \overline Y_i, \quad \widehat{\mathsf{ADP}}_{R,i} - \overline Y_i + \sup \mathbb{Y}_i \cdot 
	\hat \Delta \right],
\end{align*}
where $\widehat{\mathsf{ADP}}_{R,i}$ is the $i$-th component of $\widehat{\mathsf{ADP}}_R$, 
\begin{align*}
	\hat \Delta = \frac{1}{M}\sum_{m=1}^M 1\{f^*(X_m) \notin \mathbb{S}_X \}, \text{ and } \overline Y_i = \frac{1}{M}\sum_{m=1}^M Y_{i,m}.
\end{align*}
For bandwidths, we use cross-validation for each player $i=1,...,n$. Define
\begin{align}
	CV_i(b) = \frac{1}{M}\sum_{m=1}^M (Y_{i,m}- \hat \mu_{i,m,b}(X_m))^2,
\end{align}
where $\hat \mu_{i,m,b}(x)$ denotes the $i$-th component of $\hat \mu_{m,b}(x)$. We choose a minimizer of $CV_i(b)$ over a range of $b$'s as our bandwidth.

When the support of $X_m$ contains that of $f^*(X_m)$, we can simply estimate the average treatment effect as
\begin{align}
	\label{hat Delta i}
  \widehat{\mathsf{ATE}}_i \eqdef \widehat{\mathsf{ADP}}_{R,i} - \overline Y_i.
\end{align}

For the bootstrap standard errors for (\ref{hat Delta i}), we first draw $(Y_{m}^*,X_{m}^*)_{m=1}^M$, i.i.d.\ from the empirical distribution of $(Y_{m},X_{m})_{m=1}^M$. Using the bootstrap sample, we obtain $\hat \mu_{m,b}^*(x)$ (where $b$ is the bandwidth chosen from the cross-validation). Define
 \begin{eqnarray}
    \widehat{\mathsf{ATE}}_{i}^* = \frac{1}{M}\sum_{m=1}^M \hat \mu_{i,m,b}^*(f^*(X_{m}^*)) - \overline Y_{i}^*,
 \end{eqnarray}
 where $\overline Y_{i}^*$ is the sample mean of $\{Y_{i,m}^*\}_{m=1}^M$, and $Y_{i,m}^*$ denotes the $i$-th component of $Y_{m}^*$. In order to construct a bootstrap standard error, we can adopt the bootstrap interquartile range method used in \cite{Chernozhukov/Fernandez/Melly:13:Eca}. First, let $q_{0.75}$ and $q_{0.25}$ be the $0.75$ and $0.25$ quantiles of the bootstrap distribution of $\sqrt{M}(\widehat{\mathsf{ATE}}_{i}^* - \widehat{\mathsf{ATE}}_i)$, and let $z_{0.75}$ and $z_{0.25}$ be the $0.75$ and $0.25$ quantiles of the standard normal distribution. Then, the bootstrap standard error is taken to be $M^{-1/2}(q_{0.75} - q_{0.25})/(z_{0.75} - z_{0.25})$.

\subsubsection{(Near) Multicollinearity for Nonparametric Counterfactual Predictions}
\label{subsubsec: near multicollinearity}

For our empirical application, we reduced the dimension of $X_m$ due to near multicollinearity in the counterfactual environment. We now expand on this point, showing why multicollinearity in counterfactuals may be a concern even if it is absent in-sample. For simplicity of exposition, we omit the subscript $m$.

We split $X=(X_1, X_2)$ and write:
\begin{align*}
	 \mathbf{E}[Y] = \int \mu(x_1, x_2) d F_{X_1,X_2}(x_1,x_2).  
\end{align*}
This quantity is certainly identified even if $X_1$ and $X_2$ are perfectly correlated. This is because, if that were the case, we would be integrating over the values of $X_1$ and $X_2$ such that $X_1 = X_2$. 

However, this is different in a counterfactual setting. If the policy moves $(x_1,x_2)$ to combinations that do not occur under the pre-policy distribution, then $\mu$ carries no information at those combinations, and the decomposition method may fail. We have excluded these settings by assuming that the policy is such that the pre-policy and post-policy distributions of the payoff state are mutually absolutely continuous on the intersection of their supports. Below, we outline our choices for our empirical applications.

\subsection{Implementation Using a Multi-Index Structure}

We explain how to obtain an estimator of $\mathsf{ADP}_R(h)$ using a multi-index structure in Section \ref{sec:single-index-restrictions}. Here we assume a setting that leads to the identification result in (\ref{index}), and make explicit the dependence of random variables on the market index $m$. More specifically, we write $Y_{i,m}$, $X_{i,m,a}$, $X_{i,m,b}$ and $Z_{i,m}$ in place of $Y_{i}$, $X_{i,a}$, $X_{i,b}$, and $Z_i$. We also write $Y_m = (Y_{1,m},...,Y_{n,m})$, $X_{m,a} = (X_{1,m,a},...,X_{n,m,a})$, $X_{m,b} = (X_{1,m,b},...,X_{n,m,b})$ and $Z_{m} = (Z_{1,m},...,Z_{n,m})$, in place of $Y = (Y_1,...,Y_n)$, $X_a = (X_{1,a},...,X_{n,a})$, $X_b = (X_{1,b},...,X_{n,b})$ and $Z = (Z_1,...,Z_n)$. 

For each player $i=1,...,n$, and market $m=1,...,M$, define 
\begin{align*}
	m_i(h;z,x_a) = \mathbf{E}\left[ h(Y_{i,m},X_{i,m,a}) \mid X_{m,a} = x_a, Z_m = z \right],
\end{align*}
for some map $h$. Then, we can write
\begin{align*}
	\mathsf{ADP}_{R,i}(h) = \mathbf{E}\left[ m_i(h;Z_m, g(X_{m,a})) 1\left\{ (g(X_{m,a}),Z_m)\in \mathbb{S}_{X_a,Z}\right\} \right].
\end{align*}

We now discuss estimation of $m_i$ and $\mathsf{ADP}_{R,i}(h)$. We begin by assuming that $X_{m,a}$ is a discrete random vector. We first estimate coefficients $\theta_i$ in 
\begin{align*}
    Z_{i,m} = X_{i,m,b}'\theta_i,
\end{align*}
using methods in the literature on multi-index models. Here we assume that $Z_m$ is a continuous random vector, and adopt the proposal by \cite{Ahn/Ichimura/Powell/Ruud:18:JBES} which is based on the pairwise differencing approach of  \cite{Ahn/Powell:93:JOE}. First, we define
\begin{align*}
	\gamma_{i,m} = m_i\left(h;X_{1,m,b}' \theta_1,...,X_{n,m,b}' \theta_n, X_{m,a} \right), \quad i=1,...,n.
\end{align*}
For each fixed value of $X_{m,a}$, the vector $\gamma_m = (\gamma_{1,m},...,\gamma_{n,m})$ depends on $X_{m,b}$ only through the indices $Z_m$, which is the setting of \cite{Ahn/Ichimura/Powell/Ruud:18:JBES}. Their approach invokes the invertibility of this system of equations in the indices. Accordingly, we assume that for each value $x_a$ of $X_{m,a}$, there exist functions $\varphi_i(\csdot\,; x_a)$, $i=1,...,n$, such that
\begin{align*}
	X_{i,m,b}'\theta_i = \varphi_i(\gamma_{1,m},...,\gamma_{n,m}; X_{m,a}), \quad i=1,...,n.
\end{align*}
Furthermore, we assume that $X_{i,m,b} = [ X_{i,m,b,A}, X_{i,m,b,B}']'$, and the coefficient of the first random variable $X_{i,m,b,A}$ is not zero, and normalized to be one. By reparametrizing $\theta_i = [1, - \beta_i']'$, we can rewrite the previous equation as
\begin{align*}
	X_{i,m,b,A} = X_{i,m,b,B}' \beta_i + \varphi_i(\gamma_m ; X_{m,a}), 
\end{align*}
where $\gamma_m = (\gamma_{1,m},...,\gamma_{n,m})$. Hence when $\gamma_m \approx \gamma_k$ and $X_{m,a} = X_{k,a}$, we have
\begin{align*}
	X_{i,m,b,A} - X_{i,k,b,A} \approx (X_{i,m,b,B} - X_{i,k,b,B})' \beta_i. 
\end{align*}
This observation suggests the following estimator of $\beta_i$:\footnote{We could use an optimal weighting matrix here to improve efficiency as explained in \cite{Ahn/Ichimura/Powell/Ruud:18:JBES}. Here we choose a simple form for the estimator for the sake of expositional simplicity.}  
\begin{align*}
	\hat \beta_i &= \left( \sum_{m=1}^M \sum_{k=1, k \ne m}^M (X_{i,m,b,B} - X_{i,k,b,B})(X_{i,m,b,B} - X_{i,k,b,B})' \hat \Gamma_{m,k} \right)^{-1}\\
	  & \quad \quad \times \left( \sum_{m=1}^M \sum_{k=1, k \ne m}^M (X_{i,m,b,B} - X_{i,k,b,B})(X_{i,m,b,A} - X_{i,k,b,A}) \hat \Gamma_{m,k} \right),
\end{align*}
where $\hat \gamma_m = [\hat \gamma_{1,m},...,\hat \gamma_{n,m}]'$,
\begin{align*}
	\hat \Gamma_{m,k} &= K\left( \frac{\hat \gamma_m - \hat \gamma_k}{b}\right) 1\{X_{m,a} = X_{k,a}\}, \text{ and }\\
	\hat \gamma_{i,m} &= \frac{\displaystyle \sum_{k=1, k \ne m}^M 1\{X_{k,a} = X_{m,a}\} L\left(\frac{X_{m,b,B} - X_{k,b,B}}{b}, \frac{X_{m,b,A} - X_{k,b,A}}{b} \right) h(Y_{i,k},X_{i,k,a})}{\displaystyle \sum_{k=1, k \ne m}^M 1\{X_{k,a} = X_{m,a}\} L\left(\frac{X_{m,b,B} - X_{k,b,B}}{b}, \frac{X_{m,b,A} - X_{k,b,A}}{b} \right)},
\end{align*}
and $K$ is an $n$-variate kernel and $b$ is a bandwidth, $X_{m,b,A} = [X_{1,m,b,A},...,X_{n,m,b,A}]'$ is the $n$-dimensional vector, $X_{m,b,B} = [X_{1,m,b,B}',...,X_{n,m,b,B}']'$ is the $nd_2$-dimensional vector, with $d_2$ denoting the dimension of $X_{i,m,b,B}$, and $L$ is an $n(1+d_2)$-variate kernel.\footnote{In constructing $\hat \gamma_{i,m}$ using a kernel regression estimator, we are assuming $X_{m,b}$ is a continuous random vector. When $X_{m,b}$ contains a discrete subvector, we can obtain $\hat \gamma_{i,m}$ as in (\ref{hat mu}).} Then, for each $i=1,...,n$, we set $\hat \theta_i = [1, - \hat \beta_i']'$ and construct 
\begin{align*}
	\hat m_{i,m}(h;z,x_a) =  \frac{\displaystyle  \sum_{k=1,k \ne m}^M 1\{X_{k,a} = x_a\} H\left(\frac{\hat Z_k - z}{b}\right) h(Y_{i,k},X_{i,k,a})}{\displaystyle \sum_{k=1, k \ne m}^M 1\{X_{k,a} = x_a\} H\left(\frac{\hat Z_k -z}{b}\right)},
\end{align*}
where $H$ is an $n$-variate kernel, and $\hat Z_k = [X_{1,k,b}'\hat \theta_1,...,X_{n,k,b}'\hat \theta_n]'$. Using $\hat \theta_i$ and $\hat m_i$, we can obtain $\widehat{\mathsf{ADP}}_{R,i}(h)$ as
\begin{align}
	\label{hat ADP} 
	\widehat{\mathsf{ADP}}_{R,i}(h) = \frac{1}{M}\sum_{m =1}^M \hat m_{i,m}(h; \hat Z_m,g(X_{m,a})) 1 \left\{ (g(X_{m,a}), \hat Z_m) \in \mathbb{S}_{X_a,Z}\right\}.
\end{align} 

When the support of $X_{m,a}$ contains that of $g(X_{m,a})$, we can estimate the average policy effect as in (\ref{hat Delta i}). The bootstrap standard error for this effect can be constructed as before, constructing $\widehat{\mathsf{ADP}}_{R,i}(h)$ using the bootstrap sample.

 \subsection{Details on the Empirical Applications}
 
In this empirical application, each game $m$ corresponds to a market $m$ defined by a route (airport `A' to airport `B'). Regarding the additional data collection for the out-of-sample comparison, we used the same data sources to collect 2015 data as the authors did for their original (pre-2009) dataset. In particular, we collected all 2015 quarterly data from the DB1B Market and Ticket Origin and Destination Dataset of the U.S. Department of Transportation and then aggregated them to the yearly level. We also followed the same data cleaning choices: most notably, we set entry equal to 0 for markets with fewer than 20 passengers, and dropped any markets with 6 or more market coupons. See Section S.1 in the Supplemental Note of \cite{Ciliberto/Tamer:09:Eca}. Then, we merged the 2015 data to the same 2,742 markets considered in their work. These data are used only for the out-of-sample comparison, as our policy estimates use the same pre-repeal data as the authors.

The choice of covariates in our application follows \cite{Ciliberto/Tamer:09:Eca}, pp.1808-1811. Recall that, for firm $i$, their covariates $X_{i,m}$ are given by (1) $i$'s market presence in $m$ (the share of markets served by airline $i$ among all of those starting from $m$'s endpoints), (2) $m$'s ``cost'' (captured by two distances: the origin-to-destination distance when routed through firm $i$'s closest hub, and the nonstop origin-to-destination distance), (3) whether market $m$ is affected by the Wright Amendment, (4) Dallas market (whether this route includes an airport in Dallas), (5) market size (the geometric mean of population size at the route's endpoints), (6) average per capita income (average income of the cities at market endpoints), (7) income growth rate (similarly defined), (8) market distance (nonstop distance between the airports defining this market), (9) distance to the center of the U.S. (sum of the distances from each endpoint of the route to the geographical center of the U.S.), and (10) close airport (minimum of the distances from each airport to the closest alternative a passenger could use). Covariates (1)-(2) vary at the firm-market level, while covariates (3)-(10) vary only at the market level. 

Our interest is in changing the distribution of the Wright Amendment variable, keeping the others constant. The role of the remaining variables is to control for a variety of confounders (costs, returns from serving those markets, outside options to consumers, geographical characteristics). 

Since the conditional expectation in (\ref{cond exp}) must include all $X_{j,m}$ for every $j$, we are left with possibly 8 (market specific) plus $2 \times 4$ (firm level) covariates to be included in the analysis. This is a large number for nonparametric estimation, although our linear results are robust to using the full set. As described in the main text, we perform a dimension reduction to facilitate nonparametric estimation. First, 3 of those 16 variables (the binary Dallas market, market size, and income growth rate, numbered (4), (5), (7) above) were highly collinear or did not provide enough variation for nonparametric estimation (in the sense explained in Appendix \ref{subsubsec: near multicollinearity} above). We dropped such variables from our analysis. To see this: every market affected by the Wright Amendment was already in Dallas. There are only 110 observations (out of 2,742) with different values for the Dallas market variable and the Wright Amendment one. However, these 110 observations refer to markets using Dallas Fort Worth Airport, the hub for American Airlines.  As a result, this variation is already captured by measures of market income and market presence. Regarding market size and income growth rate, they appear to be nonlinearly predicted from market income and market presence.\footnote{Including one of these three variables generated numerical instability in nonparametric estimation. See Section \ref{subsubsec: near multicollinearity} for further discussion.} Intuitively, the latter variables already (nonlinearly) reflect economic prosperity of those markets. Finally, we make use of a causal structure argument: we do not include the proxy for a firm's cost (covariate (2) above) because it is a function of market distance and market presence by construction. As a result, we use the remaining 9 covariates in the specifications.

We now describe the implementation of our estimators. Denote the set of markets affected by the Wright Amendment in the original dataset by $\mathcal{M}$, which has cardinality $|\mathcal{M}| = 81$. Our counterfactual covariates are denoted by $f^*(X_{m})$. They are equal to $X_{m}$ for all firms and markets, except for the Wright Amendment variable, which is set to 0 for all markets $m \in \mathcal{M}$ and for all firms.

For the linear estimator, we simply parametrize $\mathbf{E}[Y_{i,m}| X_{m} = x] = x'\gamma_i$ and estimate this model on the full dataset using Ordinary Least Squares. We then use the recovered estimate $\hat{\gamma}_i$ with the counterfactual covariates, $f^*(X_{m})$. Our linear estimator of $\mathsf{ATE}_i$ for firm $i$ is then given by:
\begin{eqnarray}
\widehat{\mathsf{ATE}}_i = \frac{1}{|\mathcal{M}|}\sum_{m \in \mathcal{M}} f^*(X_{m})'\hat \gamma_i - \overline Y_i, \label{parametric_cf_est}
\end{eqnarray}
where $\overline{Y}_i$ is the average of $Y_{i,m}$'s over the markets in $\mathcal{M}$.

For the nonparametric case, we follow the previous section in estimating $\mathbf{E}[Y_{i,m} \mid X_m = x] = g(x)$ and use a quartic kernel for continuous covariates. Standard errors for both estimators are computed by the bootstrap approach outlined in the previous section.

\subsection{A Second Empirical Application: The Threat of Southwest Airlines on Competitors' Entry Behavior}
\label{subsec:Southwest}

We can pursue further counterfactual exercises beyond those in \cite{Ciliberto/Tamer:09:Eca}. One salient example is to study competition effects in the airline industry. A prominent case is the rise of Southwest Airlines and its impact on competitors' behavior, which has drawn attention because of Southwest's emergence as a low-cost entrant with novel organizational strategies (see, for example, \cite{Knorr/Arndt}). \cite{Goolsbee/Syverson:08:QJE} studied one angle to this question, focusing on whether Southwest Airlines' threat of entry affected established airlines' (e.g., American, Delta, United) pricing behavior. In this section, we use the decomposition-based prediction within the same entry game above to extend their analysis beyond pricing. 

We use the same dataset and the same model specification from the last section. This includes the entry game environment with complete information. It also includes the definition of market entry/operation (i.e., a route between two airports, irrespective of direction or layovers). Our only change is regarding the policy of interest: we replace `Wright Amendment' by a binary variable equaling 1 if Southwest operates in both endpoints of a market $m$, and 0 otherwise. This means that the variable is 1 if Southwest has entered markets that include airports in both endpoints of $m$. This is the same variable as in \cite{Goolsbee/Syverson:08:QJE}, which they interpret as a ``discontinuous'' threat of Southwest entry. However, despite its name as a ``threat of entry'', it is actually interpretable as a cost shock.\footnote{Entry decisions in other markets are independent of those in market $m$ due to the maintained assumption of independence of $\varepsilon_{i,m}$. Hence, it can be used as a policy variable in this exercise. However, we cannot condition on actual entry in a certain market, whether by Southwest or another airline, since that is a simultaneous decision in this environment.} By operating in markets that share both endpoints with market $m$, Southwest has a lower cost of operating market $m$ (since it has established facilities, labor, etc. on both endpoints of a route). However, a smaller cost for Southwest to operate in market $m$ does not restrict Southwest, or any other airline, from entering market $m$ or any other market.

Our outcomes of interest are whether the threat of Southwest entry changes established airlines' actual entry behavior (as opposed to pricing, as in \cite{Goolsbee/Syverson:08:QJE}) and whether such effects vary across smaller or larger markets (as found in \cite{Ellison/Ellison:11:AEJ} and \cite{Tenn/Wendling:14:ReStat} in the pharmaceutical industry).\footnote{Since our framework is static, the interpretation of our results also differs from those cited above. In a static environment, there is no dynamic choice of capacity and there are no incumbents, so the latter cannot ``accommodate'' entry over time. Nevertheless, we think the present exercise is still informative about whether a state in which Southwest is likely to enter induces differential (strategic) behavior by established airlines.} As in the first empirical application, $\Delta = 0$ in this setting: all covariates $X$ beyond Southwest's ``threat of entry'' remain the same post-policy, retaining the same support. Meanwhile, the support of the policy variable is $\{0,1\}$ pre-policy, and $\{0\}$ post-policy: a strict subset. Hence, the probability that the policy takes $X_i$ beyond its pre-policy support (across firms and markets) is zero.

\subsubsection{Results}

Our counterfactual policy sets the Southwest Threat of Entry variable to 0 for all markets. Therefore, we identify whether American, Delta and United are more/less likely to enter markets when Southwest is no longer a threat to entry (i.e., when it no longer has a lower cost of operating in such markets). The arguments for the validity of our decomposition-based prediction are analogous to those in the previous exercise: (i) the policy is observable to players, since Southwest's routes are observable, and (ii) the policy is within the support of the data, as there are many markets where Southwest does not operate in both endpoints. Hence, under the invariance condition on the equilibrium selection rules, Corollary \ref{example1_corollary} can be applied. Estimation and inference on these effects follow those in the previous section, detailed in Appendix \ref{App:Implementation of Decomposition Methods}. The results are presented below. 

\begin{table}[t]
	\begin{centering}
		\small
		\caption{\small \cite{Goolsbee/Syverson:08:QJE} Revisited: Effects of Removing the Threat of Southwest Airlines Entry on Other Airlines' Probability of Entry}
		\label{southwest_decomp}
		\resizebox{\columnwidth}{!}{%
			\begin{tabular}{ccccccccc}
				\hline 
				\hline 
				\tabularnewline
				& &\multicolumn{6}{c}{Outcome: \footnotesize Change in the Entry Prob. after Removing Southwest Threat of Entry} \\
				\\
				\cline{2-8}
				\tabularnewline
				&   & \multicolumn{2}{c}{All Markets} & \multicolumn{2}{c}{Small Markets} & \multicolumn{2}{c}{Large Markets} \\
				&   &  Linear & Nonparametric &  Linear & Nonparametric&  Linear & Nonparametric \\
				\tabularnewline
				\hline
				\multicolumn{1}{c}{} & &  &  &  & & & \\
				American Airlines & & -0.080 & -0.074 &  -0.153 & -0.189 & -0.006 & 0.052 \\
				& & (0.023) & (0.021) & (0.034) &  (0.031) & (0.033) &  (0.027)\\
				\tabularnewline   
				Delta Airlines &  & -0.079 & -0.078  &  -0.109 & -0.134 & -0.037 & -0.015  \\
				&		& (0.024)& (0.021) & (0.037) & (0.028) & (0.034) & (0.028) \\
				\tabularnewline    			
				United Airlines &  & -0.056 & -0.047  & -0.073  &  -0.079 & -0.056 & 0.006 \\
				& &(0.023) &(0.020) & (0.029) & (0.025) & (0.032) & (0.026) \\
				\\
				\hline 
				\multicolumn{1}{c}{} & &  &  &  &\\
			\end{tabular}
		}
		\par\end{centering}
	\parbox{\textwidth}{\footnotesize
		\textit{Notes}: We report the estimated counterfactual changes to the entry of major airlines (American, Delta, United) after the removal of the threat of Southwest Airlines entry. The effect is estimated on the markets originally subject to that threat, as defined in \cite{Goolsbee/Syverson:08:QJE}. The first two columns show the effects for all such markets, for both linear and nonparametric estimates. Columns 3-4 show the results for markets affected by such a threat, but below the median market size, while the last two show the effects for markets larger than the median. Standard errors for these columns are computed by the bootstrap, following the approach in the Online Appendix with $B=999$ replications. A negative coefficient represents a decrease in entry if there were no threat of Southwest entry, relative to there being such a threat.}
\end{table}

As we can see from Table \ref{southwest_decomp}, a threat of Southwest entry \textit{increases} the average entry probability of American Airlines (7.4-8\%), Delta Airlines (7.8-7.9\%) and United Airlines (4.7-5.6\%) in such markets (i.e., removing a threat of Southwest entry, thereby increasing Southwest's cost, decreases competitors' average entry probabilities). The results suggest that in markets where a Southwest entry is likely, the established firms (with larger market presence) will be more likely to enter. 

To further understand our results, we follow \cite{Ellison/Ellison:11:AEJ} and \cite{Tenn/Wendling:14:ReStat} and check whether such effects depend on market size by re-estimating the model for markets below and above the median market size. Consistent with those papers, we find nonmonotonic effects of the Southwest threat of entry on its competitors' decisions. In particular, we find that the increased entry due to Southwest's threat is driven by small markets. In small markets (i.e., below the median market size), where profits are more limited, the threat of Southwest entry induces other firms to enter as well. This is consistent with the change in the state variable (Southwest entry) changing the equilibrium (airlines coordinating on entry as an equilibrium ``deterrence'' to Southwest beyond pricing). In contrast, the threat of Southwest does not induce entry in larger markets. This is consistent with (a static interpretation of) ``accommodation'' in large markets: when profit is large enough, all firms enter for that state even if others are likely to enter, as suggested in \cite{Tenn/Wendling:14:ReStat}.\footnote{Table \ref{southwest_decomp2} in the Online Appendix shows the results for the second term in the average aggregate decomposition (i.e., the role of market characteristics, keeping the same threat of Southwest Airlines entry).}

\subsubsection{Heterogeneous Effects Depending on the Number of Airlines Threatening Entry}

We can extend the previous exercise beyond Southwest Airlines. For instance, researchers may be interested in whether the competition effects differ across the number of potential entrants (i.e., number of airlines with lower costs). This can also be answered within our framework.

Indeed, we can redefine our policy variable as the number of airline $i$'s competitors that threaten entry in market $m$. We keep the same definition of threat of entry as in the previous section: a competitor operates flights out of each endpoint of a route, thereby lowering its costs, but not the route itself. Since our emphasis is on the same four airlines (American, Delta, Southwest, United), each firm $i$ may face $\{0,1,2,3\}$ competitors threatening entry in each market. We conduct three separate counterfactual exercises to study how a decrease in the number of airlines threatening entry affects $i$'s choice to enter. Each exercise decreases the number of potential entrants by one (i.e., making markets with three entrants have only two entrants, etc.). The results are shown in Table \ref{ct_number} in the Online Appendix.\footnote{The different exercises are not generally comparable, because the effects are calculated over different markets (i.e., those markets with two competitors threatening entry for American are likely to be very different from those with only one threat).}

As we can see, airlines would generally increase entry if they had no potential competitors rather than one. After all, the airline is more likely to benefit from the market's profits and less likely to compete in such a market. However, decreasing the number of $i$'s competitors from three to two decreases average entry in such markets, consistent with the previous section's results.

\subsection{Additional Empirical Results}

\subsubsection{The ``Other Term'' in Aggregate Decomposition: The Role of Market (Observable) Characteristics}\label{second_term}

Table \ref{ct_decomp} presents the results for the (average) effect of the repeal of the Wright Amendment on the set of markets affected by the policy. However, this is only one of two components in the typical aggregate decomposition (e.g., Oaxaca-Blinder; see \cite{Fortin/Lemieux/Firpo:11:Handbook} for an overview). The other term in such exercises is the ``observable effect'': i.e., how the characteristics of markets affected and not affected by the repeal differ.

To see this, we write the payoff state as $X_m = (X_{m}^{\text{Wright}}, \tilde X_m)$, which makes explicit which variables are related to the policy ($X_m^{\text{Wright}}$) and which other states are firm-market specific ($\tilde X_m$). Let $\tilde X_m^1$ and $\tilde X_m^0$ denote the (observable) payoff states for firm-markets that have $X_{i,m}^{\text{Wright}} = 1$ or $0$, respectively. Then, the difference in the (average) probability of entry between firms not subject to the Wright Amendment and those subject to the policy (i.e., the effect of a repeal of the amendment) can be written as\footnote{This has a flipped sign relative to standard aggregate decompositions because we study the repeal of the Wright Amendment, rather than imposing such a policy. Hence, the comparison is between markets not affected and those affected, rather than the other way round.}:
\begin{align*}
& \mathbf{E}[Y_{i,m} | \tilde X_m^0, X_{i,m}^{\text{Wright}} = 0] - \mathbf{E}[Y_{i,m} | \tilde X_m^1, X_{i,m}^{\text{Wright}} = 1]  \\
& = \underbrace{\mathbf{E}[Y_{i,m} | \tilde X_{m}^0, X_{i,m}^{\text{Wright}} = 0] - \mathbf{E}[Y_{i,m} | \tilde X_{m}^1, X_{i,m}^{\text{Wright}} = 0]}_{\text{Observable Effect}} + \underbrace{\mathbf{E}[Y_{i,m} | \tilde X_{m}^1, X_{i,m}^{\text{Wright}} = 0] - \mathbf{E}[Y_{i,m} | \tilde X_{m}^1, X_{i,m}^{\text{Wright}} = 1]}_{\text{Policy Effect}}. 
\end{align*}

The second term on the right-hand side is the effect of the policy (repeal of the amendment), keeping fixed the distribution of observable characteristics for firms/markets subject to the Wright Amendment. Its estimates are shown in Table \ref{ct_decomp}. However, the other term shows the average difference in entry across markets that are not affected/affected by the amendment, which is due solely to differences in observable characteristics (as the policy variable does not change). 

It is straightforward to estimate the new term. It is composed of two separate terms. The first is the mean entry for firms in markets that are not affected by the Wright Amendment. It can be estimated using the empirical mean for such markets. The second is the counterfactual entry for markets subject to the Wright Amendment, if the policy were to be repealed. This has already been estimated for Table \ref{ct_decomp}, as it is a component of the policy effect. Hence, differences between the ``linear'' and ``nonparametric'' estimates of the observable effect are due solely to differences in this second term (the counterfactual entry for affected markets), which are already present in Table \ref{ct_decomp}. The results for this second term of an aggregate decomposition for the Wright Amendment application are shown in Table \ref{ct_decomp2}, while the results for the application of Southwest Airlines' Threat of Entry are shown in Table \ref{southwest_decomp2} below.

\begin{table}[t]
	\begin{centering}
		\small
		\caption{\small \cite{Ciliberto/Tamer:09:Eca} Revisited: The Effects from Differences in Market Characteristics}
		\label{ct_decomp2}
		\resizebox{0.7\columnwidth}{!}{%
			\begin{tabular}{cccc}
				\hline 
				\hline 
				\tabularnewline
				& &\multicolumn{2}{c}{Outcome: Change in Probability of Entry in Dallas Love Markets} \\
				\\
				\cline{2-4}
				\tabularnewline
				&   &  Linear Model & Nonparametric Model \\
				\tabularnewline
				\hline
				\multicolumn{1}{c}{} & &  &  \\
  			American Airlines & & 0.431 & 0.272  \\
				& & (0.037) & (0.027)\\
				\tabularnewline   
				Delta Airlines &  & 0.565 & 0.369   \\
				&		& (0.039)& (0.022) \\
				\tabularnewline    
				Southwest Airlines &  & -0.406 & -0.350   \\
				& & (0.029)&(0.051) \\
				\tabularnewline   
				United Airlines &  & 0.292 & 0.241  \\
				& &(0.033) &(0.009) \\
				\\
				\hline 
				\multicolumn{1}{c}{} & &  &\\
			\end{tabular}
		}
		\par\end{centering}
	\parbox{\textwidth}{\footnotesize
		\textit{Notes}: We report the estimated counterfactual changes to the entry of major airlines into Dallas Love Field markets following the repeal of the Wright Amendment, only due to observable (non-policy) characteristics. The specification follows Table \ref{ct_decomp}, while estimation/inference is described above. Column 1 uses a linear model, while Column 2 reports a nonparametric estimate. Standard errors for these columns are computed by the bootstrap, following the approach in the Online Appendix with $B=999$ replications.}
\end{table}

The effects in Table \ref{ct_decomp2} are much larger for American, Delta and United relative to Table \ref{ct_decomp}. For Southwest, the effect is 20\% lower than those in Table \ref{ct_decomp}. This can be explained by the characteristics of markets subject to the Wright Amendment. Such markets, which have Dallas Love Field as an endpoint, are strictly in the South, are often smaller than markets elsewhere, and have a lower market presence of the major airlines. Hence, one would expect airlines to be less likely to enter such markets, even if the Wright Amendment were repealed (or had never been present). Hence, a naive comparison of entry across different markets would overstate the effects of the policy, with effects of over 50\% for most airlines (found by aggregating the estimates in Tables \ref{ct_decomp} and \ref{ct_decomp2}). However, we estimate that the repeal itself would generate entry into affected markets as seen in Table \ref{ct_decomp}.\bigskip 

\begin{table}[t]
	\begin{centering}
		\small
		\caption{\small \cite{Goolsbee/Syverson:08:QJE} Revisited: The Effects from Differences in Market Characteristics}
		\label{southwest_decomp2}
		\resizebox{\columnwidth}{!}{%
			\begin{tabular}{cccccc}
				\hline 
				\hline 
				\tabularnewline
				& &\multicolumn{2}{c}{Outcome: Change in Probability of Entry After Removal of Southwest Threat} \\
				\\
				\cline{2-4}
				\tabularnewline
				&   & \multicolumn{2}{c}{Decomposition Method}  \\
				&   &  Linear Model & Nonparametric Model \\
				\tabularnewline
				\hline
				\multicolumn{1}{c}{} & &  &  \\
  			American Airlines &  & -0.043 & -0.048  \\
				& & (0.024) & (0.024)\\
				\tabularnewline   
				Delta Airlines &  & 0.115 &  0.115  \\
				&		& (0.024)& (0.024) \\
				\tabularnewline   
				United Airlines &  & 0.000 & -0.009  \\
				& &(0.022) &(0.020) \\
				\\
				\hline 
				\multicolumn{1}{c}{} & &  &  &  &\\
			\end{tabular}
		}
		\par\end{centering}
	\parbox{\textwidth}{\footnotesize
		\textit{Notes}: We report the estimated counterfactual changes to the entry of major airlines when Southwest Airlines no longer threatens entry into such markets. The effect presented here is due only to observable (non-policy) characteristics, complementing the policy results from Table \ref{southwest_decomp}. The specification follows the one for Table \ref{southwest_decomp}, while estimation/inference is described above. Column 1 uses a linear model, while Column 2 reports a nonparametric estimate. Standard errors for these columns are computed by the bootstrap, following the approach in the Online Appendix with $B=999$ replications.}\bigskip
\end{table}\bigskip

\subsubsection{Additional Results Referenced in the Main Text}

Table \ref{ct_number} provides an additional set of results referenced in Section \ref{subsec:Southwest} in the main text. It reports the effects from decreasing the number of competitors threatening entry on each airline's entry decision. There are three such exercises, reflecting the heterogeneous effects depending on the number of airlines currently threatening entry. This complements the main results presented in Table \ref{ct_decomp}.

\begin{table}[!h]
	\begin{centering}
		\small
		\caption{\small How Decreasing the Number of Competitors Threatening Entry Affects Airlines' Behavior}
		\label{ct_number}
		\resizebox{\columnwidth}{!}{%
			\begin{tabular}{cccccccc}
				\hline 
				\hline 
				\tabularnewline
				& &\multicolumn{6}{c}{Policy: Decreasing Number of Competitors Threatening Entry by 1} \\
				\\
				\cline{2-8}
				\tabularnewline
				&   & \multicolumn{2}{c}{Decrease from 1 to 0} & \multicolumn{2}{c}{Decrease from 2 to 1} & \multicolumn{2}{c}{Decrease from 3 to 2} \\
				\tabularnewline
				&   &  Linear & Nonparametric &  Linear & Nonparametric &  Linear & Nonparametric \\
				\tabularnewline
				\hline
\tabularnewline
  			American Airlines & &-0.026 &0.071 & -0.025 & -0.248 & -0.063 & -0.094 \\       
& &(0.016) & (0.042) & (0.015) & (0.028) & (0.015) & (0.022) \\
				\tabularnewline 
				Delta Airlines& & -0.021  & 0.153 &   -0.020 & 0.088 & -0.041 & -0.129 \\
				&	&(0.015)	& (0.026)& (0.012) & (0.028) & (0.013) & (0.023)\\
				\tabularnewline   
				Southwest Airlines& & 0.002  &  -0.029 & 0.041 & 0.020 & 0.005 & -0.039   \\
				&		&(0.008)& (0.038)& (0.009) & (0.024) & (0.006) & (0.013)\\
				\tabularnewline   
				United Airlines & &0.012 & -0.024 & 0.007 & -0.068 & -0.016 & 0.010\\
				& &(0.017) &(0.019) &(0.012) & (0.035) & (0.014) & (0.018) \\
				\\
				\hline 
				\multicolumn{1}{c}{} & &  &  &  &\\
			\end{tabular}
		}
		\par\end{centering}
	\parbox{\textwidth}{\footnotesize
		\textit{Notes}: We report the estimated counterfactual changes to the entry of major airlines after the number of competitors threatening entry is decreased by one. There are three such exercises: reducing the number of potential entrants in markets that had three such competitors to two; those that had two to one; and those that had one potential entrant to having none. The results across exercises are not directly comparable, as they are based on different markets. Following Table \ref{ct_decomp}, we present results for linear and nonparametric specifications. Standard errors for these columns are computed by the bootstrap, following the approach in the Online Appendix with $B=999$ replications.}\bigskip
\end{table}\bigskip

\FloatBarrier
\clearpage

\section{Example of a Game that Fails the Decomposition-Based Approach}
\label{App:Examples that Fail the Decomposition-Based Predictions}

Our main result that shows the validity of the decomposition-based prediction for counterfactual predictions requires that the policy changes only the publicly observable component of the payoff state. We give an example where the decomposition-based prediction fails when this requirement is not met. Our example is based on the entry game of Section \ref{example1_detail} with two players, $i=1,2$, but with $W = (\varepsilon_1,\varepsilon_2)$, where $T_i = \varepsilon_i$, i.e., each payoff type $\varepsilon_i$ is private information. We assume that the payoffs are given by: with $i=1,2$,
\begin{align}
	\label{payoff3}
	u_i(y,\varepsilon_i) = y_i (\delta y_{-i} + \varepsilon_i), 
\end{align}
\noindent where $\varepsilon_i$ is standard normally distributed, $\varepsilon_1$ and $\varepsilon_2$ are independent, and $\delta <0$. The counterfactual policy changes the payoff state $\varepsilon_i$ to $f(\varepsilon_i)=\alpha+\varepsilon_i$ for $\alpha>0$. Thus, this policy violates Assumption \ref{assump: information structure and policy} in the main text, because the policy changes the private information.

We consider a symmetric threshold Bayes Nash equilibrium of the game, characterized by $i$'s strategy of entering the market if $\varepsilon_i>\overline{w}$, where the threshold $\overline{w}$ is such that firm $i$ is indifferent between entering and not entering the market. For simplicity, we will consider an equilibrium selection rule that always chooses this BNE in the pre-policy game or its analogue in the counterfactual. This means that the lack of validity of the decomposition-based prediction in this example is not driven by the equilibrium selection rule.  

\subsection{Characterization of the Threshold $\overline{w}$}

The equilibrium described above is defined by firm $i$ playing ``Enter'' if $\varepsilon_i>\overline{w}$ and ``Not Enter'' if and only if $\varepsilon_i\leq \overline{w}$, where the (symmetric) threshold $\overline{w}$ is such that firm $i$ is indifferent between entering and not entering the market. This is the BNE counterpart to the BCE in \cite{Magnolfi/Roncoroni:22:ReStud} (i.e., where a mediator suggests this strategy). However, we cannot use the same threshold $\overline{w}$ from their example because we assume that $\varepsilon_i$ follows a standard normal distribution instead of a uniform distribution.\footnote{We use a normal distribution assumption because it allows for our counterfactual $f(\varepsilon_i) = \alpha + \varepsilon_i$ to have the same support as $\varepsilon_i$, and to also follow a normal distribution.}  To make sure this equilibrium is well defined, we now show that $\overline{w}$ in our setting exists and that it is unique.

The thresholds $\{\overline{w}_i\}_{i=1,2}$ that equalize the expected utility from entering the market to that from not entering for firm $i$ are the solutions to: 
\begin{align*}
\overline{w}_{1} + \delta(1-\Phi(\overline{w}_{2})) = 0, \text{ and } 
\overline{w}_{2} + \delta(1-\Phi(\overline{w}_{1})) = 0,
\end{align*}
where $\Phi$ is the CDF of $N(0,1)$. As we seek a symmetric equilibrium, we guess a symmetric threshold $\overline{w}=\overline{w}_1=\overline{w}_2$ such that $\overline{w} + \delta(1-\Phi(\overline{w})) = 0$. The function $h(w) = w + \delta(1-\Phi(w))$ is continuous in $w$, negative for $w<0$ (since $\delta<0$) and positive for $w>-\delta$. A solution $\overline{w}$ such that $h(\overline{w})=0$ exists by the Intermediate Value Theorem. This threshold is unique because $h(\csdot)$ is strictly increasing, since $h'(w)=1-\delta \phi(w)>0$, as $\delta<0$ and $\phi(\csdot)>0$, where $\phi$ is the density function of $N(0,1)$.

\subsection{Failure of Incentive Compatibility in the Counterfactual Game}

The decomposition-based prediction of the entry decisions after the policy $\varepsilon_i \mapsto \varepsilon_i + \alpha$ extrapolates from the pre-policy game equilibrium strategy so that it predicts that firm $i$ enters if and only if $\varepsilon_i + \alpha > \overline{w}$. We show that this prediction cannot be a valid prediction in the counterfactual game.

The expected utility for $i$ in the post-policy game from following this strategy, conditional on the competitor also doing so, is given by:
\begin{align*}
1\{\varepsilon_i + \alpha > \overline{w}\}(\delta (1-\Phi(\overline{w}-\alpha))+ \varepsilon_i + \alpha).
\end{align*}

We now show that there exists a profitable deviation from this strategy for $i$ in the post-policy game. Consider the following possible deviation from the recommendation for firm $i$: ``Enter'' if $\varepsilon_i + \alpha> \overline{w} + \kappa$, where $\kappa$ is a constant such that $0< \kappa < -\delta(1-\Phi(\overline{w}-\alpha))-\overline{w}$.  Such a $\kappa$ exists because:
\begin{eqnarray*}
	0 = \overline{w}+\delta(1-\Phi(\overline{w}))>\overline{w}+\delta(1-\Phi(\overline{w}-\alpha)),
\end{eqnarray*} 
where the inequality holds because the map $(\overline w,a) \mapsto \overline w+\delta(1-\Phi(\overline{w}-a))$ is strictly decreasing in $a$, and $\alpha>0$.

The difference in expected utilities from deviating relative to following the recommendation is given by:
\begin{eqnarray}
\left(1\{\varepsilon_i + \alpha > \overline{w} + \kappa\} - 1\{\varepsilon_i + \alpha > \overline{w}\}\right) (\delta (1-\Phi(\overline{w}-\alpha))+\varepsilon_i+\alpha). \label{diff_util}
\end{eqnarray}
This difference is equal to 0 except for the event $\{\overline{w} < \varepsilon_i + \alpha < \overline{w}+\kappa\}$, in which case the first term of (\ref{diff_util}) is equal to $-1$. We focus on this subset.  Then, deviating to the alternative strategy is a profitable deviation as long as $\delta(1-\Phi(\overline{w}-\alpha))+\varepsilon_i+\alpha<0$. This is true if and only if
\begin{eqnarray*}
\varepsilon_i+\alpha<-\delta (1-\Phi(\overline{w}-\alpha)).
\end{eqnarray*}
However, by the assumption on $\kappa$ above, we have that 
\begin{eqnarray}
\overline{w}+\kappa< -\delta(1-\Phi(\overline{w}-\alpha)). \label{eps3}
\end{eqnarray}
But if $\overline{w}< \varepsilon_i + \alpha < \overline{w}+\kappa$, using equation (\ref{eps3}), we find that
\begin{eqnarray*}
\varepsilon_i + \alpha < \overline{w}+\kappa<-\delta (1-\Phi(\overline{w}-\alpha)).
\end{eqnarray*}
This is exactly the expression required for there to be a profitable deviation for player $i$. It follows that the extrapolated strategy profile (i.e., entering if and only if $\varepsilon_i + \alpha > \overline{w}$) is not a BNE in the post-policy game.

It remains to compare the two predictions themselves. The post-policy game has a unique symmetric threshold BNE by the argument of the previous subsection: its threshold $c^*$ solves
\begin{align*}
	h_\alpha(c) = c + \delta(1-\Phi(c-\alpha)) = 0,
\end{align*}
where $h_\alpha$ is continuous and strictly increasing, since $h_\alpha'(c) = 1 - \delta \phi(c-\alpha) > 0$, negative for $c \le 0$ and positive for $c > -\delta$, so that $c^*$ exists and is unique by the Intermediate Value Theorem. Moreover, $h_\alpha(\overline{w}) = \overline{w} + \delta(1-\Phi(\overline{w}-\alpha)) < 0$ by the strict inequality displayed above, so $c^* > \overline{w}$. The equilibrium selection rule fixed at the beginning of this section chooses this BNE in the post-policy game, so the equilibrium-based prediction of firm $i$'s entry probability is $1 - \Phi(c^* - \alpha)$, whereas the decomposition-based prediction is $1 - \Phi(\overline{w} - \alpha)$. Since $c^* > \overline{w}$ and $\Phi$ is strictly increasing,
\begin{align*}
	1 - \Phi(c^* - \alpha) < 1 - \Phi(\overline{w} - \alpha),
\end{align*}
so the two predictions differ: the decomposition-based prediction strictly overstates the post-policy entry probability.

\newpage
\putbib[counterfactual2]
\end{bibunit} 
\end{document}